\documentclass[useAMS,usenatbib]{mnras}
\usepackage{multirow}
\usepackage{rotating}
\usepackage{graphicx}	
\usepackage{rotate}
\usepackage{amsmath}
\usepackage{amssymb}
\usepackage{wasysym}
\usepackage{color}

\newcommand{\ms}{\mathcal{M}_{\star}}

\DeclareMathOperator{\atantwo}{atan2}

\def\msun{\mbox{$M_\odot$}}

\def\lcdm{{$\Lambda$CDM}}

\def\mathnew{\mathsurround=0pt} 
\def\simov#1#2{\lower .5pt\vbox{\baselineskip0pt 
    \lineskip-.5pt\ialign{$\mathnew#1\hfil##\hfil$\crcr#2\crcr\sim\crcr}}}   
  
\def\'#1{\ifx#1i{\accent"13\i}\else{\accent"13#1}\fi}  

\title[Testing the fossil record method with mock observations from simulations]{Optical Integral Field Spectroscopy observations applied to simulated galaxies: Testing the fossil record method}
\author[Ibarra-Medel et al.]{H\'ector J. Ibarra-Medel$^{1}$\thanks{E-mail: hibarram@astro.unam.mx, hibarram@illinois.edu}\thanks{Current address: 1002 W Green St, Urbana, Illinois, 61801, USA.},  
Vladimir Avila-Reese,$^{1}$
Sebasti\'an F. S\'anchez,$^{1}$
\newauthor 
Alejandro Gonz\'alez-Samaniego,$^{2}$
Aldo Rodr\'iguez-Puebla.$^{1}$
\\$^{1}$Instituto de Astronom\'ia, Universidad Nacional Aut\'onoma de M{\'e}xico, A.P. 70-264, 04510 CDMX, M\'exico
\\$^{2}$Instituto de Radioastronom\'ia y Astrof\'isica, Universidad Nancional Aut\'onoma de M{\'e}xico, Box 70-3, Morelia, Michoac\'an, M{\'e}xico
}

\begin{document}


\pagerange{\pageref{firstpage}--\pageref{lastpage}} \pubyear{2018}

\maketitle

\begin{abstract} 
By means of post-processed cosmological Hydrodynamics simulations we explore the ability of the fossil record method to recover the stellar mass, age gradients, and global/radial star formation and mass growth histories of galaxies observed with an optical Integral Field Unity. We use two simulations of representative Milky Way-sized galaxies, one disk dominated and another bulge-dominated. We generate data cubes emulating MaNGA/SDSS-IV instrumental and observational conditions, and analyze them using the {\sc Pipe3D} pipeline. For optimal MaNGA instrumental/observational setups, the global masses, ages, and dust extinctions of both galaxies are well recovered within the uncertainties, while for the radial distributions, the trend is to get slightly flatter age gradients and weaker inside-out growth modes than the true ones. The main bias is in recovering younger stellar populations in the inner, older regions, and slightly older stellar populations in the outer, younger regions, as well as lower global stellar masses. These trends are enhanced as lower is the spatial sampling/resolution (less number of fibers), more inclined and dust-attenuated is the observed galaxy, lower is the signal-to-noise ratio, and worse is the seeing. The recovered properties and histories are more affected for the disk-dominated galaxy with a prominent inside-out growth mode than for the bulge-dominated galaxy assembled earlier and more coherently. We make publicly available the code and the mock IFU datacubes used in this paper.
\end{abstract}


\section{Introduction}
Integral field spectroscopy (IFS) has largely improved our ability to study galaxies. The information provided by this observational technique allows us to spatially dissect galaxies, and study the local and global properties of their stellar populations and ionized gas, as well as the stellar/ionized gas kinematics \citep[see ][for a recent review]{Sanchez+215IAU}. In such a way, we may learn, for instance, about the local/global star formation \citep[][]{Cano-Diaz+2016,Gonzalez-Delgado:2016aa, Hsieh+2017,Rowlands+2018,Pan+2018,lopfer18}, radial/global stellar mass growth \citep[e.g.][]{Perez+2013,Ibarra-Medel+2016,Garcia-Benito:2017aa,Gonzalez-Delgado:2017aa}, the radial dependency of the star formation rate \citep[e.g.][]{Gonzalez-Delgado:2016aa,Ellison:2018aa}, age/abundance gradients \citep[e.g.][]{GonzalezDelgado+2014,Sanchez:2014aa,Gonzalez-Delgado:2015aa,Goddard+2017}, the enrichment histories \citep[e.g.][]{Gonzalez-Delgado:2014ab,Tissera+2016}, the distribution of dust extinction of galaxies, as well as the dynamical mass distribution \citep[e.g.][]{Bershady+2010,Gonzalez-Delgado:2015aa,Aquino-Ortiz+2018} and the dynamical processes of galaxies \citep[e.g.][]{Cappellari+2012}.

In the last years, several optical IFS surveys of local galaxies were completed or are in the process of being completed \citep[e.g.,][]{Bacon:2001aa,Cappellari+2011,Croom:2012aa,Sanchez:2012aa,Cid-Fernandes:2013aa,Bundy+2015}. The most recent IFS surveys are the Calar Alto Legacy Integral Field Area survey \citep[CALIFA,][]{Sanchez:2012aa} that has observed $\approx 900$ galaxies, the Sydney-AAO Multi-object Integral field spectrograph survey \citep[SAMI,][]{Croom:2012aa} that has observed $\approx 2300$ galaxies, and the Mapping Nearby Galaxies at the Apache Point Observatory survey \citep[MaNGA,][]{Bundy+2015} that has observed $\approx 4800$ galaxies up to date with the goal of observing about $10000$ galaxies \citep{SDSS:2016aa}.

By applying the fossil record method to the optical IFS observations, one may reconstruct the local and global star formation histories (SFHs) or stellar mass growth histories (MGHs) of galaxies. This powerful method \citep[e.g,][]{Tinsley:1980aa,Buzzoni:1989aa,Bruzual:2003aa,Kauffmann:2003aa,Kauffmann:2003ab,Cid-Fernandes:2005aa,Gallazzi:2005aa,Tojeiro+2007,Walcher:2011aa} has been applied to the CALIFA \citep[][]{Perez+2013,Gonzalez-Delgado:2017aa,Garcia-Benito:2017aa,lopfer18} and MaNGA \citep[][]{Ibarra-Medel+2016,Goddard+2017,sanchez18a,sanchez18b} surveys to constrain the ``archaeological'' stellar MGHs in different radial bins along the galaxies as well as the age gradients. Thus, we can explore how galaxies assembled their stellar masses as a function of their radii.

The unique fossil record inferences from IFS observations of galaxies are, however, limited by many observational and methodological aspects. The radial archaeological MGHs can be affected, for instance, by the galaxy inclination, the number of spaxels that cover and resolve spatially the galaxy, the dithering and spatial binning procedures, the signal-to-noise ratio (SNR) of the spectra. On the other hand, the  optical spectra, even with high SNR, are degenerated for old stellar populations, making inferences of the early MGHs highly uncertain \citep[see for discussions and references,][]{Leitner2012,Ibarra-Medel+2016}. 

The aim of this paper is to explore how reliable are the fossil record method inferences of the radial/global MGHs and the age gradients, taking into account the several limiting aspects mentioned above. For this aim, we will use state-of-the-art N-body + Hydrodynamics simulations of two galaxies taken from the suite of zoom-in cosmological simulations presented in \citet{Colin+2016}. The MW-sized galaxies from this set present realistic properties and span a morphological range from disk- to spheroid-dominated ones. In \citet{Avila-Reese+2017}, the radial stellar mass assembly of the simulated galaxies were studied in detail. For the present work, we have chosen a disk-dominated galaxy, with an extended and highly inside-out formation history, and a bulge-dominated galaxy, with an early and more homogeneous formation history. The nominal spatial resolution of these simulations, 109$h^{-1}$ pc, is much higher than the typical resolutions attained by the observations to be emulated, $\sim 1-2$ kpc. 

By using the $z=0$ stellar particles and gas cells properties we post-process the data to assign spectro-photometric properties and dust extinction to the simulation. Then, from the post-processed simulation we generate datacubes in the optical bands (3800--10000 \AA) for different IFU bundle configurations, point spread function (PSF) dispersion, SNRs, and observe the simulated galaxies at different distances and inclinations. 
It is important to mention that the full diversity of IFU surveys/instruments (MaNGA, CALIFA, SAMI, MUSE to give some examples) use different bundle configurations or other IFU schemes (like lens arrays or image slicers) and dithering modes to reconstruct the spectral cubes.  In this paper, we focus on the MaNGA IFU configurations in specific. 
The obtained mock datacubes are analyzed with the specialized {\sc Pipe3D} software \citep{Sanchez:2016aa,Sanchez:2016ab} to infer the archaeological global/radial MGHs and the respective luminosity and mass weighted ages radial profiles, as well as the dust attenuation. This way we are able to study how well we can recover the global/radial mass assembly of the galaxies under several IFU setting and observational conditions, and using the same spectro-photometric analysis software as with real observations. Similar approaches have been previously presented by e.g., \citet[][and more references therein]{guidi18}. However, in here we describe in more detail the accuracy of the recovered parameters, making emphasis on the radial MGHs.

{\sc Pipe3D} is a pipeline optimized to analyze optical spectra from IFS observations for reconstructing the stellar populations and kinematics of galaxies. In the case of MaNGA it has been used to generate a Value Added Catalog for DR14 \citep{abol18}, presented in \citet{sanchez18a}. It is based on the {\sc FIT3D} fitting code,\footnote{\url{http://www.astroscu.unam.mx/~sfsanchez/FIT3D/}} which is able to fit both the stellar populations and emission lines of spectra of galaxies or regions of galaxies to recover their main properties. We should stress that {\sc Pipe3D} (and {\sc FIT3D}) is just one of several approaches to recover the properties of the stellar populations based on the inversion method. The most widely used code in the community is indeed {\sc Starlight} \citep[e.g.,][]{Cid-Fernandes:2005aa}. Besides this code there are several others, as can be seen in the compilation by J. Walcher.\footnote{\url{http://www.sedfitting.org/Fitting.html}}
In particular, in the case of the MaNGA data, there are other two major pipelines, {\sc FIREFLY} \citep{Goddard:2017aa} and
the official Data Analysis Pipeline (Westfall et al., in prep.). The later makes use of the new version of {\sc pPXF} \citep{cappe17}. 
It is beyond the scope of this article to make a comparison between the different codes. This goal has been addressed elsewhere by different authors \citep[e.g.][]{Sanchez:2016aa,guidi18}. In the current article we confront the recovered properties of the stellar populations by {\sc Pipe3D} against the true ones in mock galaxies obtained from cosmological simulations. Some of the conclusions maybe valid for all inversion methods, but other are surely attached to the particular adopted method with the considered set of single stellar populations (SSP), and for the wavelength range and instrumental setups of the MaNGA datasets. We will address this issue in future studies.

This paper is organized as follows. In section \ref{method} we describe our method: the numerical simulations to be used (\S\S \ref{simulation}), their post-processing for introducing light spectra and dust in them (\S\S \ref{postprocess} and Appendix \ref{post-processing}), the IFS datacube generation from different instrumental/observational settings (\S\S \ref{mock-obs}), and finally, the way the datacubes are processed according to the fossil record method (\S\S \ref{datacube}). In Section \ref{simulation-results} and Appendix \ref{simulations}, we present and compare among them the current and archaeological stellar radial MGHs as directly measured from the simulation at several inclinations. Section \ref{reconstruction} is devoted to explore how well are recovered from the observations, and further from the fossil record method, the mass and age maps (subsection \ref{maps}), the radial SFHs (subsection \ref{GLSFH}), the radial age and extinction profiles (subsection \ref{age-gradients}), and the global/radial MGHs (\ref{SubglobalMGHs} and \ref{radialMGHs}). We explore different instrumental and observational settings. In Section \ref{conclusions}, we summarize the results of this paper and provide our main conclusions. 

\section{The method}
 \label{method}


\begin{figure*}
\begin{center}
\includegraphics[width=0.45\linewidth]{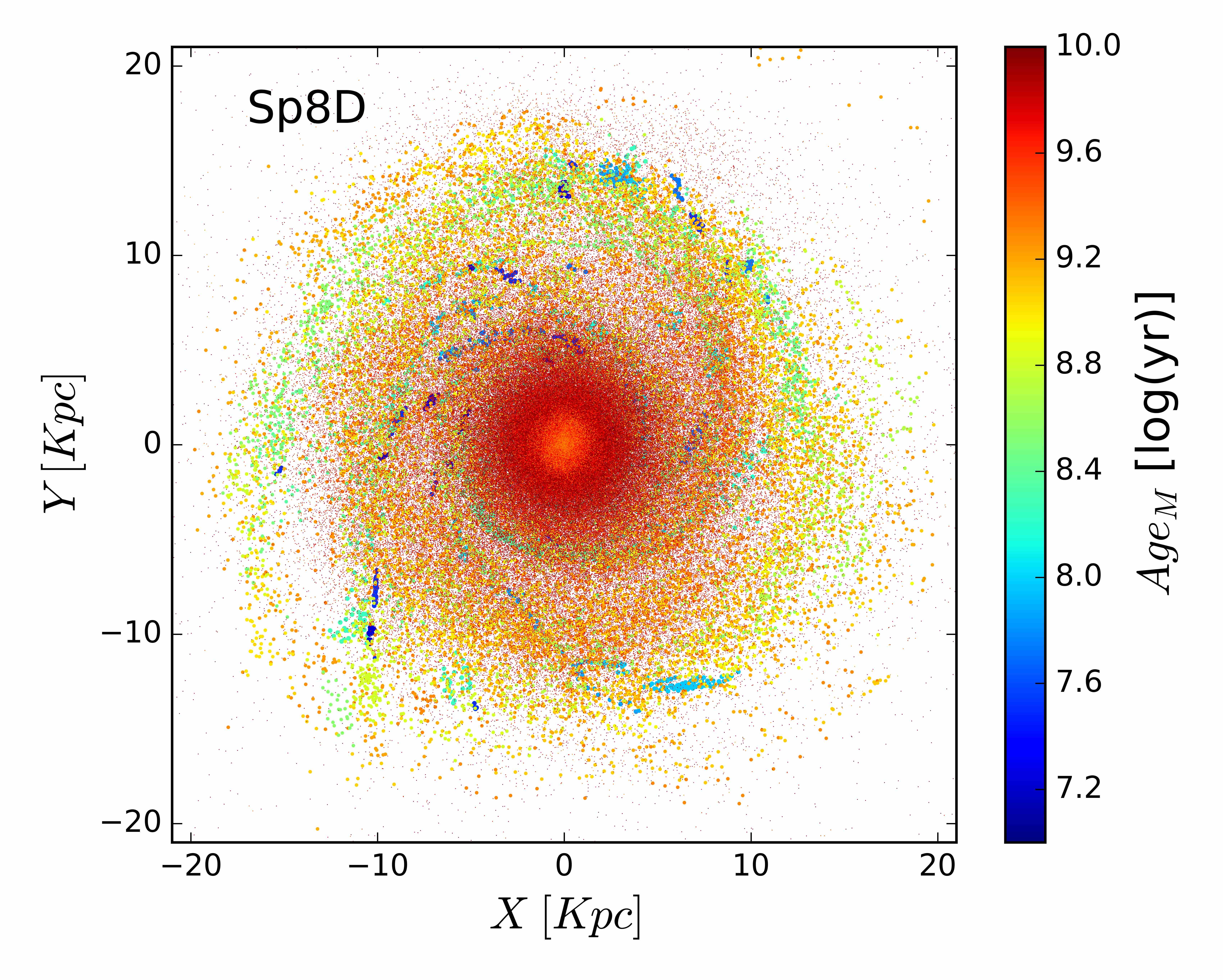}
\includegraphics[width=0.45\linewidth]{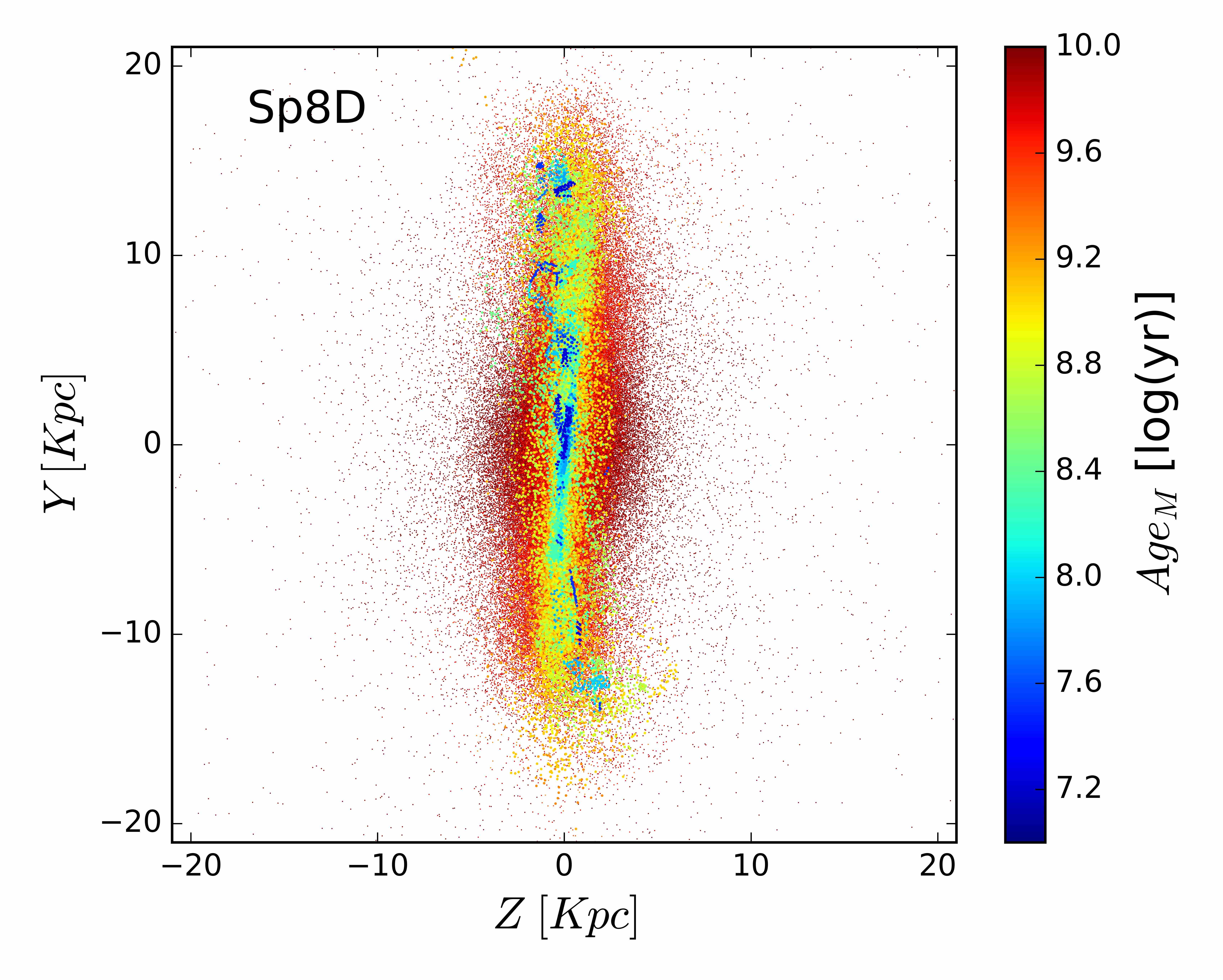}
\includegraphics[width=0.45\linewidth]{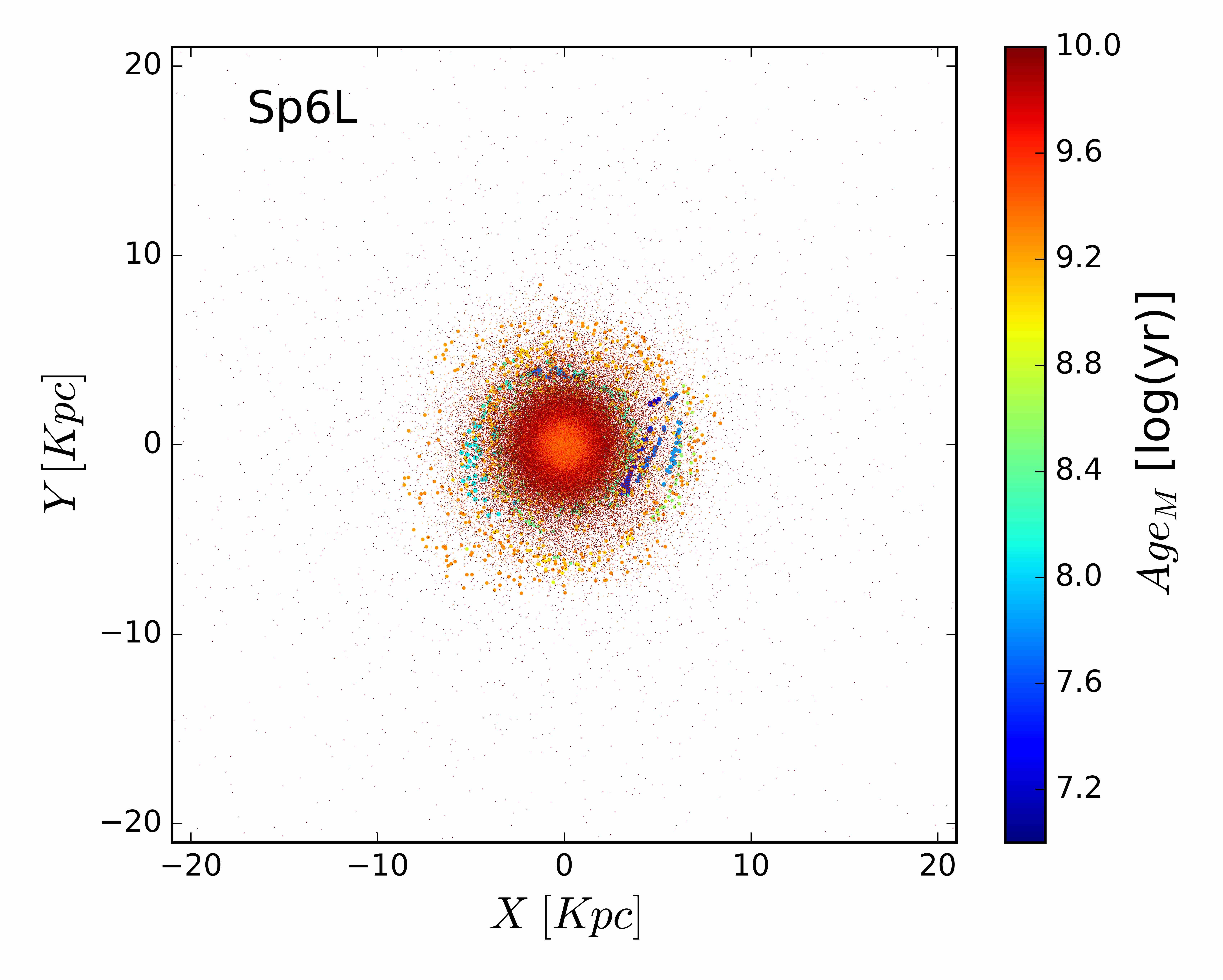}
\includegraphics[width=0.45\linewidth]{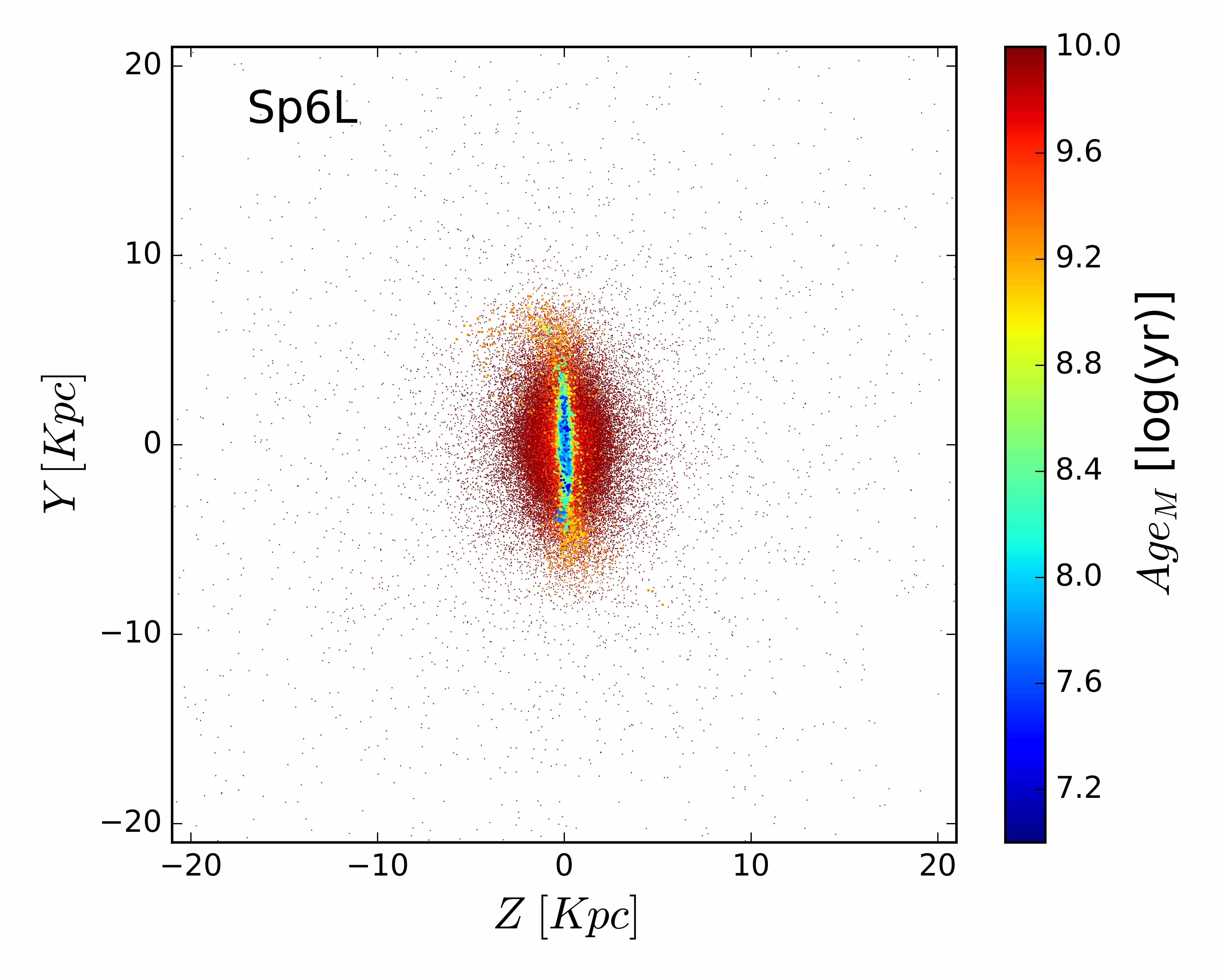}
\end{center}
\caption{Spatial stellar particle distribution of the simulated galaxies Sp8D and Sp6L at the final snapshot ($z=0$). The face-on and edge-on views are shown in the left and right panels, respectively. The color code represents the ages of the stellar particles. The size of the dots is larger when the number of particles is too small. 
}
\label{fig00}
\end{figure*}

\subsection{The simulations} 
\label{simulation}

For the main goal of this paper, we need well-resolved simulated galaxies. This is the case of the suite of eight zoom-in cosmological simulations presented in \citet[][]{Colin+2016} and \citet[][]{Avila-Reese+2017}. The simulations correspond to Milky Way-sized galaxies chosen to be in relatively isolated halos today. We use here two representative simulations from this set. The simulations were run using the N-body + Hydrodynamics Adaptive Refinement Tree (ART) code \citep[]{Kravtsov:1997aa,Kravtsov:2003aa,Kravtsov+2005}, which is an Adaptive Mesh Refinement code with Eulerian gasdynamics. ART incorporates a variety of physical processes such as gas cooling, star formation, stellar feedback, advection of metals, and a UV heating background source. In the computation of the cooling/heating rates were included Compton heating/cooling, atomic and molecular cooling, and UV heating from the cosmic background radiation \citep{HM96}.
Stellar particles are formed in gas cells with $T < 9000$ K and $n_g > 1$ cm$^{-1}$, where $T$ and $n_g$ are the temperature and number density of the gas, respectively. A stellar particle of mass $m_* = \epsilon_{\rm SF} m_g$ is placed in a grid cell every time the above conditions are simultaneously satisfied, where $m_g$ is the gas mass in the cell and $\epsilon_{\rm SF}$ is a parameter set to $0.65$. This value and those of the other subgrid parameters were found in \citet[][see also \citealp{Avila-Reese+2011} and \citealp{Roca-Fabrega:2016aa}]{Colin+2016} to be optimal for the given minimal spatial resolution (109 $h^{-1}$pc) and the integration time steps attained in their simulations. 

An ``explosive'' stellar thermal feedback recipe was used. Each stellar particle deposits into its parent cell $E_{{\rm SN+Wind}} = 2 \times 10^{51}$ erg of thermal energy for each star more massive than $M_\star= 8$ \msun\ (half of this energy is assumed to come from the type-II SNe and half from the shocked stellar winds) and a fraction $f_Z=$min(0.2,$0.01M_\star -0.06$) of their mass as metals \citep{Kravtsov:2003aa}.  
The code accounts also for SN Ia feedback, assuming a rate that slowly increases with time and broadly peaks at the population age of 1 Gyr. Each SN Ia  ejects 1.3 \msun\ of metals into the parent cell. For the assumed \citet{MS79} IMF ($M_\star$ in the range 0.1--100 \msun), a stellar particle of $10^5\ \msun$ produces 749 SNe II (instantly) and 110 SNe Ia (over several billion years). On the other hand, stellar particles return a fraction of their mass to the surrounding (stellar mass loss).

\begin{table*}
\begin{center}
\caption{Global properties (within 1.5 $R_{eff}$) of the Sp8D and Sp6L simulations at $z=0$.}\label{tab1}
\resizebox{\textwidth}{!}{
\begin{tabular*}{0.77\textwidth}{@{\extracolsep{\fill}} c c c c c c c c c c }
\hline\hline
Galaxy & $M_{\rm vir}$ & $\ms$ & $f_{g,cold}$$^a$ & $V_{max}$ & $SFR$  & $R_{1/2}$ &  $R_{eff}$$^b$ & $Age_{M}$  & $Age_{L}$ \\
 & $[10^{12}M_{\odot}]$ & $[10^{10}M_{\odot}]$ &    & [km/s] & $[M_{\odot}/yr]$ & [kpc] & [kpc] & [Gyr] & [Gyr] \\
\hline 
Sp8D & $1.20$ & $6.23$ & $0.18$ & $219.7$ & $1.39$ & $5.9$ & $7.5$ & $6.83$ & $2.31$\\
Sp6L & $0.97$ & $2.27$ & $0.15$  & $197.2$ & $0.32$ & $2.6$ & $3.2$ & $9.18$ & $3.78$\\
\hline
\end{tabular*}}
\end{center}
{$^a$ $f_{g,cold}= M_{\rm g,cold}/(\ms + M_{\rm g,cold})$, where $M_{\rm g,cold}$ is the the cold gas ($T<1.5\times 10^4$ K) mass.}\\
{$^b$ $V$-band face-on effective radius.}
\end{table*}

The simulations were performed within the $\Lambda$ Cold Dark Matter cosmology (\lcdm) with $\Omega_M = 0.3$, $\Omega_\Lambda = 0.7$, and $\Omega_B = 0.045$. The environment of the simulated galaxies can be related to what observers call the field. The maximum level of refinement was set to 12 so that the high density regions are mostly filled, at $z=0$, with cells of 109 $h^{-1}$pc per side. 
The number of dark matter particles in the high-resolution zone, where galaxies are located, ranges from about 1.5 to 2 million and the mass per particle $m_p$ is $1.02 \times 10^6$ h$^{-1}$\msun. The numbers of stellar particles in the galaxies are of several hundreds of thousands at $z=0$. 

\citet{Colin+2016} presented the main properties of the eight MW-sized simulated galaxies. The galaxies are only slightly above the $V_{\rm max}-\ms$ empirical relation but mostly within the intrinsic scatter around this relation. Their half-mass radii and cold gas fractions are also in agreement with observations given their masses. The rotation curves of all the galaxies are nearly flat. The stellar-to-halo mass ratios at $z=0$ are roughly in agreement (within the 1$\sigma$ scatter) with empirical and semi-empirical determinations. In \citet{Colin+2016}, the kinematic bulge-to-total mass ratio at $z=0$ was determined by means of two methods. The results are similar in both cases. From these results, the eight simulated galaxies can be classified as: disk dominated (4), lenticular like (2), and completely spheroid-dominated objects (2). 

For this paper, we have chosen two galaxies, one disk-dominated (run Sp8D, disk-to-total ratio at $z=0$ of $0.65$) and one lenticular like (run Sp6L, disk-to-total ratio at $z=0$ of $0.39$); the letters D and L stand for ``disk-dominated'' and ``lenticular-like'', respectively. These galaxies present different global and radial stellar mass growth regimes, one extended in time with a strong inside-out assembly (Sp8D) and the other one concentrated at early time and with a weak inside-out assembly (Sp6L). It is important to have these two different regimes to probe the fossil record method inferences in both cases.  Broadly speaking, Sp8D and Sp6L are representative of late- and early-type galaxies, respectively. Note also that the spatial resolution of the simulations is high enough as to study the resolved properties of the galaxias as observed with the MaNGA instrumental/observational setups. 

The main properties of the two selected galaxies are listed in Table~\ref{tab1}. Figure \ref{fig00} shows the stellar particle distribution of the Sp8D and Sp6L simulations at $z=0$, both face-on and edge-on. The color code indicates the age of the stellar particles. In run Sp8D there is a clear age gradient both radial and vertical, while in the Sp6L run the age gradient is more evident in the vertical direction. Note that in the innermost regions there is a mix of old and young stellar populations. In fact, the innermost region of both galaxies is almost at any epoch the most active in forming stars across the galaxy. There is a very young stellar population in the spiral arms located in the plane of the disks, specially for run Sp8D.

\subsection{Post-processing of the simulation}
\label{postprocess}

To construct the mock observations we use the last snapshot at $z=0$ from the Sp8D and Sp6L simulations. The information we use consists of: the age and initial and final masses of each stellar particle, their metallicities, positions and velocities, as well as the mass, number density, temperature, metallicity, position and velocity of each gas cell. With this information, to each stellar particle is assigned an spectrum of a SSP from a stellar library, and the synthetic observed spectrum is constructed within a given solid angle $\Delta \Omega$ towards the observer line of sight (hereafter LOS). It is important to remark that for the SSPs, we use the same stellar library and IMF employed in the {\sc Pipe3D} code \citep{Sanchez:2016aa,Sanchez:2016ab}. In this way {\it we avoid the introduction of an extra uncertainty due to the use of different stellar population libraries and IMFs for the assignation of spectra to the stellar particles and the recovery of the stellar mass from the spectra.}
To the stellar synthetic spectrum we add dust attenuation constructed with a simple prescription based on the properties of the gas cells. We also construct the gas emission spectrum within $\Delta \Omega$ towards the LOS using the gas and stellar properties within each cell. The final synthetic spectra take into account the effects of galaxy kinematics and cosmological redshift.  

A detailed description of the post-processing and construction of the final observed stellar synthetic spectrum towards each LOS within $\Delta \Omega$, $F_{\alpha,\delta,s}(\lambda)$, is given in Appendix \ref{post-processing} (Eqs. \ref{stellar_only} and \ref{stellar_ext} without and with extinction, respectively); $\alpha$ and $\delta$ are the observed angular projection positions.  The same for the gas emission spectrum, $F_{\alpha,\delta,g}(\lambda)$ (Eq. \ref{gas_ext}). We show in that Appendix the reconstructed photometric ultraviolet, optical, and infrared views of our Sp8D and Sp6L galaxies, without taking into account extinction (figures \ref{photo_sp8} and \ref{photo_sp6}).


\begin{table*}
\begin{center}
\caption{Instrumental/observational conditions and recovered properties of the Sp8D and Sp6L mock observations.}\label{tab2}
\resizebox{\textwidth}{!}{
\begin{tabular*}{1.1\textwidth}{@{\extracolsep{\fill}} c c c c c c c c c c c c c c c}
\hline\hline
(1) & (2) & (3) & (4) & (5) & (6) & (7) & (8) & (9) & (10) & (11) & (12) & (13) & (14) & (15)\\
Galaxy & $\theta ^a$ & \# Fibers & $\sigma_{lim}$$^b$ & $PSF$$^c$ & $R_{eff}$$^d$  & $\Delta M_{0}$$^e$  & $\langle SNR \rangle^f$  & $PSF_{rec}^g$ & Res$^h$ & $\Delta Age_{M}^i$ & $\Delta Age_{L}^j$ & $\langle A_V\rangle_{P}$$^k$ & $\langle A_V\rangle_{O}$$^l$ & $\langle A_V\rangle_{S}$$^m$ \\
 & $[^{\circ}]$ &  & $[\mu_{AB}]$ & $[arcsec]$ & $[kpc]$ & $[dex]$ & & $[arcsec]$ & $[kpc]$ & $[dex]$ & $[dex]$ & $[mag]$ & $[mag]$ & $[mag]$ \\
\hline 
Sp8D & $ 0.0$ & $547$ & $26.83$ & $1.43$ & $7.50$ & $-0.09$ & $25$ & $2.21$ & $1.00$ & $-0.01$ & $-0.04$ & $0.34$ & $0.20$ & $0.30$ \\
{\bf Sp8D} & {\bf  0.0} & {\bf 127} & {\bf 26.83} & {\bf 1.43} & {\bf 7.50} & {\bf -0.15} & {\bf 24} & {\bf 2.21} & {\bf 2.14} & {\bf -0.01} & {\bf -0.03} & {\bf 0.26} & {\bf 0.19} & {\bf 0.30} \\
Sp8D & $ 0.0$ & $ 61$ & $26.83$ & $1.43$ & $7.50$ & $-0.21$ & $22$ & $2.21$ & $3.00$ & $-0.10$ & $-0.09$ & $0.31$ & $0.19$ & $0.30$\\
Sp8D & $ 0.0$ & $ 19$ & $26.83$ & $1.43$ & $7.50$ & $-0.30$ & $22$ & $2.21$ & $4.60$ & $-0.14$ & $-0.09$ & $0.29$ & $0.17$ & $0.30$\\
Sp8D & $45.0$ & $127$ & $26.83$ & $1.43$ & $6.99$ & $-0.19$ & $25$ & $2.21$ & $2.14$ & $0.003$ & $0.003$ & $0.26$ & $0.20$ & $0.36$\\
Sp8D & $75.0$ & $127$ & $26.83$ & $1.43$ & $7.08$ & $-0.28$ & $19$ & $2.21$ & $2.14$ & $+0.01$ & $+0.07$ & $0.35$ & $0.39$ & $0.72$\\
Sp8D & $90.0$ & $127$ & $26.83$ & $1.43$ & $6.85$ & $-0.35$ & $15$ & $2.21$ & $2.14$ & $+0.04$ & $+0.09$ & $0.44$ & $0.48$ & $1.05$\\
Sp8D & $ 0.0$ & $127$ & $27.83$ & $1.43$ & $7.50$ & $-0.15$ & $81$ & $2.21$ & $2.14$ & $-0.03$ & $-0.04$ & $0.30$ & $0.20$ & $0.30$\\
Sp8D & $ 0.0$ & $127$ & $23.00$ & $1.43$ & $7.50$ & $-0.20$ & $ 5$ & $2.21$ & $2.14$ & $-0.14$ & $-0.10$ & $0.30$ & $0.20$ & $0.30$ \\
Sp8D & $ 0.0$ & $127$ & $26.83$ & $0.20$ & $7.50$ & $-0.14$ & $24$ & $1.99$ & $1.94$ & $-0.02$ & $-0.04$ & $0.29$ & $0.20$ & $0.30$\\
Sp8D & $ 0.0$ & $127$ & $26.83$ & $2.86$ & $7.50$ & $-0.17$ & $24$ & $3.15$ & $3.07$ & $-0.07$ & $-0.08$ & $0.31$ & $0.20$ & $0.30$\\
\hline
Sp6L & $ 0.0$ & $547$ & $26.83$ & $1.43$ & $3.20$ & $-0.11$ & $31$ & $2.21$ & $0.50$ & $-0.004$ & $-0.07$ & $0.31$ & $0.27$ & $0.40$\\
{\bf Sp6L} &  {\bf 0.0} & {\bf 127} & {\bf 26.83} & {\bf 1.43} & {\bf 3.20} & {\bf -0.10} & {\bf 27} & {\bf 2.21} & {\bf 0.95} & {\bf 0.007} & {\bf -0.04} & {\bf 0.32} & {\bf 0.25} & {\bf 0.40}\\
Sp6L & $ 0.0$ & $ 61$ & $26.83$ & $1.43$ & $3.20$ & $-0.09$ & $29$ & $2.21$ & $1.23$ & $ -0.01$ & $-0.08$ & $0.34$ & $0.24$ & $0.40$\\
Sp6L & $ 0.0$ & $ 19$ & $26.83$ & $1.43$ & $3.20$ & $-0.08$ & $28$ & $2.21$ & $1.81$ & $-0.005$ & $-0.08$ & $0.31$ & $0.23$ & $0.40$\\
Sp6L & $45.0$ & $127$ & $26.83$ & $1.43$ & $3.08$ & $-0.12$ & $27$ & $2.21$ & $0.95$ & $ +0.01$ & $-0.09$ & $0.29$ & $0.27$ & $0.45$\\
Sp6L & $75.0$ & $127$ & $26.83$ & $1.43$ & $3.29$ & $-0.19$ & $22$ & $2.21$ & $0.95$ & $ +0.02$ & $-0.01$ & $0.38$ & $0.31$ & $0.70$\\
Sp6L & $90.0$ & $127$ & $26.83$ & $1.43$ & $3.33$ & $-0.22$ & $20$ & $2.21$ & $0.95$ & $ +0.03$ & $+0.18$ & $0.32$ & $0.32$ & $0.82$\\
\hline
\end{tabular*}}
\end{center}
{\raggedright \textbf{Notes.}}
{The fiducial instrumental/observational setting is shown with bold face.}
{ $^a$ Angle with respect to the galaxy rotation axis.}
{ $^b$ RMS of the noise.}
{ $^c$ PSF size of the atmospheric seeing.}
{ $^d$ $V$-band half-light radius in kpc.}
{ $^e$ Logarithm of the Pipe3D-to-True mass ratio at $z=0$. }
{ $^f$ Effective Signal-to-Noise Ratio observed within the $V$-band.}
{ $^g$ Reconstructed PSF size of the IFU mock observation.}
{ $^h$ Effective spatial resolution in kpc within the reconstructed PSF.}
{ $^i$ Logarithm of the Pipe3D-to-True global MW age ratio. }
{ $^j$ Logarithm of the Pipe3D-to-True global LW age ratio.}
{ $^k$ Flux weighted average of the extinction recovered by Pipe3D within $1.5R_{eff}$. }
{ $^l$ Flux weighted average of the extinction from the "observed" simulation within $1.5R_{eff}$.}
{ $^m$ Light weighted average of the extinction from the simulation within $1.5R_{eff}$.}
\end{table*}

\begin{figure}
\begin{center}
\includegraphics[width=0.95\linewidth]{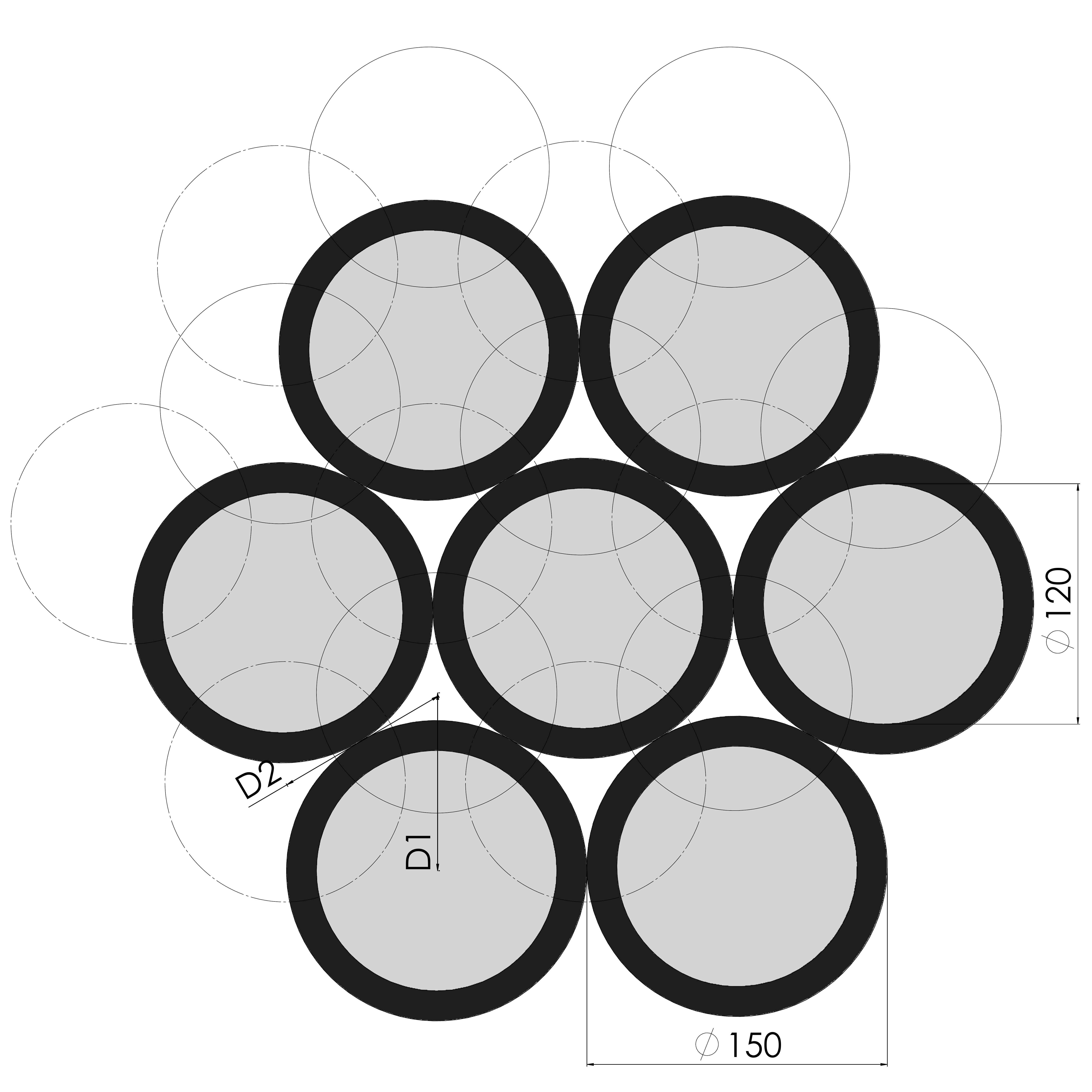}
\end{center}
\caption{IFU-bundle hexagonal configuration. The core fibers are represented with the light gray circles and the coating body with the black rings. The light-gray circles show the two extra dithering positions, following a track defined by D1 and D2. For the case of MaNGA, the core diameter is $120\mu m$, the coating diameter is $150\mu m$, and the angle between D1 and D2 is about $60^{\circ}$.}
\label{ifu}
\end{figure}

\begin{figure*}
\begin{center}
\includegraphics[width=0.23\linewidth]{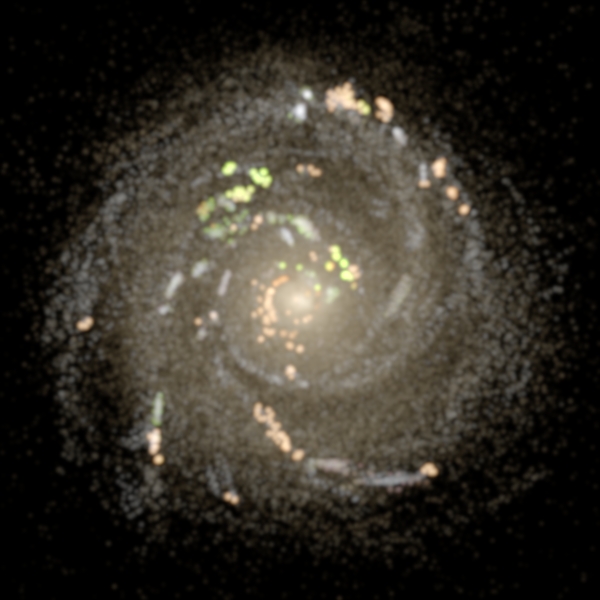}
\includegraphics[width=0.23\linewidth]{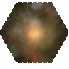}
\includegraphics[width=0.23\linewidth]{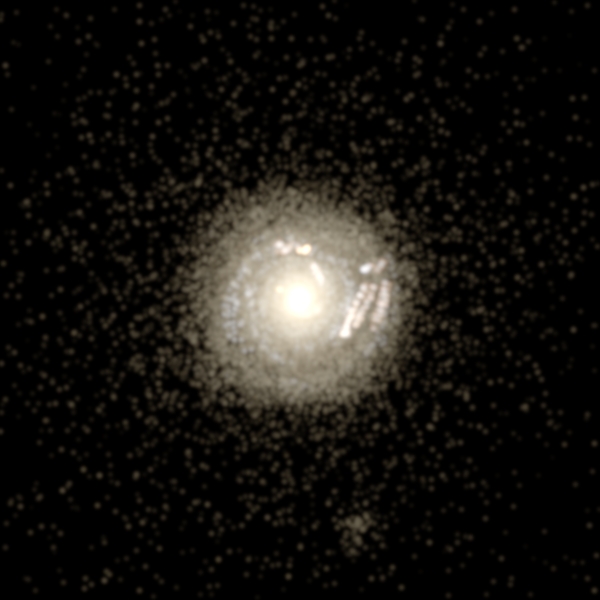}
\includegraphics[width=0.23\linewidth]{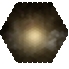}
\end{center}
\caption{Examples of mock photometric and IFU observations. {\it Left panel:} Reconstructed RGB (SDSS $g,r,i$ bands) photometric image of the face-on Sp8D galaxy. A modeled PSF (sky seeing) of 0.2\ arcsec and a spatial resolution of $0.97$ kpc/arcsec were used. {\it Middle left panel:} Reconstructed RGB (SDSS $g,r,i$ bands) from a MaNGA type IFU-cube of 127 fibers "observed" with the average APO PSF (sky seeing) of $1.43$\ arcsec and a spatial resolution of $0.97$ kpc/arcsec. The reconstructed PSF size is of $2.21$ arcsec. {Middle Right and Right panels:} As left and middle left panels but for the Sp6L galaxy. The spatial resolution for this case is $0.43$\ kpc/arcsec. In both cases, we imposed the condition that the galaxy filled the bundle up to 1.5 $R_{\rm eff}$.}
\label{cubes_photo}
\end{figure*}

\subsection{Mock observations}
\label{mock-obs}

After post-processing the simulation, we have the spectrum for each stellar particle, the properties of the spectral lines of the photoionized gas around them, and the extinction at each gas cell. With this information, we proceed to construct a set of mock IFU observations from the simulated galaxies (Sp8D and Sp6L) for any inclination, distance, atmospheric seeing, bundle configuration, SNR, PSF, and instrumental noise. Next, we describe the steps we follow to construct the mock photometric and IFS observations.

\begin{figure}
\begin{center}
\includegraphics[width=0.95\linewidth]{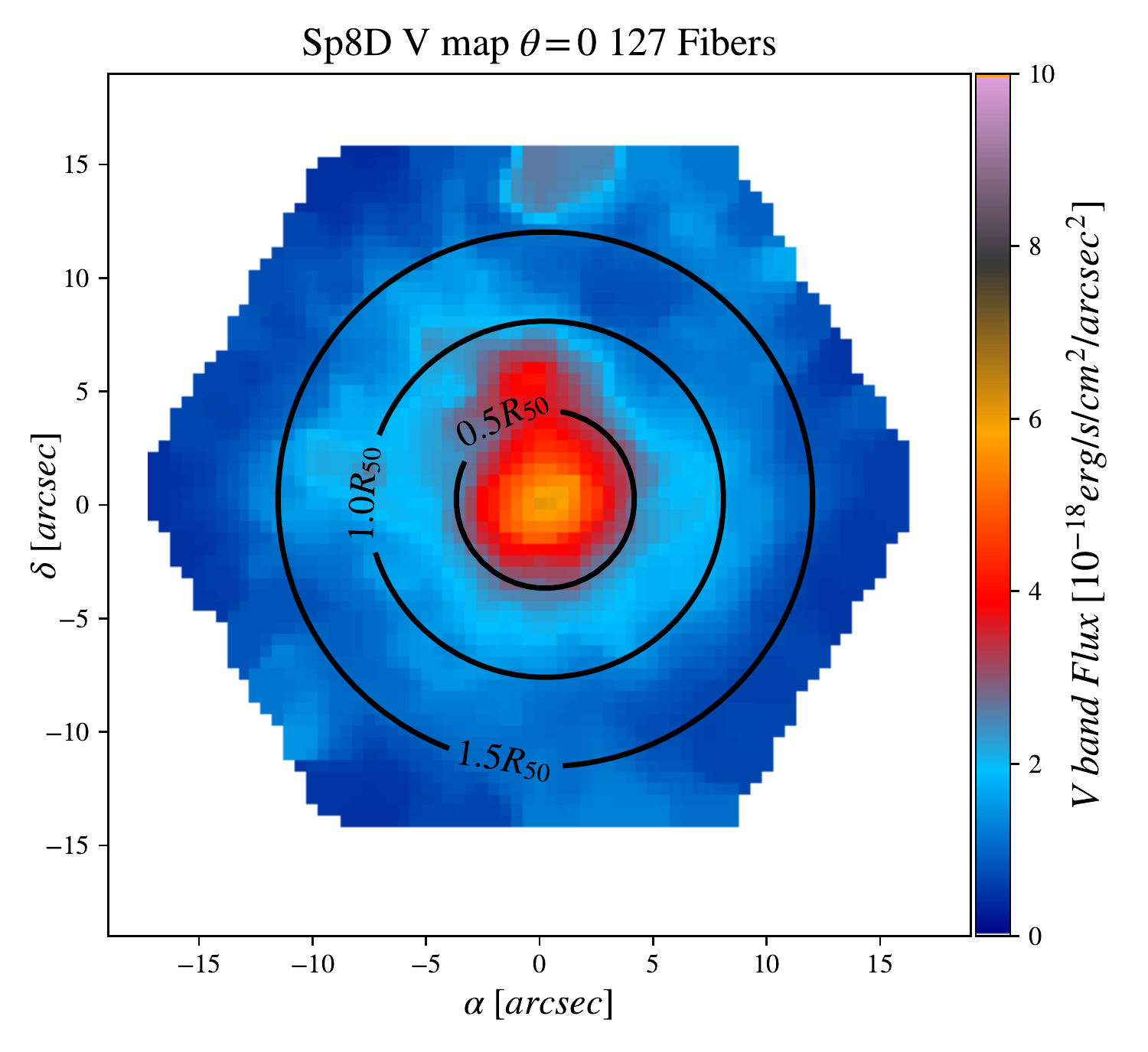}
\includegraphics[width=0.95\linewidth]{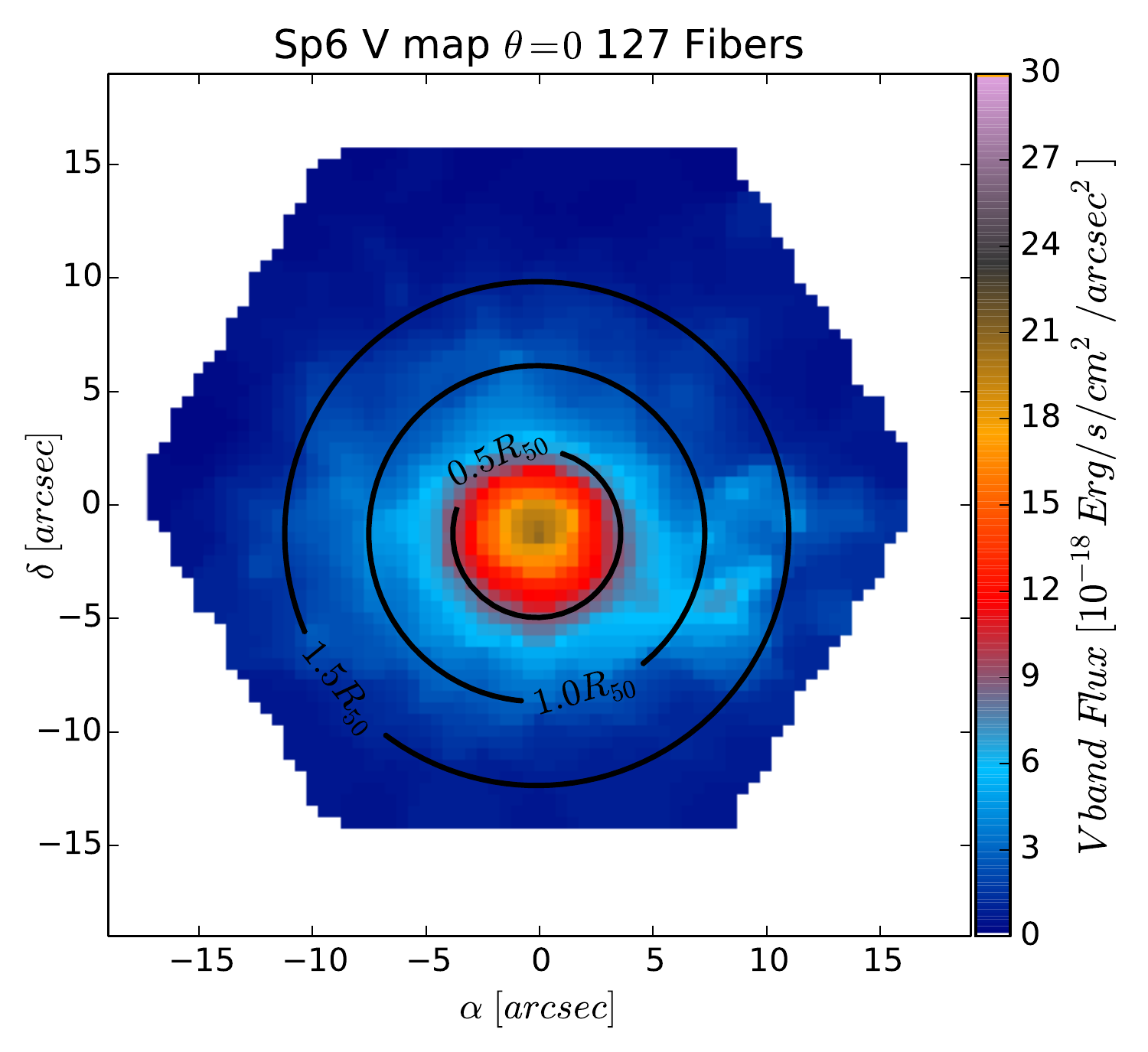}
\end{center}
\caption{Face-on ($\theta=0$) mock surface brightness maps from a MaNGA type IFU-cube of 127 fibers in the $V$ band for the Sp8D (top) and Sp6L (bottom) galaxies. We use a seeing PSF size of $1.43\ arcsec$ with a spatial resolution of 0.97\ kpc/arcsec for Sp8D and  0.43\ kpc/arcsec for Sp6L. Over-plotted are the three radial bins ($R/R_{\rm eff}<0.5$, $0.5<R/R_{\rm eff}<1$ and $1<R/R_{\rm eff}<1.5$) that we use in our study. All the IFU mock observations have a reconstructed PSF size of 2.21\ arcsec.}
\label{ifu_flux}
\end{figure}

\begin{description}
\item[{\bf 1.}]  First, we define $\Delta \Omega$ in Equations (\ref{stellar_ext}) and (\ref{gas_ext}) (see Appendix \ref{post-processing}) as the solid angle of the LOS at the position ($\alpha,\delta$) on the projected ``sky''. At this point, we need to take into account the observational effect of an imaginary atmospheric seeing, which we model as a point-spread-function (PSF) by defining the ``observed'' angular projected positions as a function of a normal random distribution, $\mathcal{N}$, with a dispersion equal to $\sigma_{PSF_{seeing}}$. Hence, to estimate the values of the stellar or gas spectrum we use the ``observed'' angular projection positions $(\alpha_{o},\delta_{o}) = \mathcal{N}(\alpha,\delta; \sigma_{PSF_{rec}})$ in Equations (\ref{stellar_ext}) and (\ref{gas_ext}).

\item[{\bf 2.}] Then, the mock spectrum at each location is calculated as the sum of the stellar spectra plus the photoionized gas spectra convolved with a Gaussian function, $\mathcal{N}(\lambda,\sigma_{\rm inst})$, that models the broadening of the instrumental dispersion\footnote{For all the mock observations in this paper, we fix the instrumental dispersion, $\sigma_{\rm inst}$, equal to 25.0 km/s.}, plus a random noise $N_{\alpha_{o},\delta_{o}}(\lambda)$: 
\begin{multline}
F_{\alpha_{o},\delta_{o}}(\lambda)=[F_{\alpha_{o},\delta_{o},s}(\lambda)+F_{\alpha_{o},\delta_{o},g}(\lambda)]*\mathcal{N}(\lambda,\sigma_{inst})\\ +N_{\alpha_{o},\delta_{o}}(\sigma_{lim},\lambda)+Nw(\lambda).
\label{full-spectrum}
\end{multline}
For only photometric mock observations, is not necessary to use $\mathcal{N}(\lambda,\sigma_{\rm inst})$. $N_{\alpha_{o},\delta_{o}}(\sigma_{lim},\lambda)$ is a random value that depends on the wavelength and is introduced to model the detector response function while $Nw(\lambda)$ models the detector reading noise. $N_{\alpha_{o},\delta_{o}}(\sigma_{lim},\lambda)$ simulates a detector more efficient in the optical wavelengths and less efficient in the ultraviolet or infrared ones, see Appendix \ref{noise} for more details. $N_{\alpha_{o},\delta_{o}}(\sigma_{lim},\lambda)$ mimics as best as possible the detector noise response function of a MaNGA IFU observation at the Apache Point Observatory (APO). This function is similar to the sensitivity spectrum presented by \citet{Law:2016aa}. Therefore, between $3600$ \AA~ and $10,300$\AA~ we define a root mean square (rms) noise, $\sigma_{lim}$, see Appendix \ref{noise}. Finally we define the SNR as $F_{V,\alpha_{o},\delta_{o}}/\sigma_{lim}$, where $F_{V,\alpha_{o},\delta_{o}}$ is the total flux of the observed mock galaxy within the Johnson V band filter, see below.

Now, using Equation (\ref{full-spectrum}), the total observed flux within a given photometric band, which has a transmission profile 
$T_{band}(\lambda)$, is:
\begin{equation}
F_{band,\alpha_{o},\delta_{o}}=\int_{\lambda_i}^{\lambda_f}F_{\alpha_{o},\delta_{o}}(\lambda)T_{band}(\lambda)d\lambda/\int_{\lambda_i}^{\lambda_f}T_{band}(\lambda)d\lambda.
\label{SB}
\end{equation} 
The photometric magnitude is then: $$m_{band,\alpha_{o},\delta_{o}}=-2.5\log_{10}(F_{band,\alpha_{o},\delta_{o}}/F_{zpoint}),$$ 
where $F_{zpoint}$ is the zero-point flux of the photometric system. In this work, we use the zero point of the $AB$ photometric system.

\item[{\bf 3.}] Next, we mimic the IFS observation with a given instrument, for instance, for the MaNGA bundles. To do so, we cover the galaxy $FOV$ with a configuration of bundles and dithering points as done with real observations. We assume an hexagonal bundle configuration, whose fibers  have a physical diameter of $150\mu m$ and a core diameter of $120\mu m$, as in the setup of MaNGA IFU \citep[see as an example Figure~\ref{ifu} for a setup of 7 fibers; see][for more details]{Bundy+2015,Law:2015aa}. We use the plate-scale of the SDSS telescope \citep[see][for details]{Gunn:2006aa}. Therefore, we use the fiber core aperture ($D_c=1.99$ arcsec in diameter) as $\Delta \Omega$ to integrate the total observed spectrum per fiber and redefine Equation~(\ref{full-spectrum}) as $F_{\alpha_{o},\delta_{o}}(\lambda)=F_{tot,i}(\lambda)$. On the other hand, the hexagonal bundle configuration has an incomplete covering factor of the FoV, with empty spaces between adjacent fibers in the bundle as illustrated in Figure~\ref{ifu}. As with the MaNGA observations, we emulate a set of three dither observations following the pattern of Figure~\ref{ifu}, to fill the unobserved sky areas. We also consider a pointing error of $0.025\ arcsec$ for each dither observation. 

\item[{\bf 4.}] Finally, to spatially re-arrange and reconstruct the final image, we use the same methodology employed in MaNGA \citep{Law:2016aa}, that is based on the interpolation method presented in \citet{Sanchez:2012aa}. This interpolation method defines the flux at each spaxel of the cube as $$F_{x,y}(\lambda)=\sum_i^{n_{fib}}w(r_i)F_{tot,i}(\lambda)/\sum_i^{n_{fib}}w(r_i),$$ where $F_{tot,i}(\lambda)$ is the ``observed'' mock spectrum per fiber and $w(r)$ is a Gaussian weight set as $w(r)=exp(-0.5(r/\sigma)^2)$. The value of $r$ is the distance from the spaxel ($x,y$) to the centre of the fiber $i$, with a boundary length of $2.5$\ arcsec. We also use a value of $\sigma=0.6$\ arcsec, giving a final spaxel scale in the data cube of $0.5$\ arcsec/spaxel, the same as MaNGA. 

\end{description}

In Figure~\ref{cubes_photo} we present two examples of the mock reconstruction of the Sp8D and Sp6L galaxies. For the IFU reconstructions, we use the MaNGA bundle of 127 fibers (which corresponds to $\sim$30\%\ of the objects observed by this survey) and impose the condition that the galaxy filled the bundle up to 1.5 $R_{\rm eff}$ (which corresponds to $\sim$70\% of the observed by MaNGA). In this case, we have used an atmospheric seeing of $1.43$ arcsec, the same as the average at APO \citep{Stoughton:2002ab}. Note that the final reconstructed PSF in an IFU observation, $ PSF_{rec}$, includes both the atmospheric seeing and the fiber core diameter $D_c$. Following the relation presented in \cite{Allen:2015aa}, we calculate this PSF as $PSF_{rec}=(D_{c}^3+PSF_{seeing}^3)^{1/3}$.  Therefore, our reconstructed PSF should have a value of $2.21$ arcsec.  

\subsection{Datacube processing}
\label{datacube}

The obtained datacubes are processed with {\sc Pipe3D} \citep[][]{Sanchez:2016aa,Sanchez:2016ab}, a pipeline optimized to analyze in an automatic way spectra from IFS observations. It is based on the {\sc FIT3D} fitting code,\footnote{\url{http://www.astroscu.unam.mx/~sfsanchez/FIT3D/index.html}} which is able to fit both the stellar populations and emission lines of spectra of galaxies or regions of galaxies to recover their main properties. For the stellar populations it decomposes them in a set of SSPs, based on an inversion method. However, contrary to other methods, {\sc FIT3D} performs multiple inversions, through a Monte-Carlo iteration, and removing those SSPs in each iteration that produce negative weights in the inversion. This approach mimics an stochastic sampling of the star-formation history. The validity and limitations of the approach was demonstrated in \citet{Sanchez:2016ab}. 

For each datacube, we perform a SNR analysis to obtain a spatial binning with the goal of increasing the SNR across the FoV of the IFU, to the penalty of a lower spatial resolution. The next step is to fit each binned spectrum to a limited stellar template to derive the kinematics properties and the dust attenuation. The stronger emission lines are fitted using a set of Gaussian functions to remove them from the binned data cubes. The obtained stellar continuum spectrum is the one {\sc Pipe3D} fits to a linear combination of (dust attenuated) SSP templates, convolved and shifted to take into account the derived kinematics. The adopted stellar library comprises 156 templates that cover 39 stellar ages (from 1 Myr to 14.2 Gyr), and four metallicities ($Z/Z_{\odot} = 0.2$, 0.4, 1, and 1.5); see \citet{Cid-Fernandes:2013aa}. {\it The age distribution is very asymmetric, with many SSP templates at very young ages and only a few templates for old ages}. This is because the optical spectra of old populations are almost indistinguishable among them for ages larger than $\sim 8-10$ Gyr \citep[e.g.,][]{sanchez18b}. For more details and information on the adopted SSP templates, see \citet[][]{Sanchez:2016ab} and \citet{Ibarra-Medel+2016}.

\section{Results from the simulations}
\label{simulation-results}

The main properties of the disk-dominated Sp8D and lenticular-like Sp6L simulated galaxies are presented in Table~\ref{tab1}. To calculate the optical effective radius, $R_{\rm eff}$, we use the surface brightness maps generated as described in Section \ref{mock-obs} (see Eq. \ref{SB}). Figure~\ref{ifu_flux} shows the face-on $V$ band maps of both simulations. 
We apply to these maps a 2D fit to a S\'ersic profile by using the GALFIT code \citep{Peng:2002aa}. Hence, we obtain the S\'ersic index, the principal angle, the ellipticity and $R_{\rm eff}$. For face-on galaxies Sp8D and Sp6L, $R_{\rm eff}=7.5$ and 3.2 kpc, respectively. These radii are $\approx 25\%$ larger than the respective half-mass radii, in agreement with results reported in \citet{Szomoru:2013aa}, who measured the mass profiles of 177 galaxies and determined that the effective radius in mass is $\approx 25\%$ smaller than the effective radius in light. These results are also in agreement with those found by \citet{GonzalezDelgado+2014} using the CALIFA data.

From the simulations we can determine global/radial stellar masses and SF rates at each snapshot, and trace this way the respective {\it current} global/radial SFHs and MGHs. We can also use only the $z\sim 0$ stellar particles and from their age distribution calculate the respective {\it archaeological} histories (as well as the first-moment of the SFHs, i.e., the mass-weighted ages). 
While for the SFHs, the masses the stellar particles had at their formation time are used, for the MGHs, these masses are corrected by the stellar mass recycled to the ISM as a function of time and metallicity. 
The archaeological histories are conceptually similar to what is obtained from IFS observations by applying the fossil record method. For a formal definition of these different histories and their respective determinations for the Sp8D and Sp6L galaxies, see Appendix \ref{simulations}. We remark that both the current (real) and archaeological radial MGHs are very similar for the simulated galaxies Sp8D and Sp6L, which shows that mergers and radial mass transport do not significantly change their radial mass distributions \citep{Avila-Reese+2017}.

In the next Section, we will present the age radial profiles and the archaeological global/radial MGHs as obtained directly from the simulations (green dashed lines/symbols in all the figures therein), compare them with those measured from the mock observations, and finally, with those recovered with the fossil record method (Pipe3D). Besides the absolute MGHs, we will study the {\it normalized} MGHs, i.e., when a given (global or radial) MGH is divided by its mass attained at $z=0$ to show this way the relative mass growth rate at different epochs. In  Appendix \ref{simulations} we present these normalized MGHs from the simulations, and how they are affected by projection effects. As expected, the inclination tends to diminish the differences among the radial MGHs and make flatter the age gradients. However, significant differences are seen only for inclinations larger than $\sim 70$ degrees. In the next Section, among other questions, we will explore the further effects of inclination after applying the fossil record method. 

\label{maps-figs}
\begin{figure*}
\begin{center}
\includegraphics[width=0.9\linewidth]{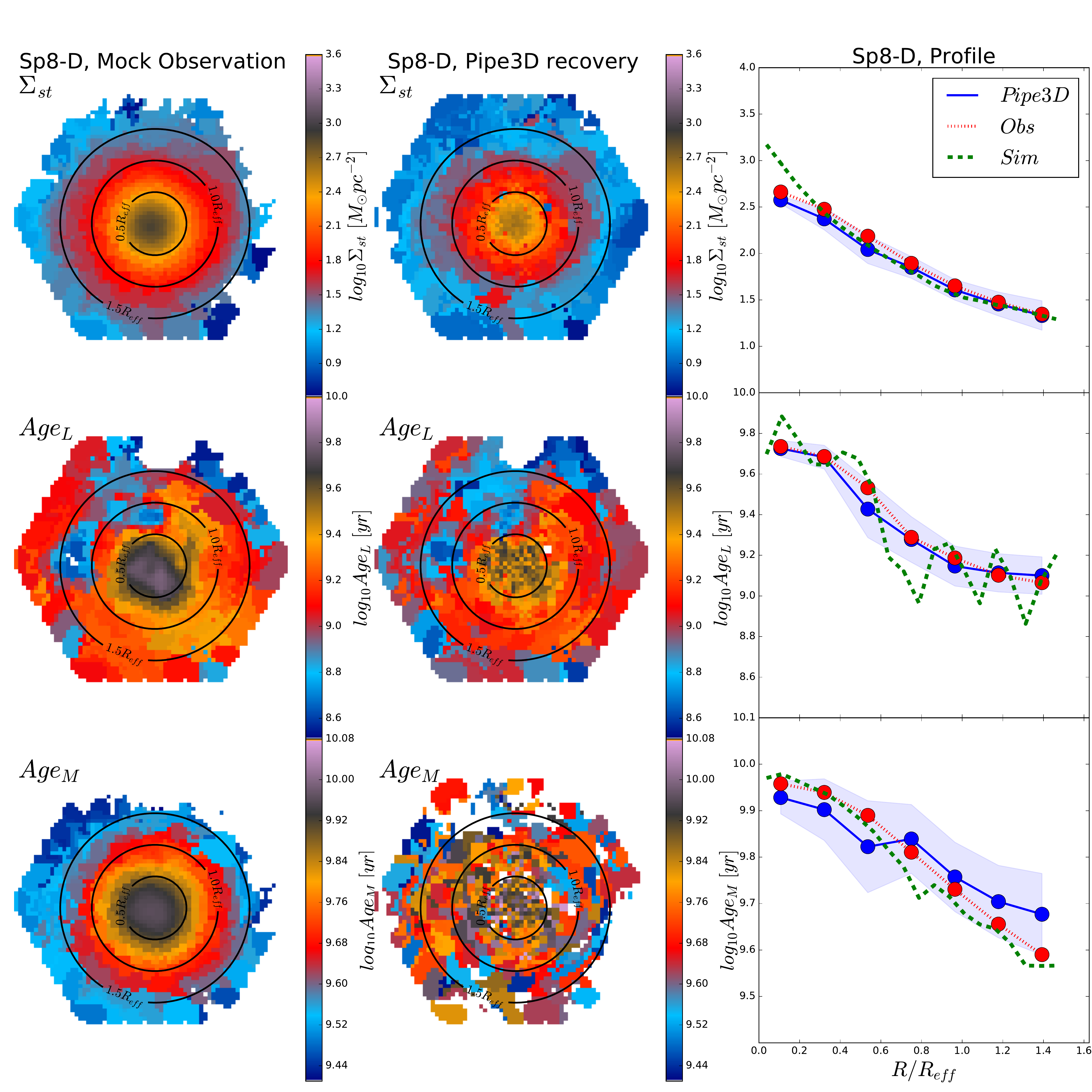}
\end{center}
\caption{Stellar mass, LW age, and MW age maps for the galaxy Sp8D observed under the fiducial setting. Left panels show the true maps after the instrumental/observational setting, and right panels show the {\sc Pipe3D} recovered maps. The right panels show the respective azimutally-averaged radial profiles. The dashed green lines are for the direct determinations from the simulation. The red dots connected by the dotted line are for measures from the simulation after applying the instrumental/observational setting, including the dithering (they correspond to the left panels). The blue dots connected by the solid line are from the {\sc Pipe3D} inferences (they correspond to the medium panels). } 
\label{Sp8D-maps}
\end{figure*}

\begin{figure*}
\begin{center}
\includegraphics[width=0.9\linewidth]{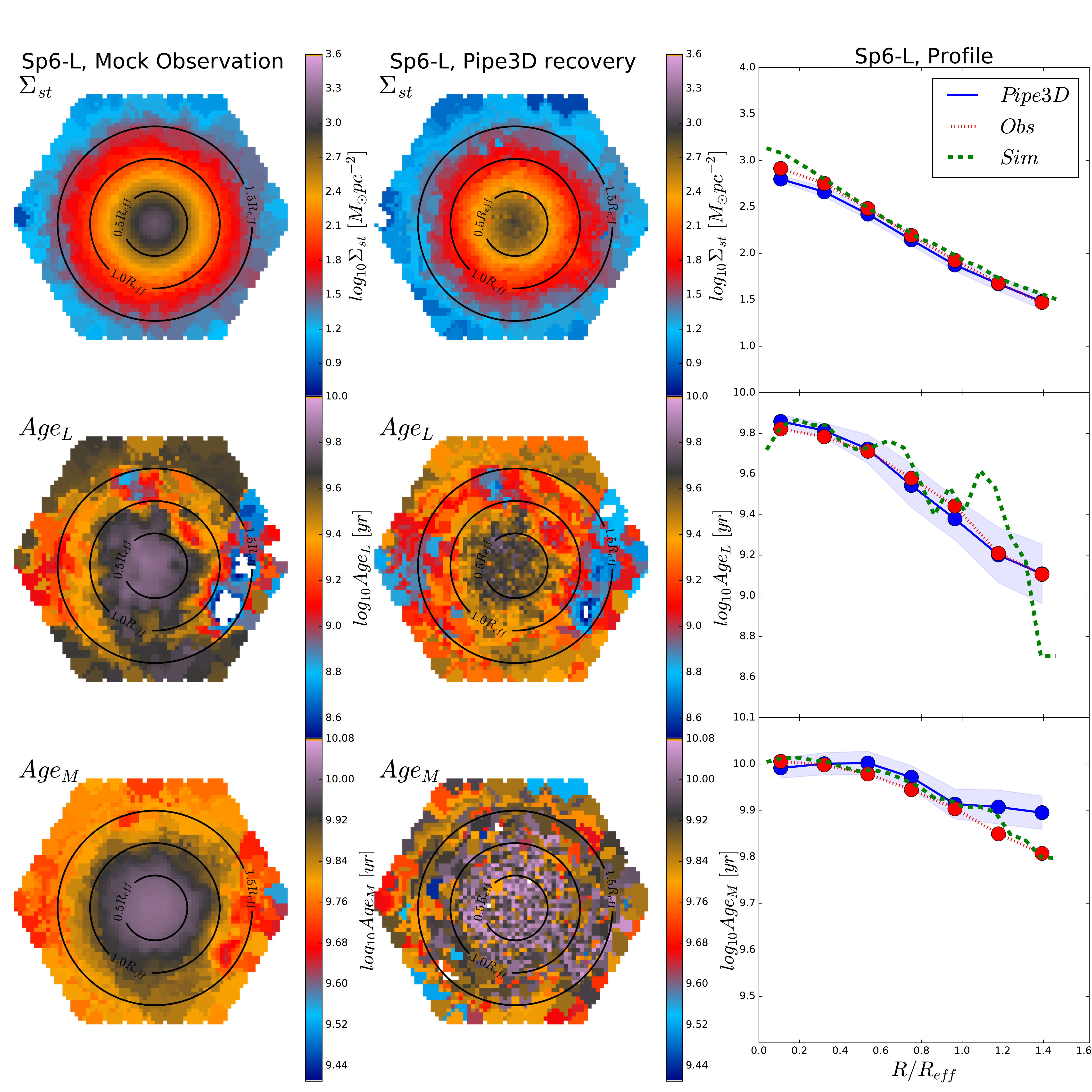}
\end{center}
\caption{As Figure \ref{Sp8D-maps} but for the galaxy Sp6L. } 
\label{Sp6L-maps}
\end{figure*}

\section{Fossil record inferences from the mock observations}
\label{reconstruction}

Our aim is to apply the {\sc Pipe3D} fossil record method \citep{Sanchez:2016ab,Ibarra-Medel+2016} to the IFU data cubes generated from the simulated and post-processed galaxies under several instrumental/observational settings. This way we will be able to explore how well the true properties of the galaxies are recovered. 

{\bf The Strategy:} 
To keep as close as possible to the observations, for instance those from MaNGA, we impose the condition that the given galaxy covers the IFU bundle FOV up to 1.5 $R_{\rm eff}$.\footnote{The goal of MaNGA is to have $\sim 70\%$ of the observed galaxies covered up to 1.5 $R_{\rm eff}$ \citep[the Primary+ sample;][]{Bundy+2015}.} Then, we ``observe'' the galaxies with four different MaNGA-like bundle sizes containing: 547, 127, 61, and 19 fibers of diameter $D$ in an hexagonal configuration (see subsection \ref{mock-obs}).\footnote{The MaNGA bundles contain 127, 91, 61, 31, and 19 fibers; the case of 547 fibers is just for exploring how much gain we would have with such a very high spatial sampling. Approximately 30\% of the galaxies from the Primary+ sample are expected to be observed with the bundle of 127 fibers.} Hence, to cover the same 1.5 $R_{\rm eff}$ of the galaxies with these different bundle apertures, they are located at different distances, which implies that they have different spatial resolutions. In this way, we will explore the effects of the spatial sampling and resolution on the recovered fossil record information. For a given bundle configuration, we explore further the effects of inclination, PSF dispersion size, and partially, the effects of the SNR.
The latter is a quantity that we actually can not control since it depends on the integration time (the noise level $\sigma_{\rm lim}$), the distance, the inclination and dust extinction, the instrumental noise, etc. However, if we fix the received flux, then by varying the noise level, $\sigma_{\rm lim}$, we attain different total SNRs (see subsection \ref{mock-obs}) and can explore how its variation affects our inferences by means of the fossil record method.  In Table \ref{tab2} we summarize the different settings explored in this work.

{\bf The fiducial setting:} For comparison purposes, we establish a {\it fiducial} setting of instrumental/observational conditions: an hexagonal MaNGA bundle of 127 fibers covering the galaxy in their face-on position up to 1.5 $R_{\rm eff}$ (this implies spatial resolutions of $0.97$ kpc\ arcsec$^{-1}$ and $0.43$\ kpc\ arcsec$^{-1}$ for Sp8D and Sp6L, respectively), a seeing dispersion of 1.43 arcsec with a reconstructed (total) PSF dispersion of $2.21$ arcsec, and a $\sigma_{\rm lim}$ of $26.83\ AB \ arcsec^{-2}$.  Under these conditions, the physical resolution (taking into account the reconstructed PSF) for the Sp8D and Sp6L galaxies is 2.14 and 0.95 kpc, respectively. Note that the original spatial resolution of our simulations is far better than these values. The average SNRs are 24 and 25, respectively. Note that the fiducial and the other settings include already the dithering process described in subsection \ref{mock-obs}.

In the following, we will report quantities determined directly from the simulations (the ``true ones''; green dashed lines/symbols) and measured as an observer (red dotted lines/symbols). On the other hand, we will report the {\sc Pipe3D} recovered quantities (blue solid lines/symbols), which by construction include the instrumental/observational setting (with the respective datacube generation), and the ulterior application of the inversion method. In most of our explorations, we explicitly separate the effects of (i) the instrumental/observational setting and (ii) the {\sc Pipe3D} inversion method.  


\begin{figure*}
\begin{center}
\includegraphics[width=0.32\linewidth]{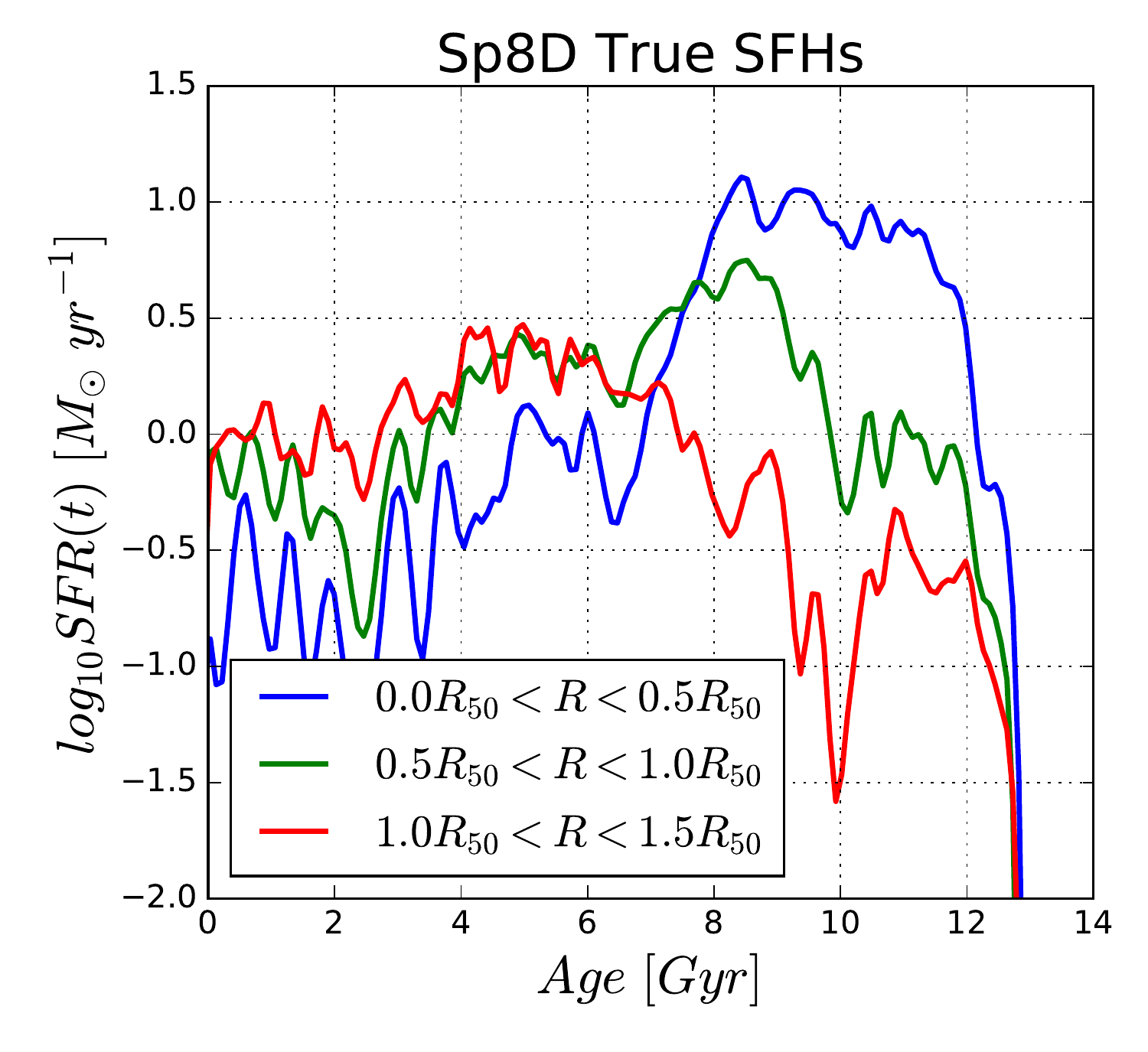}
\includegraphics[width=0.32\linewidth]{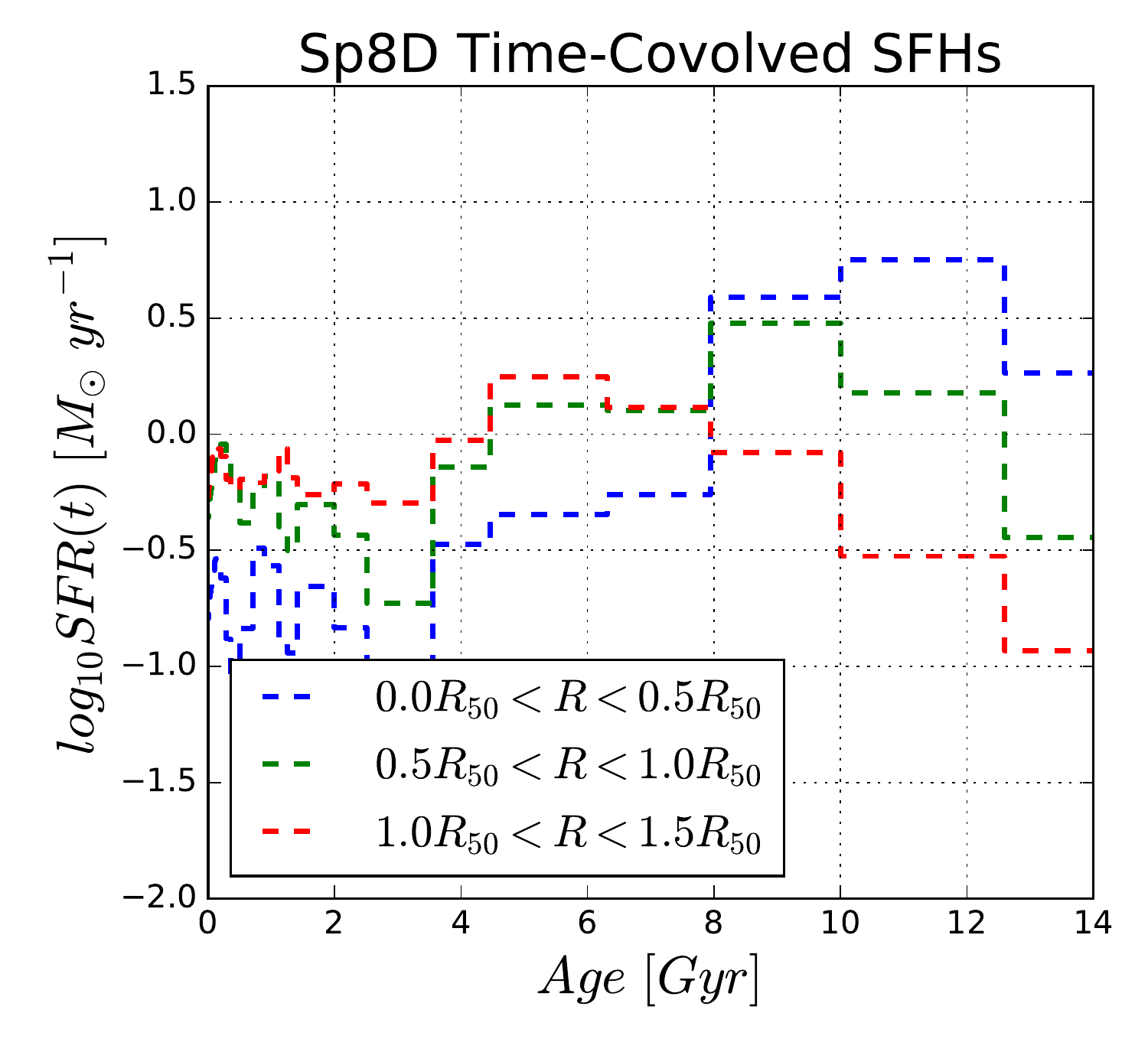}
\includegraphics[width=0.32\linewidth]{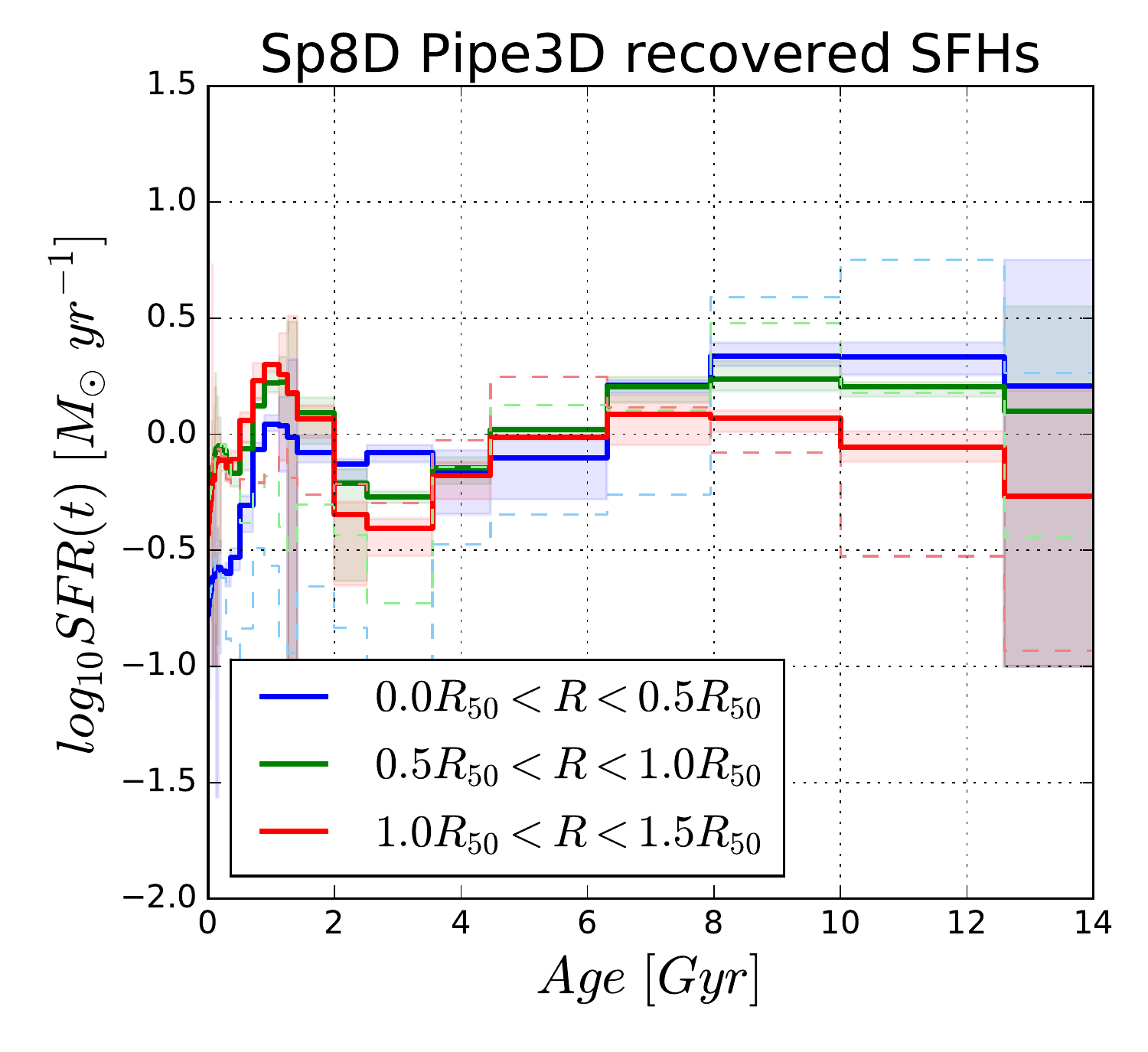}
\includegraphics[width=0.32\linewidth]{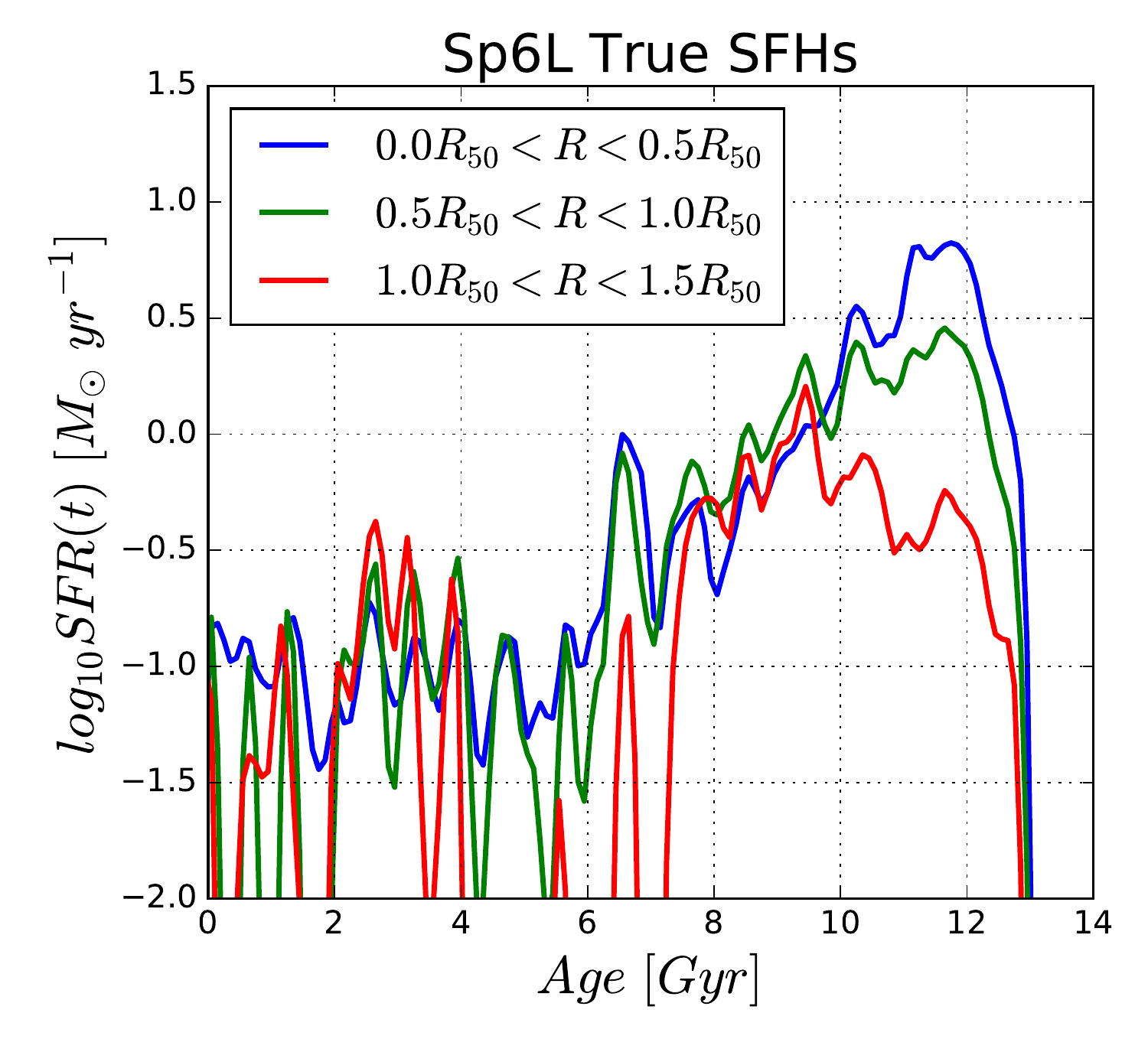}
\includegraphics[width=0.32\linewidth]{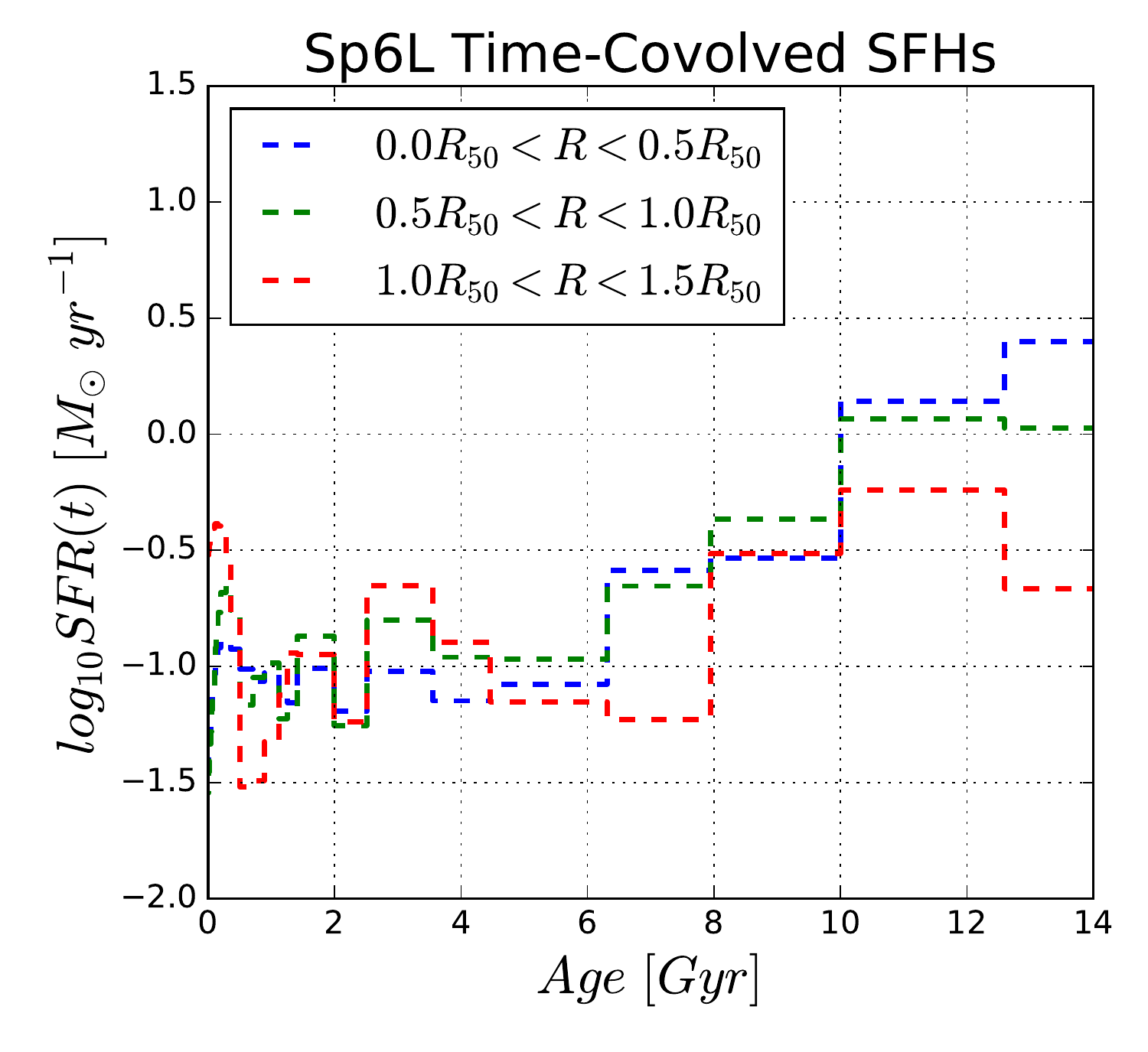}
\includegraphics[width=0.32\linewidth]{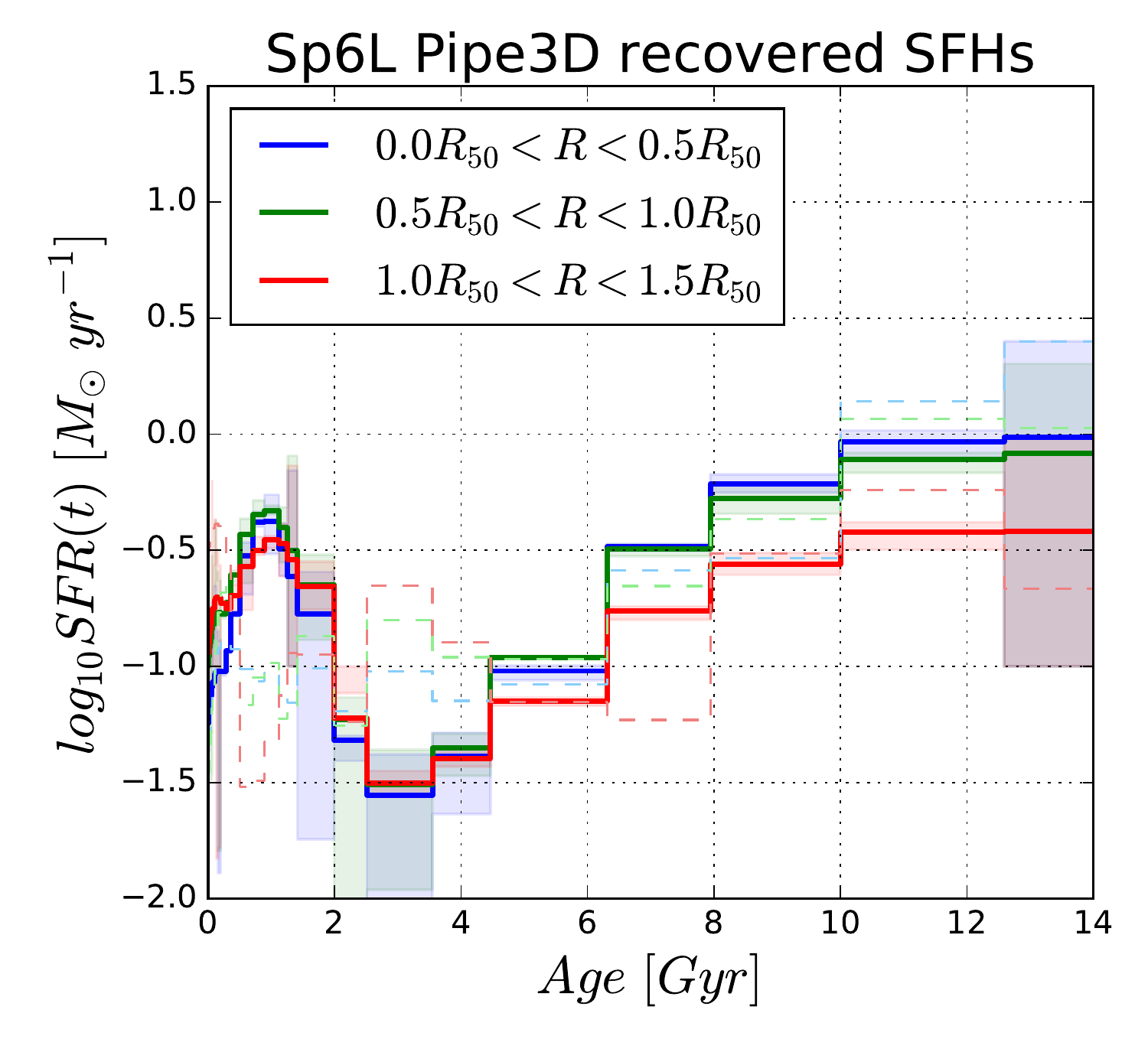}
\end{center}
\caption{Average archaeological SFHs within the radial bins indicated in the insets for the Sp8D and Sp6L galaxies (upper and lower panels, respectively). {\it Left panels:} The radial archaeological SFHs as measured directly from the simulation. {\it Middle panels:} The radial SFHs as measured from the simulation once convolved by the age resolution of the {\sc Pipe3D} stellar library. {\it Right panels:} The radial SFHs recovered from the fossil record analysis with {\sc Pipe3D} (solid lines). The corresponding shaded regions around the solid lines are the estimated 1-$\sigma$ uncertainties from the inversion method.  The short-dashed lines reproduce the SFHs from the medium panels.}
\label{sfh_sp8}
\end{figure*}

\begin{figure*}
\begin{center}
\includegraphics[width=0.48\linewidth]{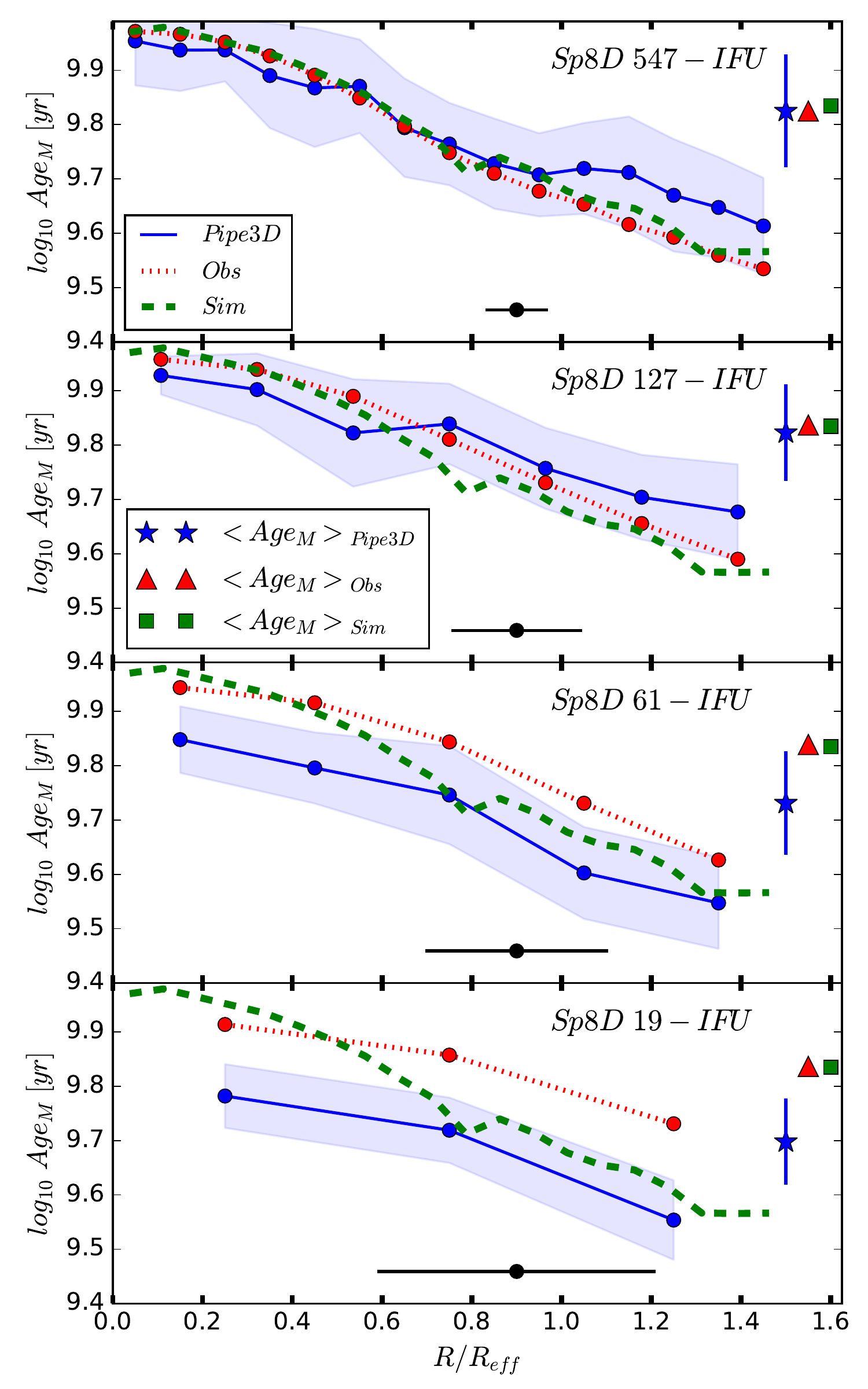}
\includegraphics[width=0.48\linewidth]{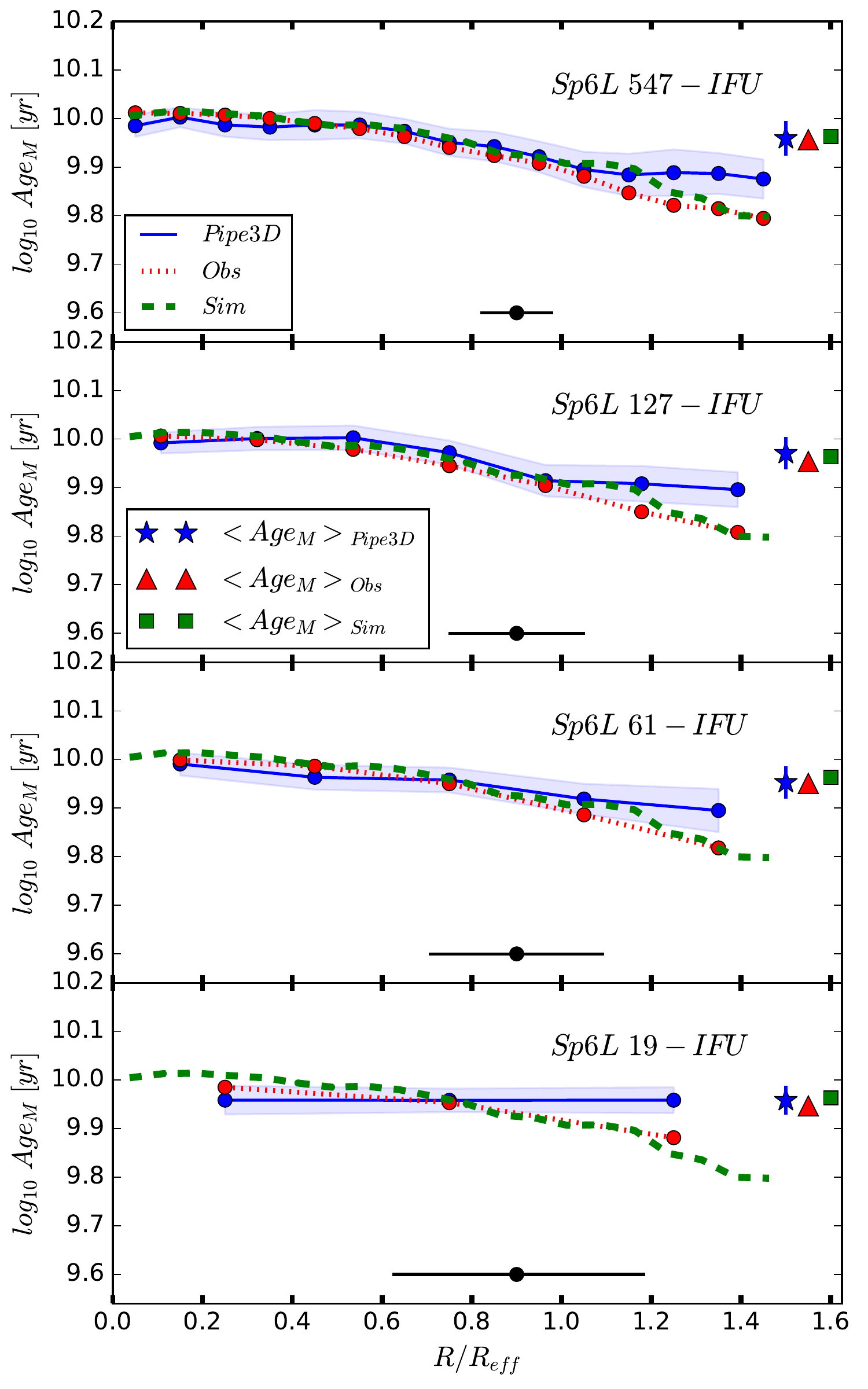}
\end{center}
\caption{Mass-weighted (MW) age radial profiles as obtained directly from the simulations (dashed green line), after applying the instrumental/observational setting (red dotted line), and as recovered with {\sc Pipe3D} (blue solid line). Right and left panels are for the Sp8D and Sp6L simulated galaxies; note that the vertical axes cover slightly different age ranges. The second upper panels correspond to our {\it fiducial} setting: 127 fibers, face-on view, reconstructed PSF size of 2.21 arcsec, $\sigma_{lim}=26.83$ arcsec$^{-2}$, and average SNR=24-25). From top to bottom, the instrumental/observational setting is as for the fiducial case but varying the number of fibers in the hexagonal bundle, from 547 to 19 fibers. The black horizontal bars show the respective reconstructed PSF sizes in units of $R_{\rm eff}$, and they determine the spatial resolution of the ``observed'' and recovered radial profiles. The symbols in the right side show the corresponding global average (within $1.5R_{\rm eff}$) MW ages.  }
\label{age_m-res}
\end{figure*}

\subsection{Stellar mass and age maps}
\label{maps}

The left columns of Figures \ref{Sp8D-maps} and \ref{Sp6L-maps} show the maps of stellar mass, luminosity-weighted (LW) age, and mass-weighted (MW) age from the post-processed Sp8D and Sp6L simulations, respectively, after imposing the instrumental/observational setup, which includes the dithering process. The setup corresponds to the {\it fiducial} case (see Table \ref{tab2}). The medium column shows the respective maps as reconstructed with {\sc Pipe3D} from the datacubes. 
The MW and LW ages are calculated as follows:
\begin{eqnarray}
\log(Age_{M_{ssp}})=\sum_j^{n_{ssp}}\log(Age_{ssp,j})m_{ssp,j}/\sum_j^{n_{ssp}}m_{ssp,j},\\
\log(Age_{L_{ssp}})=\sum_j^{n_{ssp}}\log(Age_{ssp,j})L_{ssp,j}/\sum_j^{n_{ssp}}L_{ssp,j},
\label{ages}
\end{eqnarray}
where $j$ runs over each SSP with a specific age ($Age_{ssp,j}$), luminosity ($L_{ssp,j}$), and mass ($m_{ssp,j}$).
The right column shows the corresponding azimuthally-averaged radial profiles. 
The blue shaded regions show the estimated 1-$\sigma$ uncertainties in the {\sc Pipe3D} inversion method inference (see below).

The mass surface densities in most of the spaxels are well recovered by {\sc Pipe3D}. For galaxy Sp8D, which has a wider range of stellar populations, there are more local spatial differences than for galaxy Sp6L. These differences are mostly up to $\pm 0.1$ dex, with a very slight trend towards an average under-estimation in the intermediate and central regions. For Sp6L, the differences are even smaller on average. The {\sc Pipe3D} recovered radial mass surface density profiles become shallower than the true ones (dashed green line) in the central regions of both Sp8D and Sp6L galaxies. However, this central flattening of the profiles is mostly due to the resolution limitations of the instrumental/observational setting rather than due to the fossil record method, as demonstrated when comparing with the ``observed'' ones (blue line vs. red dotted one).

For the LW age maps, the spatial distribution is well recovered by {\sc Pipe3D} for both galaxies. For Sp8D, the recovered ages in spaxels within the inner 0.5 $R_{\rm eff}$ tend to be younger than those from the mock observations, mostly by $\sim 0.1$ dex, while in the outer regions there is not a systematical trend but the spaxel-by-spaxel fluctuations are larger.  For Sp6L, the recovered ages in a large fraction of spaxels are on average slightly younger than those from the mock observations. In fact, most of the local and (azimuthally-averaged) radial differences in the recovered ages for both galaxies are within the 1-$\sigma$ uncertainties in the age determination with {\sc Pipe3D}. The radial LW age profiles recovered with {\sc Pipe3D} follow well the true ones (dashed green lines), in general, though they are not able to recover the small-scale radial fluctuations. This is due to the limited spatial resolution from the observational setups rather than due to the fossil record method. 

Regarding the MW age maps and radial profiles from the mock observations (and from the simulations), as expected, they are smoother than the respective LW age maps/profiles (LW ages are sensitive to young stellar populations, even if their contributions to the mass are small).  However, as respects to the recovered MW ages by {\sc Pipe3D} with respect to the ``observed'' ones, they present more spaxel-by-spaxel fluctuations than the corresponding LW ages, specially for the Sp8D galaxy. This is partially because the uncertainty in the MW age inferences is larger than the one in the LW age inferences \citep[note that the first ones are weighted by the M/L ratio, that can change up to around an order of magnitude between different galaxy types, e.g.][]{bell01}. The recovered radial MW age profiles tend to be slightly flatter than the "observed" ones, though the differences are well within the uncertainties, in particular for the Sp8D galaxy.

\begin{figure*}
\begin{center}
\includegraphics[width=0.48\linewidth]{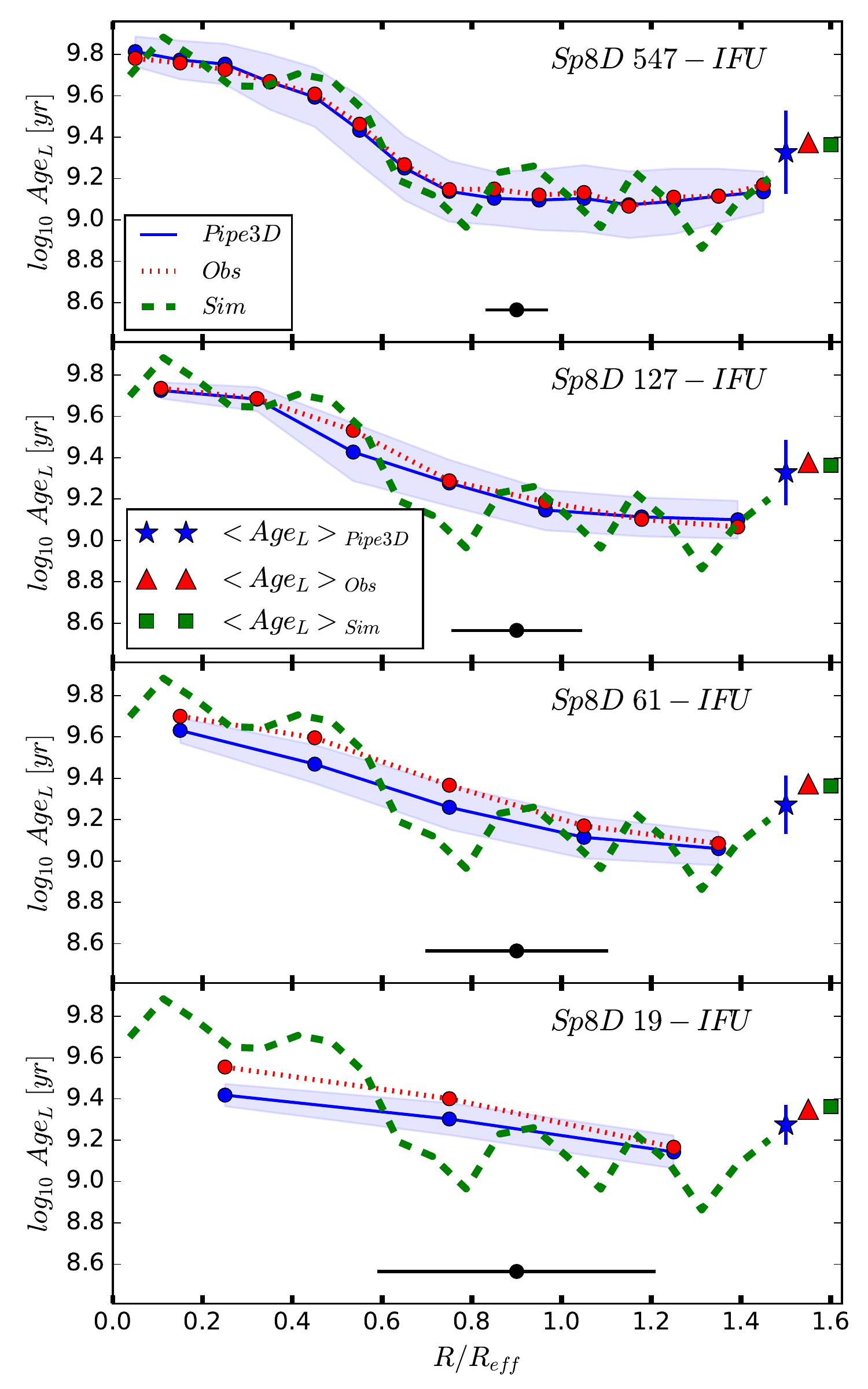}
\includegraphics[width=0.48\linewidth]{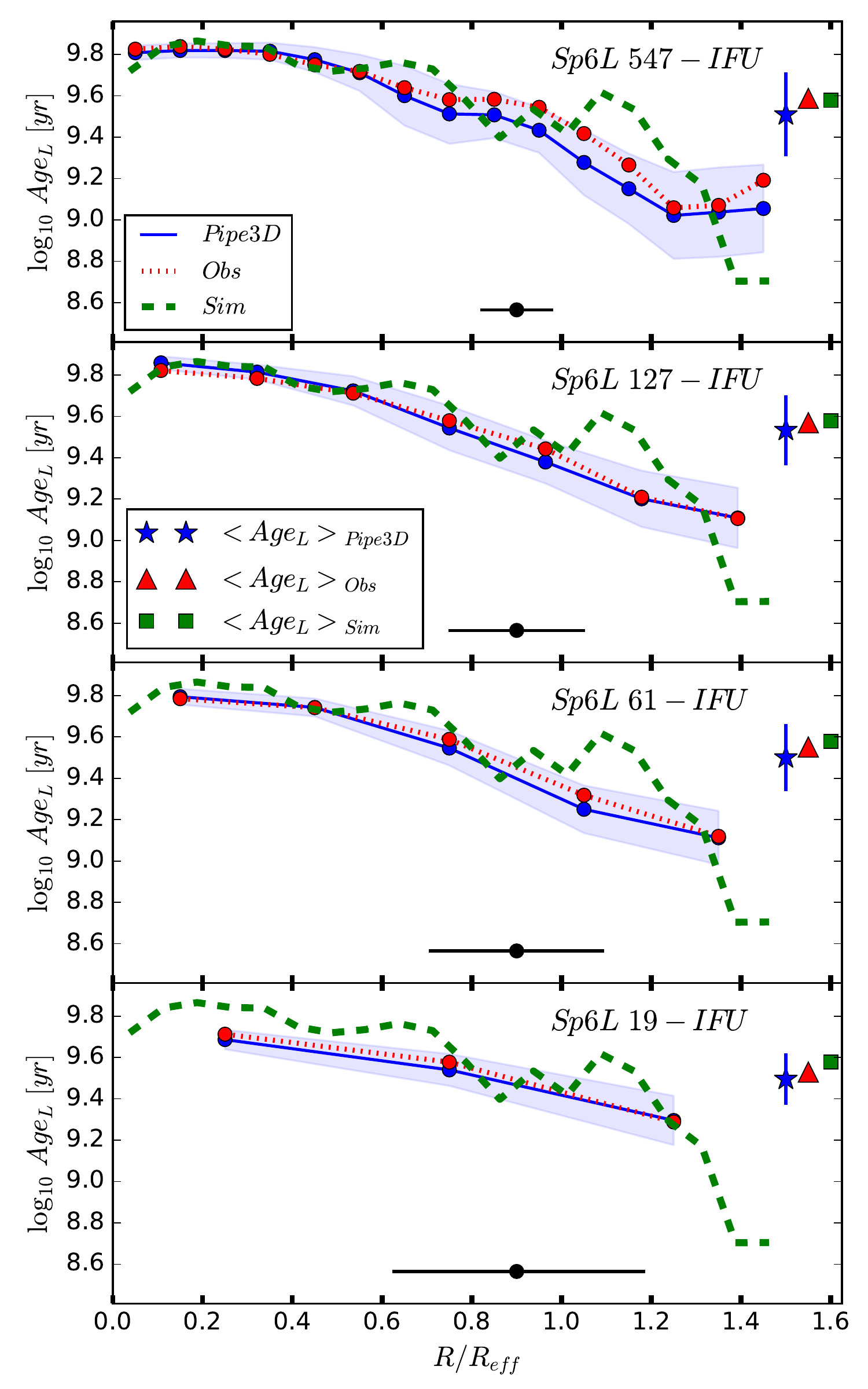}
\end{center}
\caption{As Figure \ref{age_m-res} but for luminosity-weighted (LW) ages. }
\label{age_l-res}
\end{figure*}

\begin{figure*}
\begin{center}
\includegraphics[width=0.9\linewidth]{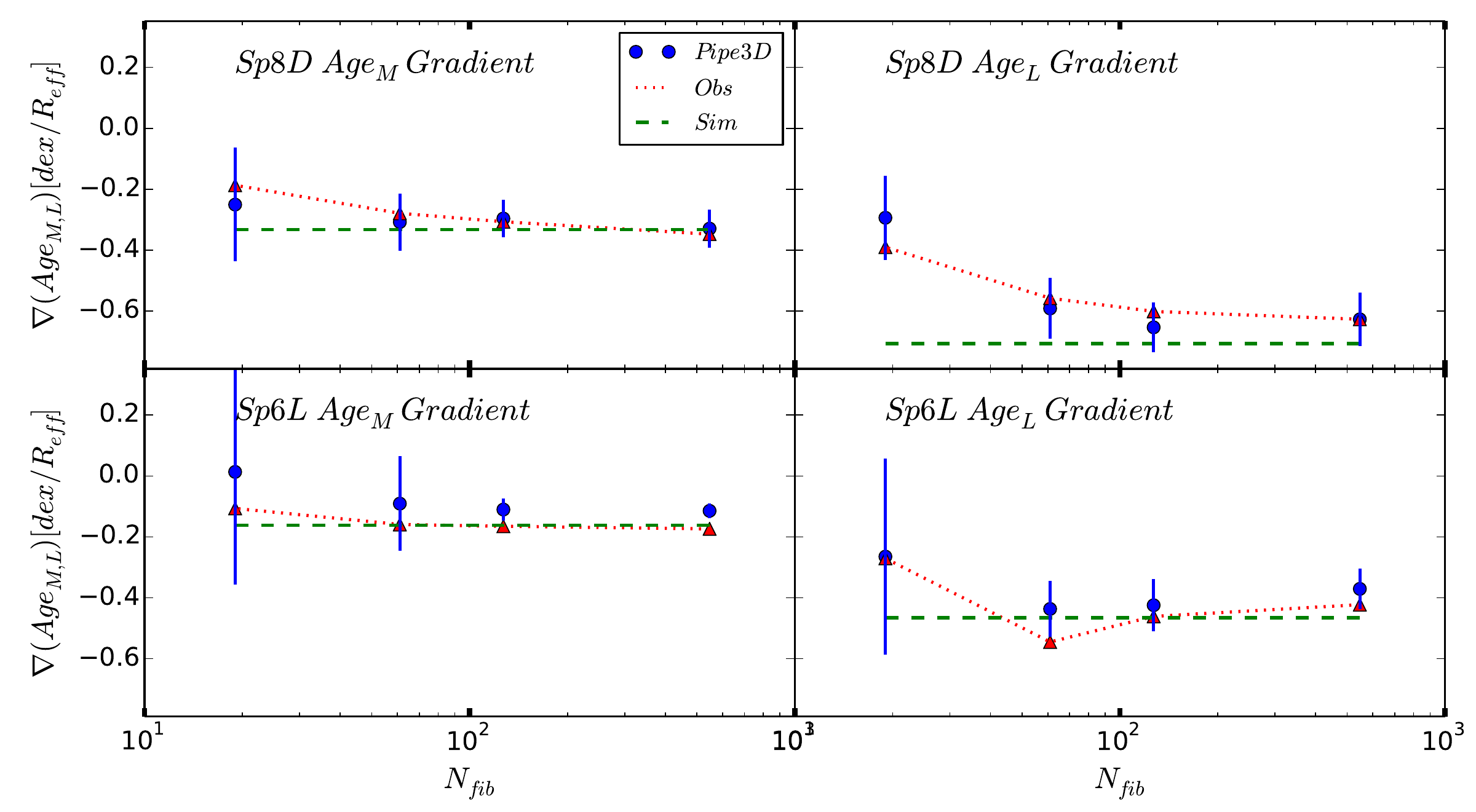}
\end{center}
\caption{Slopes in dex per $R_{\rm eff}$ (gradients) of the log-linear fits to the radial age profiles shown in Figures \ref{age_m-res} and \ref{age_l-res} as a function of the IFU number of fibers. Upper and lower panels are for the galaxies Sp8D and Sp6L, respectively. Left and right panels are for the MW and LW age gradients, respectively. The dashed green line correspond to ages obtained directly from the simulations, the red triangles are for ages determined after applying the instrumental/observational setting, and the blue dots with error bars are for the ages recovered with {\sc Pipe3D}, taking into account the uncertainties in the radial age profiles.}
\label{grads}
\end{figure*}

\begin{figure*}
\begin{center}
\includegraphics[width=0.48\linewidth]{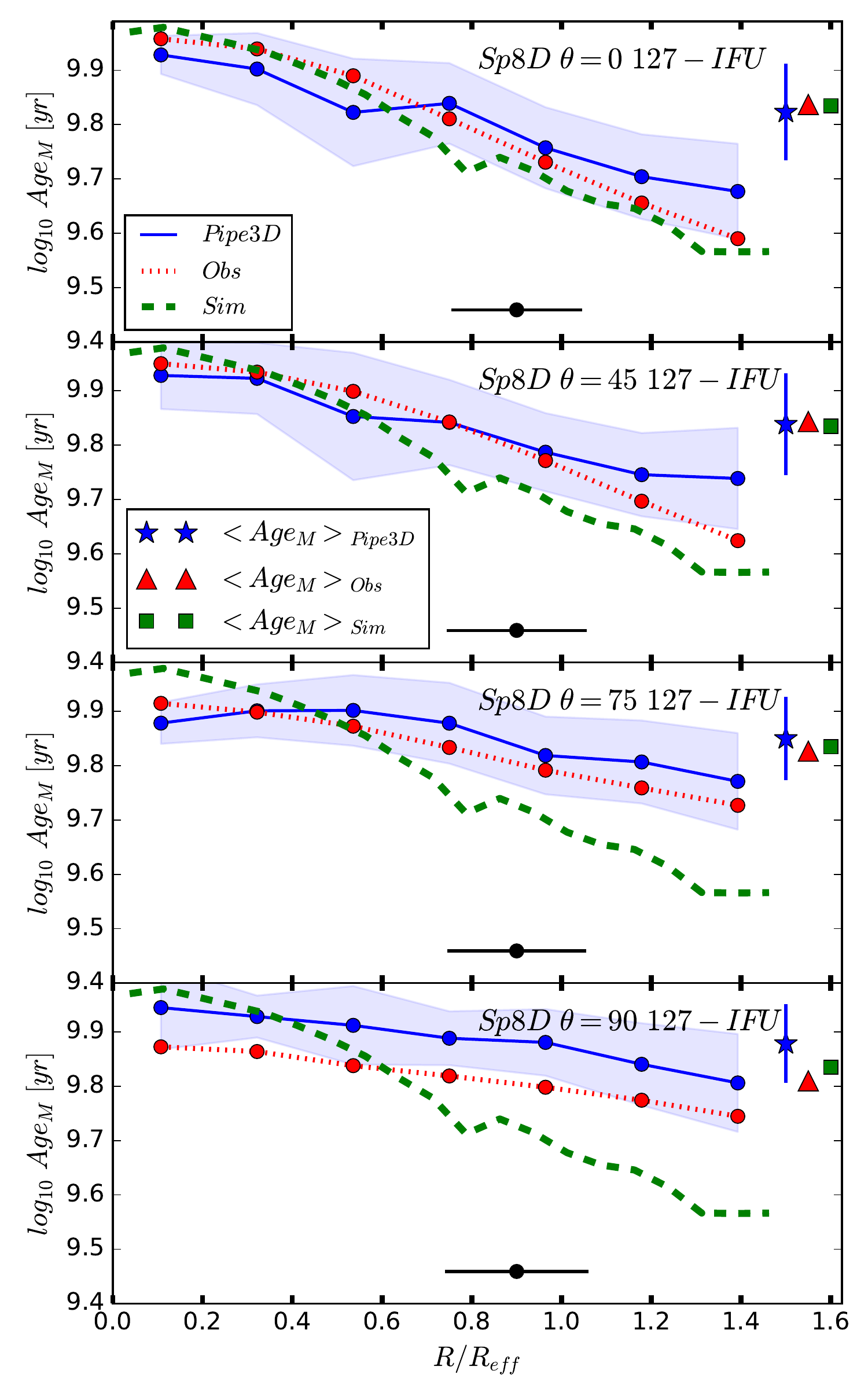}
\includegraphics[width=0.48\linewidth]{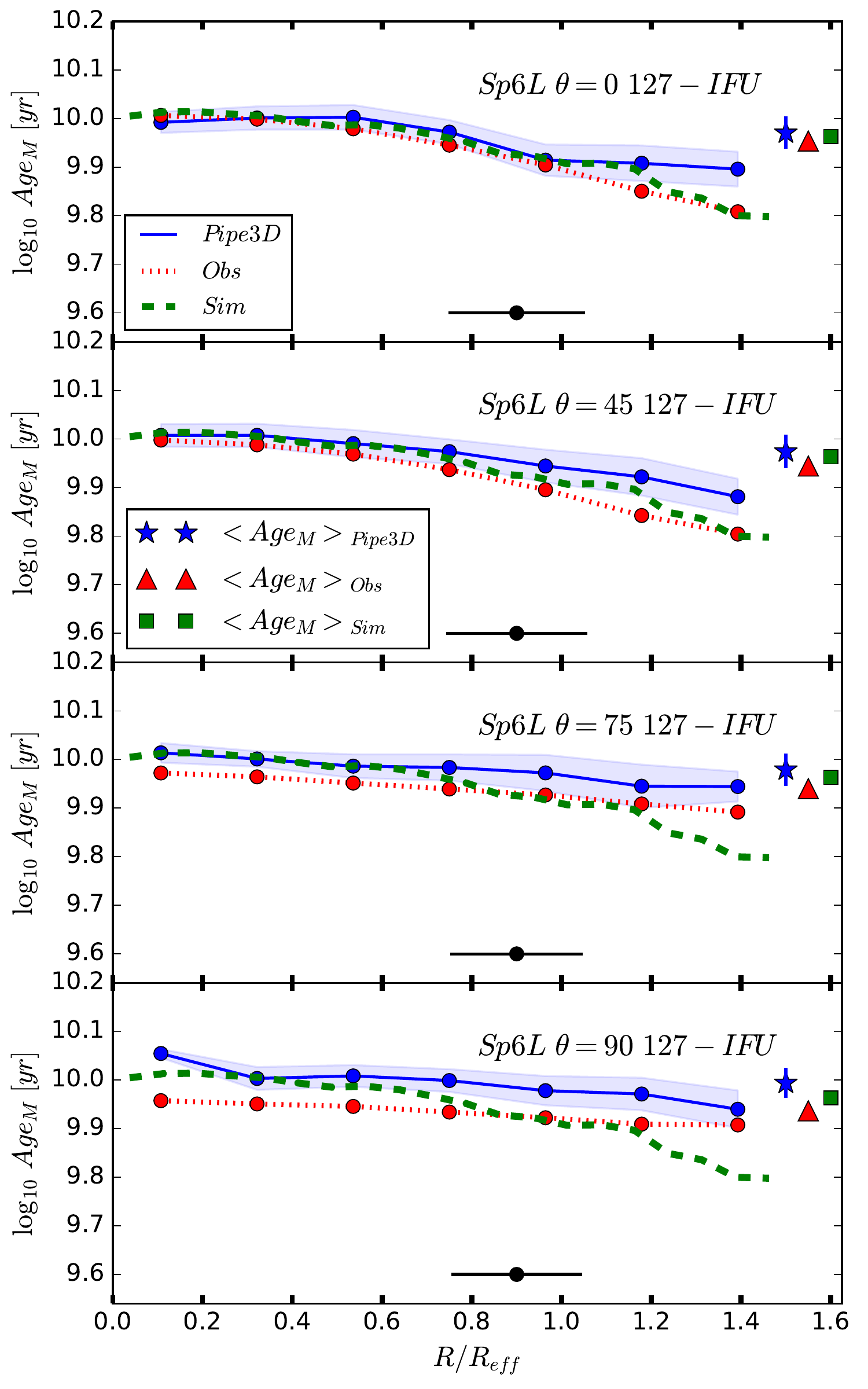}
\end{center}
\caption{Mass-weighted (MW) age radial profiles as in Figure \ref{age_m-res} but now fixing the IFU number of fibers (127) and varying the inclination angle of the galaxies. From top to bottom the angles are $\theta=0^{\circ}, 45^{\circ}, 75^{\circ}$ and $90^{\circ}$. Note that the vertical axes in the left and right panels cover slightly different age ranges.}
\label{age_m-incl}
\end{figure*}

\subsection{Radial Star Formation Histories}
\label{GLSFH}

The MW ages discussed in the previous subsection are actually the first moment of the SFHs \citep[e.g.,][]{Zibetti:2017aa}. We explore here how well the fossil record method recovers the radial SFHs from the mock observations of our simulated galaxies. This exploration will help us to understand the differences between the {\sc Pipe3D} inferred MW age radial gradients (see also subsection \ref{age-gradients}) and the radial MGHs (subsection \ref{radialMGHs}) with the true ones. 
{ Note that the SFH is directly proportional to the SSP stellar mass fraction distribution by age, i.e., the distribution constrained from the spectrum by means of the inversion method or measured from the stellar particle ages in the simulations. The SFH is calculated indeed from the stellar masses of the SSPs (corrected by the mass loss at the age of its corresponding SSP time) divided by the temporal width of each SSP \citep[for more details see][]{Ibarra-Medel+2016,Sanchez:2016aa}.} 
In this subsection we limit the exploration to the {\it fiducial} observational setting. 

In the left panels of Figure \ref{sfh_sp8}, we present the SFHs measured directly from the simulated Sp8D and Sp6L galaxies in three radial bins defined in Section \ref{simulation-results} and used for the radial MGHs. The medium panels show these SFHs but convolved with the age binning of the stellar libraries used in {\sc Pipe3D}.  Finally, the right panels present the three radial SFHs as inferred from the mock observations by the inversion method (solid lines); for comparison, the SFHs plotted in the medium panels are reproduced here with dashed lines. 

First, we notice that the SFHs are smoothed-out, more as older are the ages, just due to the {\it temporal binning of the used stellar libraries}, which is very fine for the youngest ages and very coarse for old ages (see subsection \ref{datacube}). 
Therefore, strong peaks and detailed features in the early SFH of galaxies can not be recovered by the inversion method.  Further, we see that the inversion method tends in general to spread out the recovered SFHs (or equivalently, the mass fraction distributions by ages) with respect to the original ones. These two effects combine to produce radial SFHs that result less different among them than the true ones, as seen in the right panels of Figure \ref{sfh_sp8}. This explains the general trends of getting flatter age radial profiles and radial MGHs more similar among them than the true ones, as it will be discussed in subsections \ref{age-gradients} and \ref{radialMGHs}, respectively. These radial effects are more remarkable for Sp8D, a galaxy with a strong inside-out growth mode and with large radial gradients, than for Sp6L, a galaxy assembled more uniformly as a function of radius.

Second, we see that the spread out in the inferred SFH (or in the constrained population age distribution) depends on the shape of the true SFH. For example, for the innermost SFH of the Sp8D galaxy (blue line), which is peaked at old ages, the spread out is biased to younger ages, while for the outermost SFH (red line), which is more extended and with a peak at intermediate ages, the spread out tends to be biased to older ages. The more constant SFH of the intermediate radial bin (green line), is better recovered by the inversion method; its age distribution is broad so there is no too much room for spreading out more it. For the Sp6L galaxy, the three radial SFHs are peaked at old ages, more the innermost one. Consequently, there is a trend of spreading out the SFHs more towards young ages, specially for the innermost bin. 

However, note that, in general, the radial SFHs of both galaxies use to be more complex than smooth peaked functions, so that the SFHs recovered by the inversion method suffer the combination of different biases. Besides, the fossil record method is applied to the local resolved regions across the galaxy, which are composed, at a given radial bin, by regions dominated by different stellar populations and with different fractions of dust. This implies also that the recovered radial SFHs (and consequently MGHs and age gradients) will depend on the spatial sampling/resolution and inclination.

Finally, note that the inversion method systematically retrieves a local peak in the radial SFHs composed of stellar populations of ages $\sim 1$ Gyr. This bias in the recent SFRs was already noticed by \citet{lopfer16}, and seems to be related to the limited wavelength coverage of optical spectroscopic data in the blue/UV regimes, more sensible to young stellar populations. This creates unrealistic fluctuations in light weights of the youngest stellar populations considered in the SSP library.

\subsection{Age radial profiles}
\label{age-gradients}

  In Figure \ref{age_m-res} we plot the MW age radial profiles for the face-on  Sp8D (left panels) and Sp6L (right panels) galaxies for the bundle configuration of 547, 127, 61, and 19 fibers presented in Table \ref{tab2}.  The symbols and lines are as in the right panels of Figures \ref{Sp8D-maps} and\ref{Sp6L-maps}. In the case the fossil record recovers the MW ages perfectly, the blue solid lines (Pipe3D age profiles) should approach to the red dotted lines (age profiles measured from simulations but after applying the instrumental/observational setting, including the dithering process).  The shaded areas around the blue lines are the 1-$\sigma$ uncertainty in the age inferences by Pipe3D.  The radial binning in the ``observational'' determinations is for radii larger than the effective spatial resolution, which is conditioned by both the distance and the reconstructed PSF size. For a given reconstructed PSF size (e.g., 2.21 arcsec, see subsection \ref{mock-obs}), as more distant is located the galaxy, poorer is the effective spatial resolution, plotted as horizontal black error bars in each panel of Figure~\ref{age_m-res}. The symbols plotted in the right side of each panel show the corresponding global (within 1.5 $R_{\rm eff}$) average MW ages.  

\subsubsection{The fiducial setting}

Let us first analyze the results for the {\it fiducial} instrumental/observational setting (127 fibers, second upper panels in Fig. ~\ref{age_m-res}). For this setting, the spatial resolution/sampling and the dithering produce only minor changes in the measured MW age profiles (red dotted lines). By comparing these profiles with the inferred ones with {\sc Pipe3D} (blue solid lines), we find that the fossil record method flattens the age gradients, specially for Sp8D.
This is a consequence of the fact that the fossil record method tends in general to spread out the inferred fractional distribution of SSP ages with respect to the original one, and whether this spreading out is biased to older or younger ages depends on how is the original fractional distribution of SSP ages, mainly on the peak age and the width of the peak. The fractional distribution of SSP ages is established by the given SFH. For the Sp8D galaxy, the SFH peaks at later times as larger is the radius (pronounced inside-out mode; see Fig. \ref{sfh_sp8} above). Therefore, the light and mass distributions between different SSP's peak on average at younger ages as larger is the radius, too. As we discuss in  subsection \ref{GLSFH}, in general, the recovered distribution tends to be biased to younger ages when the true distribution is peaked at times close to the oldest SSP template (this is the case of the innermost regions of Sp8D), and to older ages for intermediate-age populations (they dominate in the outer regions of Sp8D). 

For the Sp6L galaxy, which has intrinsically a much weaker age gradient and older stellar populations than the Sp8D one, the recovered MW radial age profile is close to the ``observed'' and true ones, excepting in the outermost radii, where the inferred ages are older than the true ones. In fact, the SFH of the outermost regions of this galaxy is episodic (see Fig. \ref{sfh_sp8}), in such a way that populations of diverse ages are mixed. 

Finally, we see that the recovered global (within 1.5 $R_{\rm eff}$) MW ages of both galaxies agree very well, within the uncertainty, with the true ones (see also columns 11 and 12 of Table \ref{tab2}).

\subsubsection{The effects of spatial sampling and resolution}
\label{sampling}

According to Figure~\ref{age_m-res}, when the spatial sampling/resolution is high, as in the case of the 547 fibers bundle, the MW age radial profile is well recovered within the uncertainties by the fossil record method, both for the Sp8D and Sp6L simulated galaxies, excepting at the largest radii. As the spatial sampling/resolution becomes poorer, the measured age radial profile after the instrumental/observational settings (dotted red line) becomes flatter and with older ages in general for the galaxy Sp8D, which has an intrinsic steep age gradient. Due to the poor spatial sampling and the dithering, the ``observed'' stellar populations are mixed (smoothed) radially. For Sp6L, which has a less steep age gradient than Sp8D, the effect of radial mixing affects less the radial age profile. 

Regarding the {\sc Pipe3D} recovery, for the Sp8D galaxy, as is sampled with less fibers and is located further away, the MW age gradient becomes slightly flatter, though the main effect is that the MW ages become younger. For example, for the bundles with 61 and 19 fibers, the inferred global MW ages are $\log$[Age$_M$/yr] $= 9.72 \pm 0.1$ and $9.69 \pm 0.07$, respectively, versus the true global age of 9.83. In spite that the ages are mass weighted, recall that the inferences come from light, and young populations are intense in light in such a way that, even if they are a very small fraction, they can  ``contaminate'' the observed spectra of the spatially binned regions. As less and larger are the binned regions (poorer spatial sampling), the more probable is this contamination in the spectrum of a given region, with the respective inference of a younger MW age and lower mass-to-luminosity (M/L) ratio than the real stellar population. 
For the Sp6L galaxy, which has a much smaller fraction of young populations than Sp8D, the inferred MW age radial profile is not globally shifted to younger ages. However, the gradient becomes shallower as poorer is the spatial sampling/resolution due to the strong radial mixing of the populations.
Finally,  we should note here that the recovery of the correct ages is strongly affected by the accuracy with which the dust attenuation is derived, as we will see in a forthcoming section.

In Figure ~\ref{age_l-res} we present the corresponding LW age radial profiles. At a qualitative level, these profiles follow the same trends discussed above for the MW age profiles of both galaxies. The LW ages recovered by {\sc Pipe3D} do agree well with the true and ``observed'' ones in almost all radii for the best spatial resolutions and sampling. The accuracy in the recovery of the true values gets worse as poorer is the spatial resolution/sampling (see also column 12 in Table \ref{tab2}), though the general trend of the radial profiles is well recovered. The small-scale radial fluctuations in the LW age profiles can not be recovered even for our highest spatial resolution/sampling (547 fibers). Thus, they seem to be associated with structures of sub-kpc scale size, like spiral arms or star-forming clusters, below the spatial resolution of the mock datasets. We should note that this many not be true for IFS observations of higher spatial resolutions, like those provided by the CALIFA survey or the MUSE instrument.

In Figure~\ref{grads} we present a summary of the age gradient dependence on the bundle size/fiber number. We have fitted the different age radial profiles showed in Figures~\ref{age_m-res} and \ref{age_l-res} to a log-linear function and plot the obtained slopes in unities of dex per $R_{\rm eff}$ for each one of the four bundle configurations studied in this work. The symbols and colors are the same as in Figure~\ref{age_m-res}. The errors bars show the 1-$\sigma$ scatter of the fitted slopes taking into account the uncertainty in the radial age profiles (shadow regions in Figs. \ref{age_m-res} and \ref{age_l-res}).
For the  disk-dominated Sp8D galaxy, the MW age gradients calculated after applying the instrumental/observational setting (red triangles) become flatter than the true ones (green dashed line) as the spatial resolution decreases, i.e., as the number of fibers decrease. The corresponding mean slopes obtained with {\sc Pipe3D} after applying the observational setting (blue dots) agree well within the uncertainties with those obtained from the ``observed'' profiles and from the true profiles. 
For the bulge-dominated Sp6L galaxy, which presents a less steep gradient than the Sp8D one, the MW age gradients calculated after applying the instrumental/observational setting remain nearly the same, excepting for the lowest resolved case (19 fibers) that becomes flatter than the true one. Regarding the age gradients estimated by {\sc Pipe3D}, they are flatter in general, with a weak trend of being even flatter as smaller is the number of fibers (i.e., for worse samplings). However, the gradients from the recovered profiles agree within the uncertainties with those from the ``observed'' and true profiles. 

For the LW age gradients, which are steeper than the MW ones, we find similar trends in general, although in some cases there are small differences. The gradients from the recovered profiles agree within the uncertainties with those from the ``observed'' profiles. However, for the Sp8D galaxy, both gradients become much flatter than the true one for 61 and 19 fibers (again, the worst samplings).

\subsubsection{The effect of inclination and extinction}
\label{resolution-inclination2}

\begin{figure*}
\begin{center}
\includegraphics[width=0.9\linewidth]{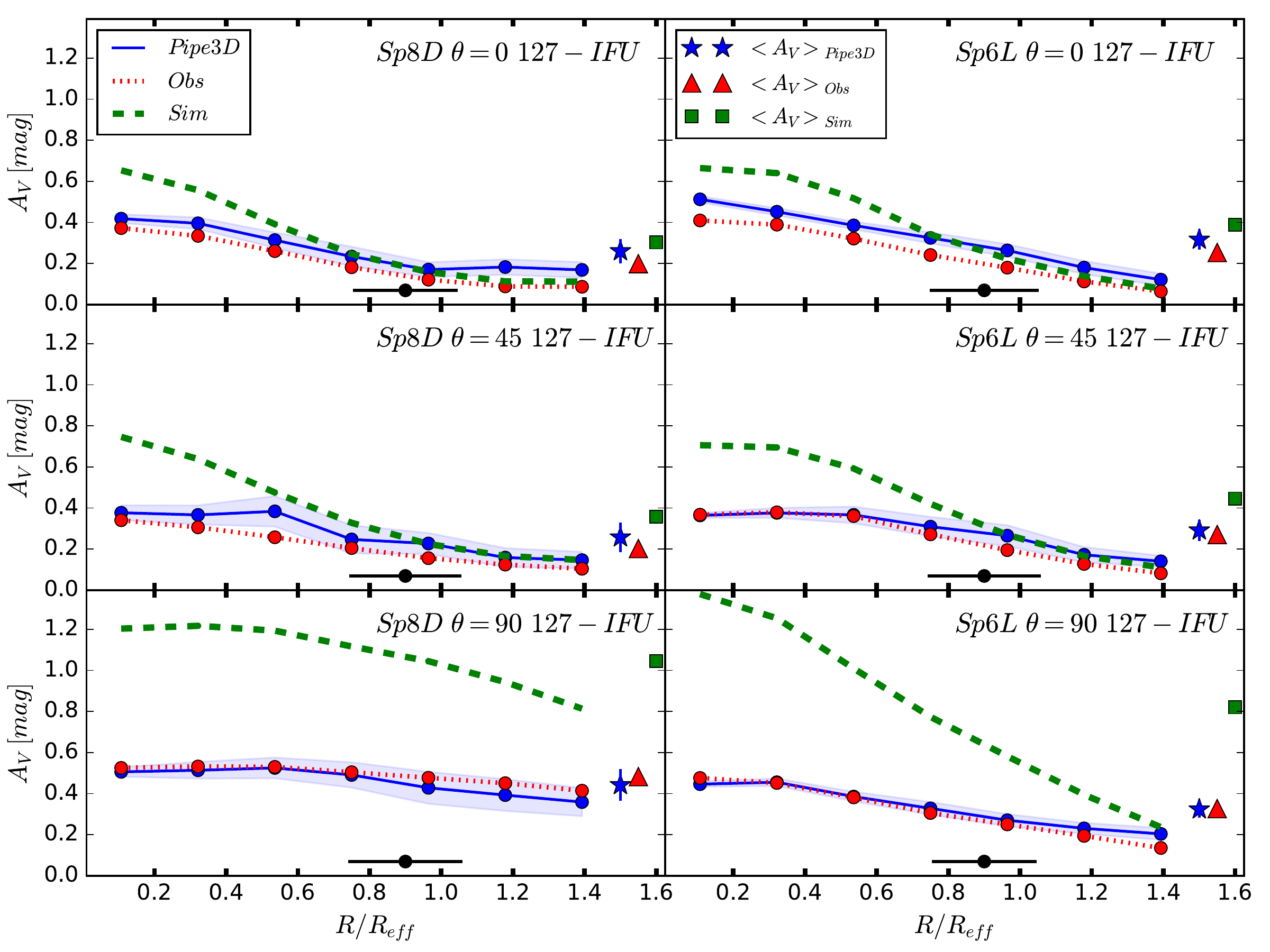}
\end{center}
\caption{Radial profiles of dust attenuation obtained directly from the simulations (dashed green line) and after the instrumental/observational setting (red dotted line), and as recovered with {\sc Pipe3D} (blue solid line). Right and left panel are for the Sp8D and Sp6L simulated galaxies. The instrumental/observational setting in the upper panels corresponds to our fiducial case, while in the medium and lower panels the setting is the same but the inclination angle is set to $\theta=45^{\circ}$ and $\theta=90^{\circ}$, respectively. The black horizontal bars show the respective reconstructed PSF sizes in units of $R_{\rm eff}$. The symbols in the right side show the corresponding average (within $1.5R_{\rm eff}$) dust attenuations.}
\label{extinctio}
\end{figure*}

\begin{figure*}
\begin{center}
\includegraphics[width=0.48\linewidth]{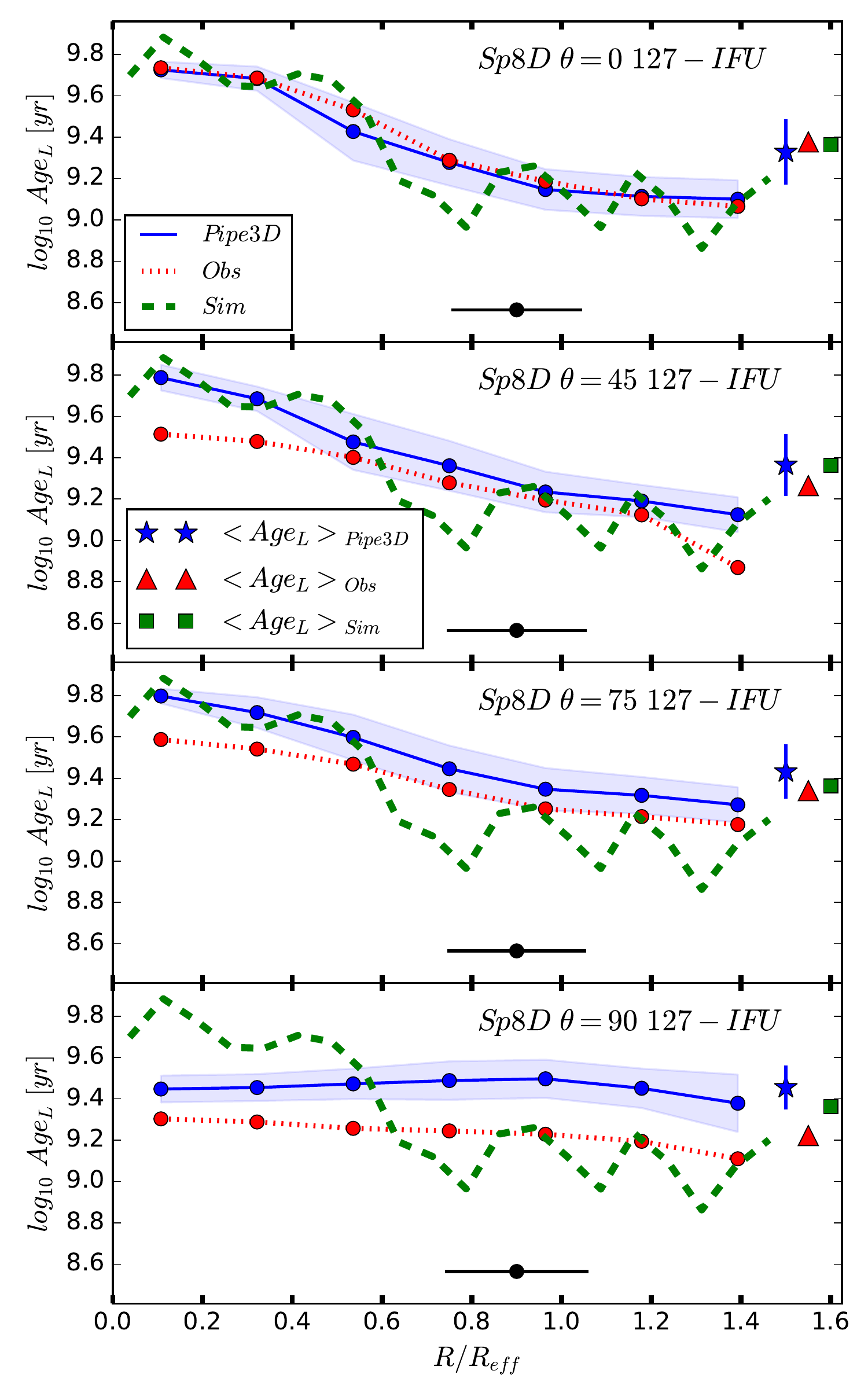}
\includegraphics[width=0.48\linewidth]{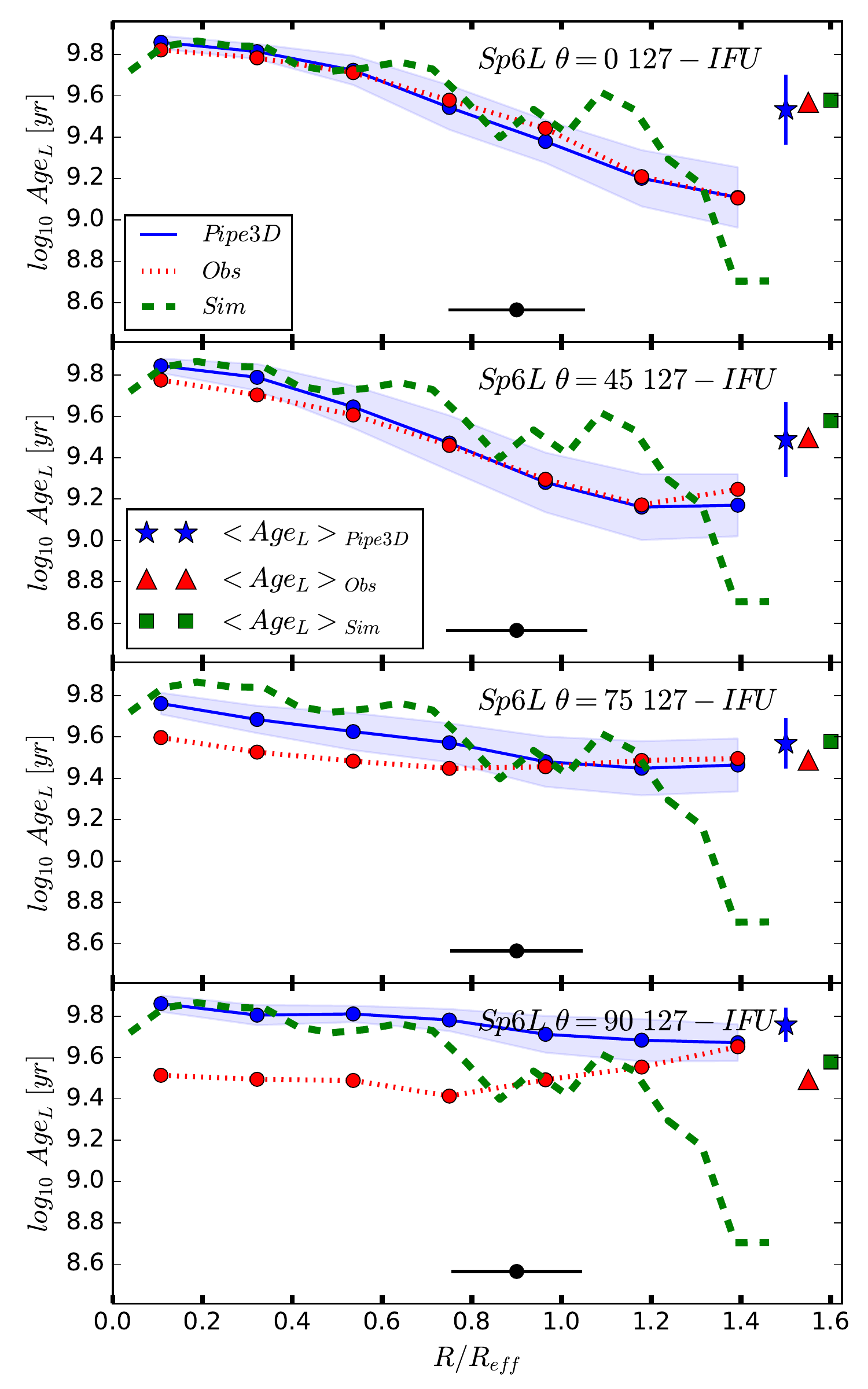}
\end{center}
\caption{As Figure \ref{age_m-res} but for luminosity-weighted (LW) ages and now fixing the IFU number of fibers (127) and varying the inclination angle of the galaxies. From top to bottom the angles are $\theta=0^{\circ}, 45^{\circ}, 75^{\circ}$ and $90^{\circ}$. }
\label{age_l-incl}
\end{figure*}

Figure \ref{age_m-incl} shows the MW age radial profiles as in Figure~\ref{age_m-res} but this time fixing the bundle configuration to 127 fibers and varying the observational inclination of the Sp8D and Sp6L galaxies. As expected, just due to a geometrical effect, the age gradient flattens as the galaxy is observed more inclined (red dotted lines). The effect is stronger for the Sp8D galaxy. The measured ages through the observational setup along the LOS towards the center are similar or younger than the true ones as higher is the inclination, while towards larger radii the ages result older.  Further, the recovered ages by {\sc Pipe3D} (blue solid lines) result older than those measured after applying the observational setup, though yet in agreement with them within the uncertainties. This effect is stronger for Sp8D. It is produced by (i) the accumulation of dust attenuation as more inclined is the galaxy, and (ii) the under-correction of this attenuation due to a geometrical effect (being under-corrected the extinction in the spectra, the recovered stellar populations will result older). Thus the ages recovered older than the true ones in the cases of high inclination are due to a geometrical effect and not by the fossil record method. 
 
This is appreciated in Fig. \ref{extinctio}, where we plot the radial profiles of the dust attenuation (extinction) as extracted directly from the simulations (green dashed line), as measured considering the instrumental/observational setting (red dotted line), and finally the recovered ones by {\sc Pipe3D} from the mock cubes (solid blue line). The extinction extracted from the simulations, $A_{\rm v,sim}$, was created by summing all the dust content along the LOS from the observer to each considered direction. On the other hand, the  observed extinction, $A_{\rm v,obs}$, was built as the average along the same LOS, weighted by the flux intensity of each particle in the $V$-band, taking into account the dust attenuation that affects each particle. Thus, the following equations were used:
\begin{eqnarray}
10^{Av,sim}=\sum_i10^{Av,i}L_i/\sum_i^nL_i,\\
10^{Av,obs}=\sum_i10^{Av,i}F_i/\sum_i^nF_i,
\end{eqnarray}
where $i$ is a running index for all particles along the LOS, $A_{\rm v,i}$ is the dust attenuation affecting the particle $i$ along the same LOS, and $F_i$ ($L_i$) is the flux intensity (luminosity) of the stellar particle in the $V$-band. This way we reproduce the effect that the observed dust attenuation is weighted by the stellar flux as observed through the gas cells along their LOS. This is indeed what {\sc Pipe3D} (and any SED fitting technique) is trying to reproduce by its estimation of the dust attenuation.

As expected, the true extinction increases with inclination. For the inclination of 45$^{\circ}$, this geometrical effect is actually small but more relevant at small galactocentric radii (where the light has to go through the bulge and thick disk). On the other hand, the observed dust attenuation is systematically below the true one, at each galactocentric distance and for each inclination. The effect is stronger in the inner regions, and for the largest inclinations, as expected since the light has to go through a larger amount of dust, and therefore the observed attenuation is more similar to the one in the nearest side of the galaxy than to the total one along the LOS. This result is in agreement with several observational and theoretical studies in this regards \citep[e.g.][]{tuff04,yip10,wild11,catalan15,batt17}. Consistently, the extinction recovered by {\sc Pipe3D} are also below the true ones, being very close to the values expected in an ideal observation through the considered instrumental/observational configuration, with differences below 0.1 mag, and without a clear radial trend. 

Summarizing, the effect of inclination and extinction on the recovered MW age radial profiles (i) is significant in the disk-dominated galaxy only for high inclinations ($\ga 70$ degrees), while for the bulge-dominated galaxy, the effect is of minor relevance, (ii) the inclination/extinction effect works in the direction of making the age radial profile flatter and older as more inclined is the galaxy,  and (iii) this bias is dominated by the geometrical effect of the inclination on the observed dust attenuation along the LOS, with a much weaker contribution of the inaccuracies from the fossil record method.

Finally, in Figure ~\ref{age_l-incl} we present the LW age radial profiles for the same cases as in Figure \ref{age_m-incl}. The trends with inclination are similar to those for the MW age profiles. For high inclinations, (i) the profiles from the observation and the recovered ones with {\sc Pipe3D} get flatter than the true ones, (ii) and the recovered ones are older than the observed ones.

\begin{figure}
\begin{center}
\includegraphics[width=\linewidth]{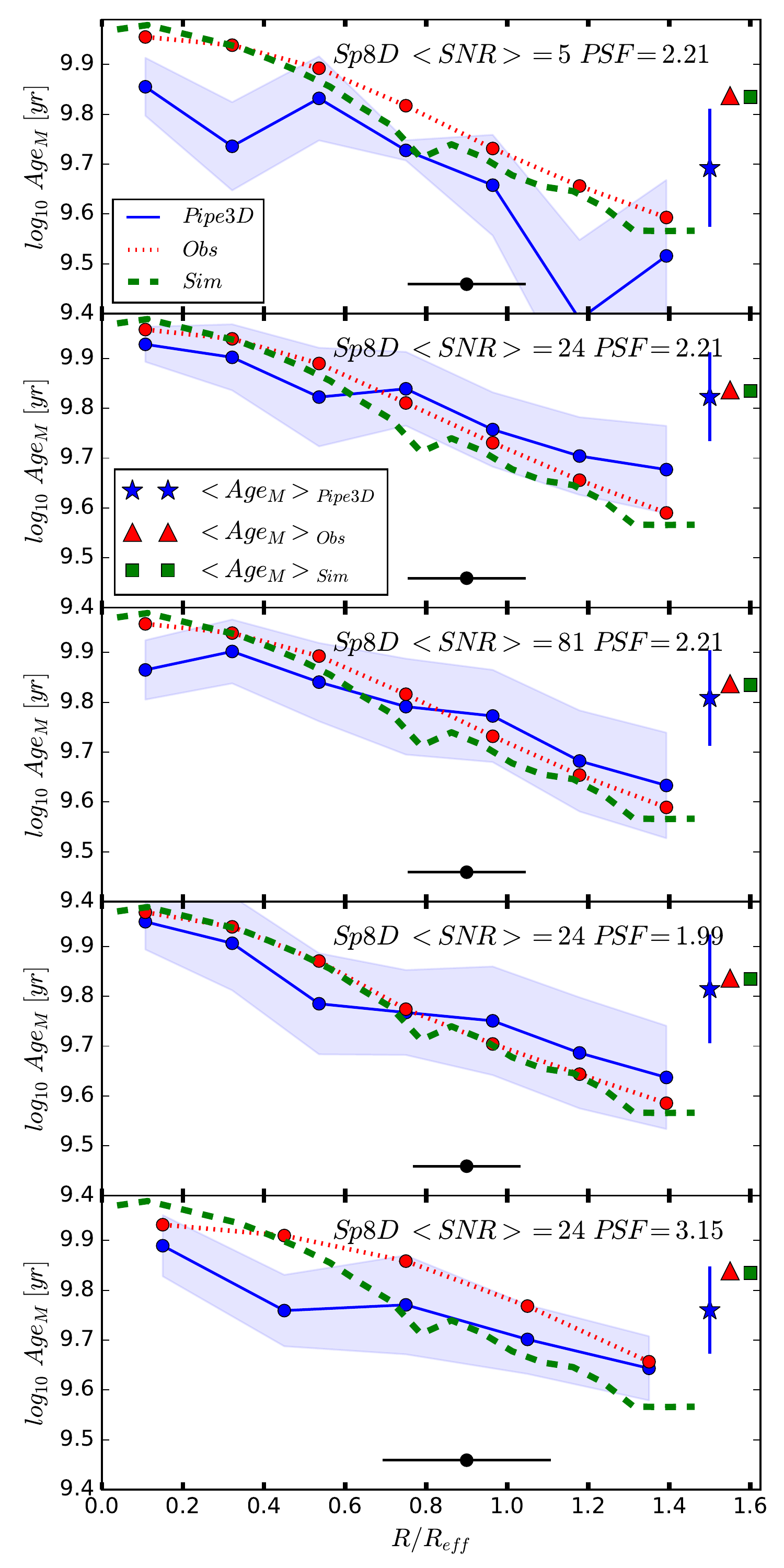}
\end{center}
\caption{Mass-weighted age radial profiles as in Figure \ref{age_m-res} (only for the Sp8D galaxy) but varying the average SNR and PSF dispersion sizes. In the first three upper panels, the average SNR is 5, 24, and 81. In the two lower panels, the reconstructed PSF size is decreased to 1.99 arcsec and increased to 3.15 arcsec with respect to the PSF size of the fiducial case (2.21 arcsec). }
\label{age_m-psf-sn}
\end{figure}


\subsubsection{The effects of the SNR and PSF}

In Figure \ref{age_m-psf-sn} we explore the effects of the average SNR and the PSF size on the inferred MW age radial profiles of the galaxy Sp8D (for Sp6L the effects are much weaker given that its age radial gradient is small). Using the same symbols and lines as in Figure \ref{age_m-res}, we plot for the bundle configuration of 127 fibers and the face-on case, the age radial profiles changing the average SNR (upper panels) and the reconstructed PSF (lower panels); see also Table \ref{tab2}. The SNR affects the fossil record inferences (blue symbols). As seen, the effect is small in the age radial profile when improving the average SNR from 24 to 81.  However, when the SNR is degraded down to 5, the recovered age radial profile results very noisy and significantly younger than the true one. Average SNRs around 20-30 seem to be a good compromise. 

Further, by changing the atmospheric seeing (see  Table \ref{tab2}), we obtain three different reconstructed PSFs and keep the average SNR equal to 24. The PSF size affects the spatial resolution of the observation, producing then an effect similar to the spatial sampling discussed in subsection \ref{sampling}. As larger is the PSF size, lower is the resolution and the measured age radial profile becomes flatter and with older ages at larger radii (red symbols). However, the fossil record method inferences make the MW ages younger as more contamination of young stellar populations enters in the resolved regions (see the discussion in subsection \ref{sampling}).

\begin{figure*}
\begin{center}
\includegraphics[width=1.0\linewidth]{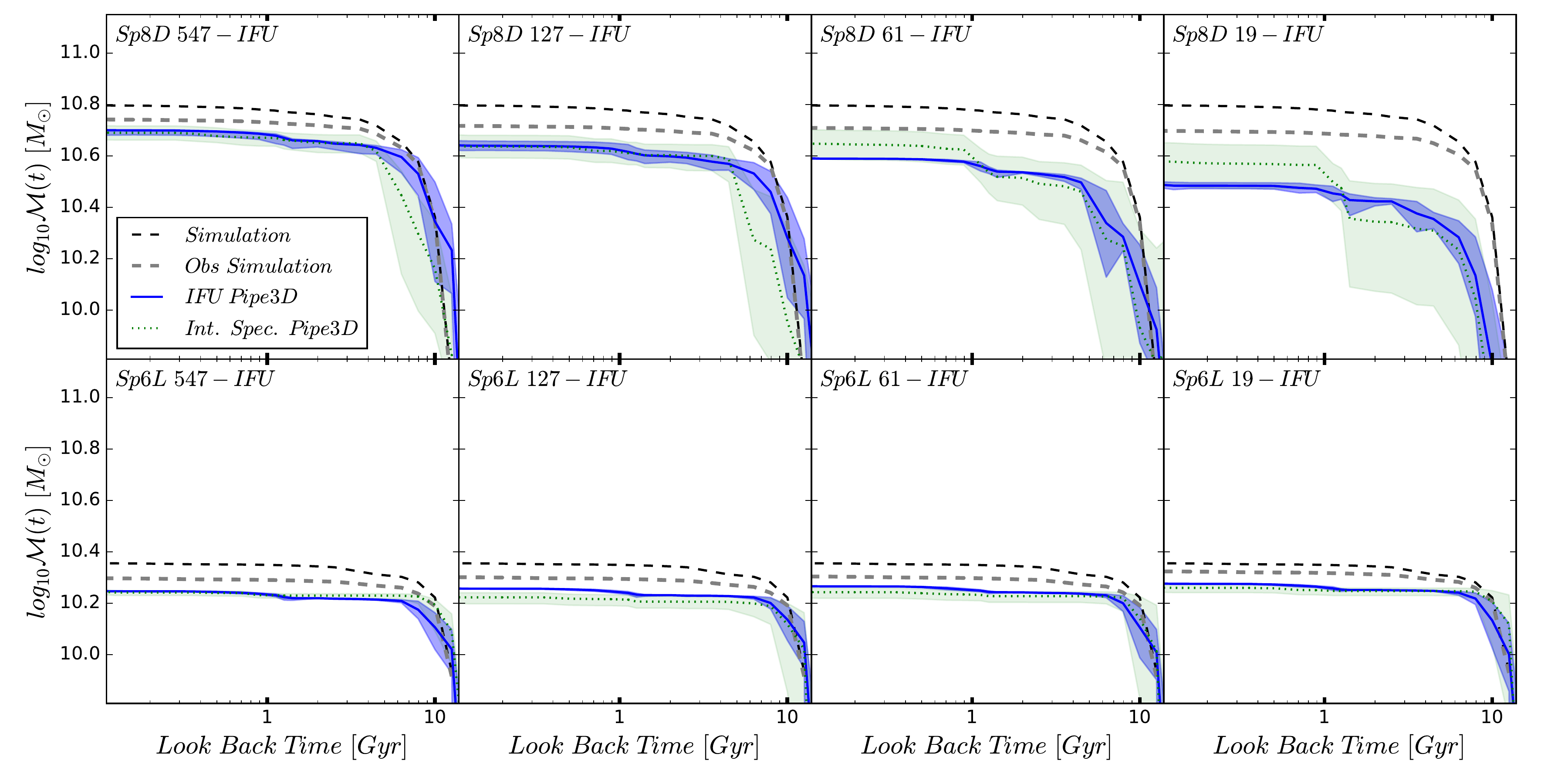}
\end{center}
\caption{Global (within 1.5 $R_{\rm eff}$) MGHs for the Sp8D (upper panels) and Sp6L (lower panels) face-on galaxies for different IFU number of fibers, and with the reconstructed PSF dispersion and SNR values corresponding to our fiducial setting. The black and gray dashed lines are for the archaeological MGHs obtained directly from the simulations and after applying the IFU and observational conditions, respectively. The solid blue and dotted lines are the MGHs recovered from the IFS mock observations with {\sc Pipe3D}. The former were calculated by summing the masses recovered with the fossil record method applied to the spatially-resolved spectra (spaxels), while the latter are from the fossil record method applied to an unique integral spectrum obtained by summing the spectra of all the resolved regions. The corresponding shaded areas around the lines are the estimated 1-$\sigma$ uncertainties (see the text for details).}
\label{globalMGHs}
\end{figure*}

\subsection{Global stellar MGHs}
\label{SubglobalMGHs}

{\bf Masses at $z=0$.-} How well the fossil record method recovers the galaxy stellar masses?  In Table \ref{tab2}, we report the differences between the recovered {\sc Pipe3D} and true global stellar masses at $z=0$, $\Delta \mathcal{M}_0=\log(\mathcal{M}_{\rm 0,Pipe3D}/\mathcal{M}_{\rm 0,true})$, for the different cases. The masses are calculated within 1.5 $R_{\rm eff}$. For our fiducial setting, the differences are of $-0.15$ and $-0.098$ dex for Sp8D and Sp6L, respectively. Thus, the recovered global masses at $z\sim 0$ are slightly underestimated by {\sc Pipe3D}. 
In fact, this underestimation is partially just due to the instrumental/observational setting, including the dithering process. By using the mass maps of the stellar particles along the LOS as obtained through the IFU bundle and data cube generation (Figs. \ref{Sp8D-maps} and \ref{Sp6L-maps}), we calculate the mass within 1.5 $R_{\rm eff}$. For our fiducial setting, the differences between the ``observed'' and true masses are of $-0.08$ and $-0.05$ dex for Sp8D and Sp6L, respectively. The rest of 
the difference between the recovered and true masses is likely because the ages are recovered slightly younger than the true ones in the inner regions (see Figure \ref{age_m-res} and discussion in subsection \ref{GLSFH}), which more contribute to the total mass. This implies lower M/L ratios, and therefore lower stellar mass surface densities in the inner regions (see Figs. \ref{Sp8D-maps} and \ref{Sp6L-maps}).  
For lower spatial resolution and sampling, the underestimation in the mass of Sp8D increases, both due to the instrumental/observational conditions and the fossil record method, while for Sp6L, remains roughly the same. For high inclinations, the underestimation increases for both galaxies, but more for Sp8D.

An interesting question is how different can be the global stellar masses inferred through the fossil record method in two different cases:  (i) by summing the spectra of all the resolved regions to get a co-added spectrum of the whole galaxy that is used then for the fossil reconstruction, and (ii) by applying the fossil reconstruction to each spectrum from the resolved regions, and summing the masses inferred for these regions. The results shown above and along this paper refer to the latter.  For the fiducial case, the differences with the true masses when using the integrated spectrum are now  of $-0.16$ and $-0.13$ dex for Sp8D and Sp6L, respectively. These differences are actually similar, within the uncertainties, to those reported above, i.e., the case of using the resolved spectra for the fossil reconstruction.

{\bf The MGHs.-} In Figure \ref{globalMGHs} we plot the global (within 1.5 $R_{\rm eff}$) face-on MGHs for four IFU configurations (four bundle sizes, see above), both for Sp8D (upper panels) and Sp6L (lower panels) galaxies. In the X axis is plotted the look-back time (hereafter LBT). The black dashed line is the true archaeological MGH (using the same temporal basis as in Pipe3D), the gray dashed line is the ``observed'' archaeological MGH obtained from the processed mass maps (see e.g., Figs. \ref{Sp8D-maps} and \ref{Sp6L-maps}), and the blue solid and green dotted lines are the recovered MGHs with {\sc Pipe3D}, for the spatially-resolved and integrated spectra, respectively. The blue shaded area around the solid line corresponds to the estimated 1-$\sigma$ uncertainty in the recovered MGH. This uncertainty is calculated by propagating the errors given by {\sc Pipe3D} for the mass at each SSP age in each spaxel, and taking into account the age resolution of the SSPs along the LBT. For the integrated spectrum, the green shaded area around the dotted line is the standard deviation from 500 fits of {\sc Pipe3D} with different seeds ({\sc Pipe3D} introduces a random noise to the observed spectrum adopting the reported variance of the data to derive the errors; see for more details \citealp{Sanchez:2016aa}).

It is appreciated that the instrumental/observational settings, including the dithering process, lead to an underestimation of the global mass (black dotted vs. grey dashed lines). The effect is stronger for lower spatial resolutions/sampling, and more evident for the Sp8D galaxy than for the Sp6L one. On top of this, the fossil records produce an additional underestimation and distortion in the shape of the MGHs.

For the {\it spatially-resolved fossil reconstruction (blue solid line)}, the global MGHs of the early-assembled Sp6L galaxy are reasonably well recovered even when the spatial sampling is poor (small number of fibers). Since Sp6L is dominated by old stellar populations with a radial distribution nearly uniform, the inferences under different spatial samplings do not change significantly. In general, the initial mass growth rate ($t_{\rm lb}>12$ Gyr) inferred with {\sc Pipe3D} appears to be slightly faster than the true one for both galaxies. This may be due to the lack of age resolution for these large LBTs and the use of a SSP template as old as 14.2 Gyr, when in the simulations the oldest populations are of $\approx 13.2$ Gyr. For Sp6L, since $\sim 10$ Gyr ago the mass growth rates are similar to the true ones. However, the total inferred stellar mass remains lower than the true one because the fossil record method assigns on average younger populations than the true ones in the inner regions (see Fig. \ref{age_m-res} and subsection \ref{GLSFH}), and these regions dominate in the mass contribution. Since younger populations have lower M/L ratios, this leads to a lower total mass. 

In the case of Sp8D, this effect is stronger. Thus, the differences between the inferred and true (or ``observed'') MGHs are larger for this galaxy. On the other hand, Sp8D has a pronounced radial gradient in the stellar populations. Therefore, one expects that the spatial sampling/resolution will affect more the recovered MGHs than in the case of Sp6L. As worse is the sampling and resolution, the more is shifted to lower LBTs the global MGH and the lower is the recovered mass.
Due to the poor spatial sampling, the inner regions, which largely contribute with old stellar populations, are smoothed out. Therefore, the global MGH resembles more the MGHs corresponding to intermediate regions, which present a later mass assembly than those of the inner regions (see subsection \ref{radialMGHs} below).  On the other hand, the old populations have high M/L ratios; if they are smoothed out due to the poor resolution, then the global recovered masses result lower than the true ones.

The MGHs calculated {\it from an unique integrated spectrum (dotted green line)} result worse than those calculated from the spatially-resolved spectra, specially for the Sp8D galaxy. The inference from only one integrated spectrum offers less possibilities to constrain the mix of SSP ages than in the case of many resolved regions. Therefore, its ``temporal resolution'' in the recovered global MGH is worst, and the ``contamination'' in the integrated spectrum by the intense light of young stellar populations, even if they are a small fraction, may lead to a MGH biased to lower LBTs. In the spatially-resolved case, as better is the spatial sampling, more spectra to infer the global MGH are available, many of them from regions, where there are not young populations. Therefore, the recovery of the global MGH is expected to be better as better sampling and resolution has the observation.

\begin{figure*}
\begin{center}
\includegraphics[width=0.95\linewidth]{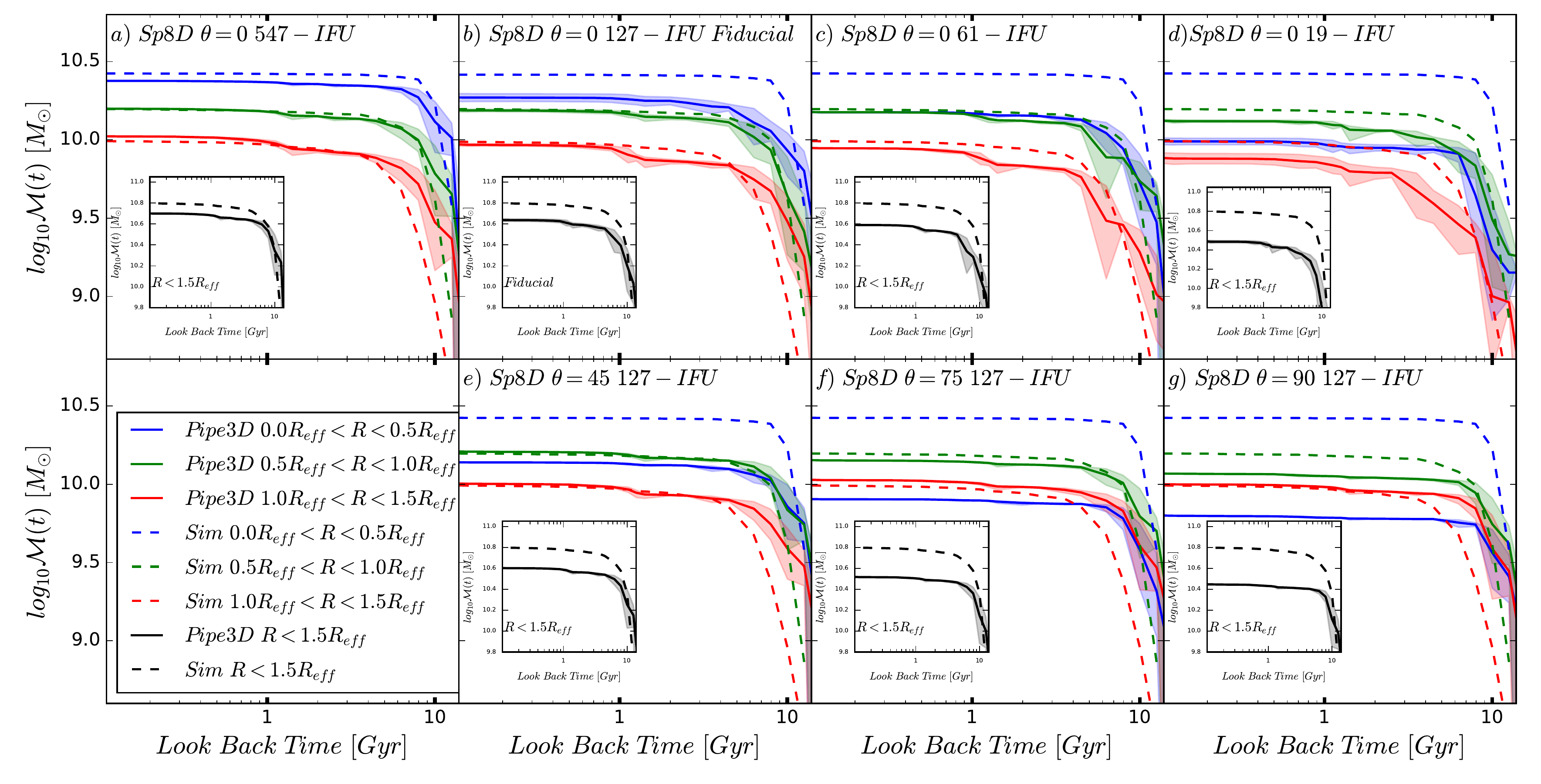}
\includegraphics[width=0.95\linewidth]{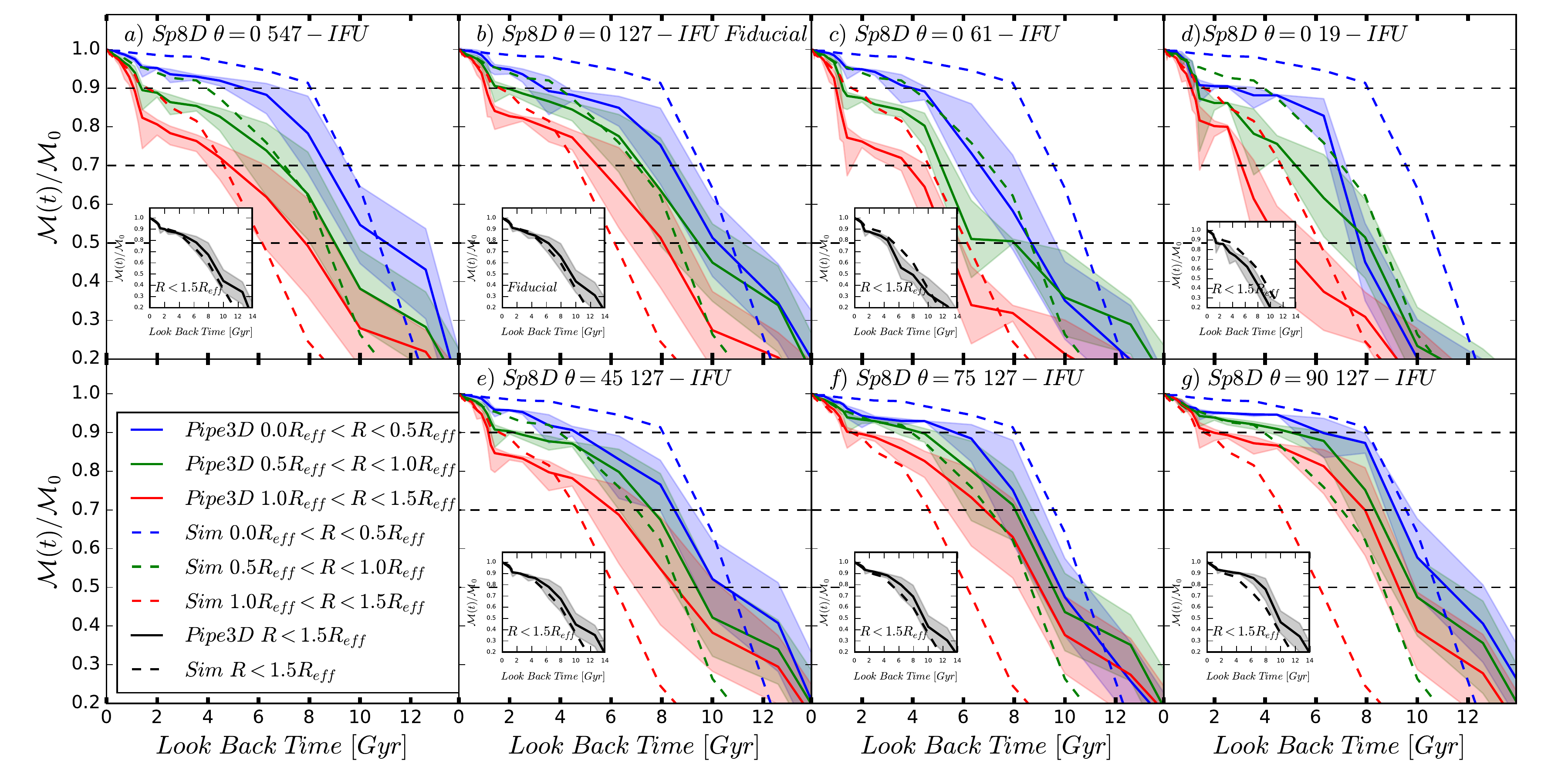}
\end{center}
\caption{Radial MGHS, $\mathcal{M}(t_{\rm lb})$ (upper block of panels) and normalized MGHs, $\mathcal{M}(t_{\rm lb})/\mathcal{M}_0$ (lower block of panels) for the Sp8D galaxy under different instrumental/observational conditions. The respective global (within 1.5 $R_{\rm eff}$) absolute and normalized MGHs are shown in the insets. The blue, green, and red dashed lines correspond to the average MGHs obtained directly from the simulations along the LOS within the radial bins $R/R_{\rm eff}<0.5$, $0.5<R/R_{\rm eff}<1$, and $1<R/R_{\rm eff}<1.5$, respectively. The respective solid lines are the MGHs recovered with {\sc Pipe3D} from the IFU data cubes.  The shaded areas show the estimated 1-$\sigma$ uncertainties from the fossil record method. 
In the upper row of each block of panels the IFU number of fibers is varied from 547 to 19, while the other instrumental/observational conditions are kept as in the fiducial case. The panels (b) correspond to the fiducial case (127 fibers, $\theta=0$).  In the lower row of each block of panels the observational inclination angle is varied from $\theta=45^{\circ}$ to $\theta=90^{\circ}$ (for $\theta=0^{\circ}$, see panels b), while the other instrumental/observational conditions are kept as in the fiducial case. }
\label{sp8_re_mgh}
\end{figure*}

\begin{figure*}
\begin{center}
\includegraphics[width=0.95\linewidth]{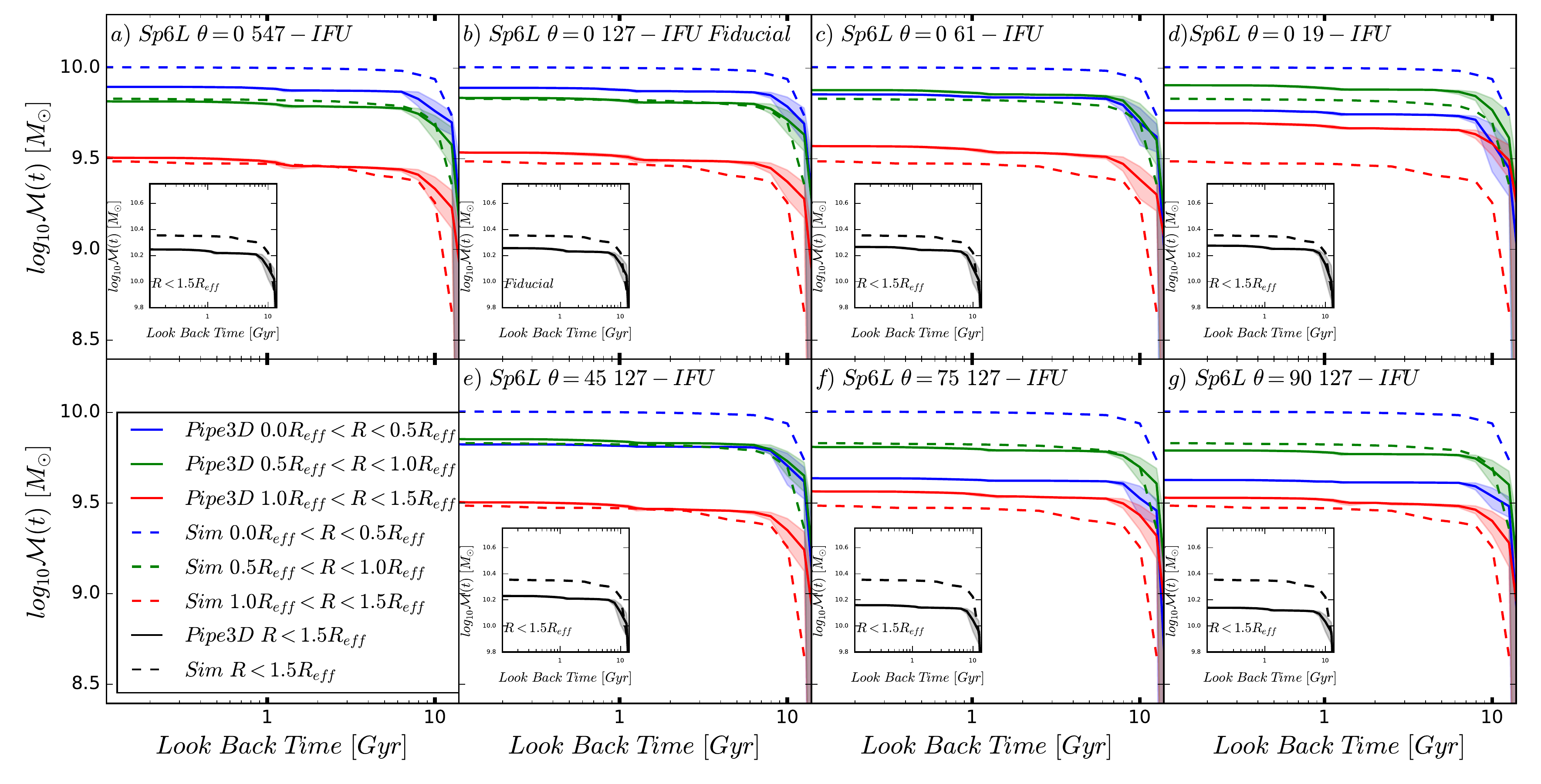}
\includegraphics[width=0.95\linewidth]{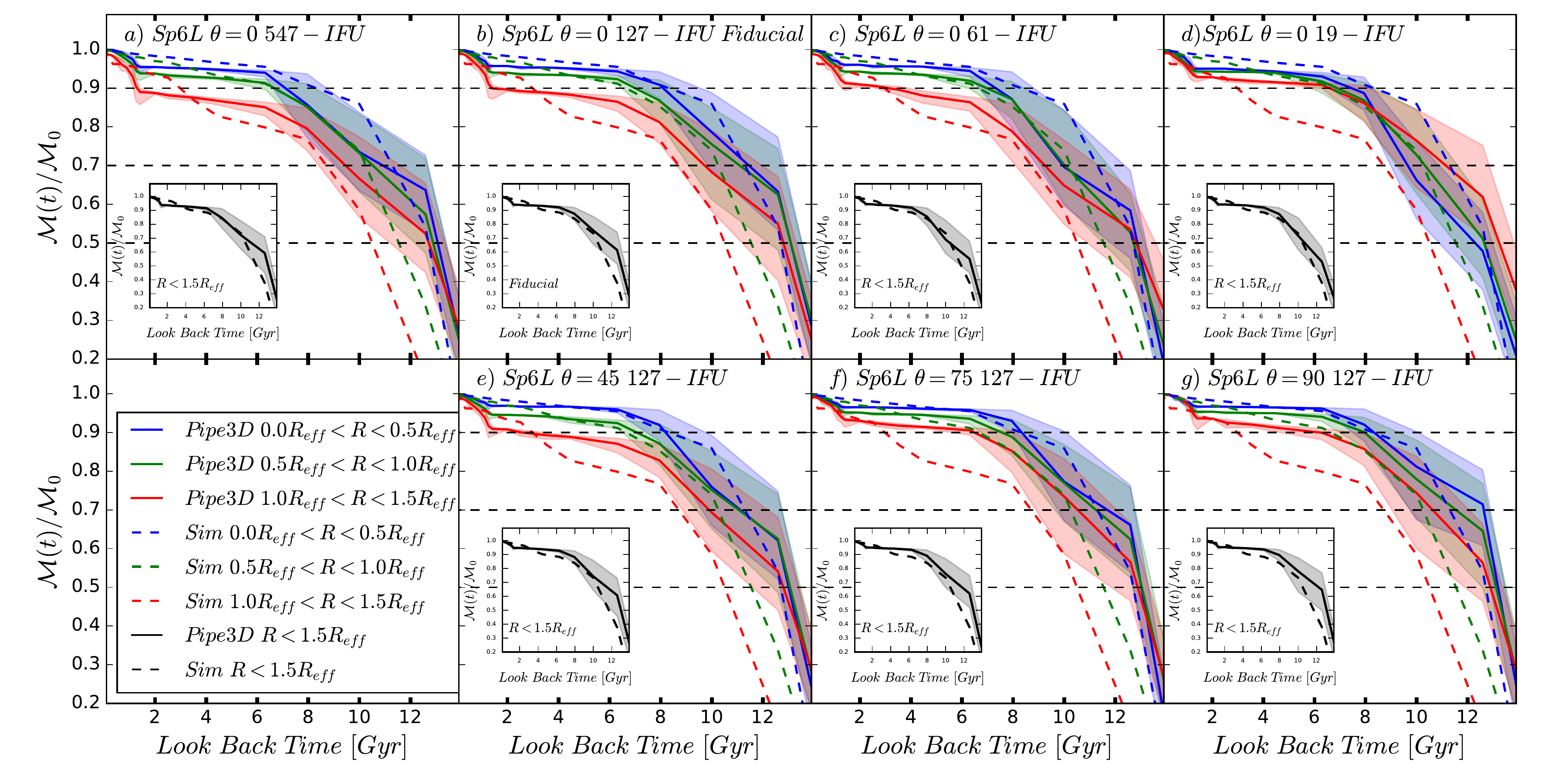}
\end{center}
\caption{Same as Figure \ref{sp8_re_mgh} but for the Sp6L galaxy.
}
\label{sp6_re_mgh}
\end{figure*}

\subsection{Radial stellar MGHs}
\label{radialMGHs}

The archaeological MGHs for three radial bins of the galaxies Sp8D and Sp6L inferred with {\sc Pipe3D} for different instrumental/observational settings, are plotted in the {\it upper block} of panels of Figures \ref{sp8_re_mgh} and \ref{sp6_re_mgh}, respectively. The shaded regions around the solid lines are the estimated 1-$\sigma$ uncertainties (see above). The dashed lines correspond to the (true) archaeological MGHs directly measured from the simulations in these radial bins.  The corresponding global MGHs, within 1.5 $R_{\rm eff}$, are  shown in the insets. In the {\it lower block} of panels, we present the respective {\it normalized} MGHs. Note that in this case both axes are lineal. The normalized MGHs represent the relative rate of mass growth at different epochs, and allow to compare these rates as a function of the radius, in such a way that we may evaluate, for instance, whether the radial growth is inside-out or outside-in. The normalized radial MGHs were studied extensively in observational works \citep[e.g.,][]{Perez+2013,Ibarra-Medel+2016}.    

Let us first analyze the {\it fiducial setting} (panels b in Figs. \ref{sp8_re_mgh} and \ref{sp6_re_mgh}; see the beginning of this Section and Table \ref{tab2}). 
The mass contained within the radial bins of both galaxies is smaller as larger is the radius at all epochs, more for Sp6L, which is a more compact galaxy than Sp8D. Since early times, the fossil record method recovers MGHs lower in mass than the true ones for the inner, earlier formed regions, while for the outer, later formed regions the MGHs are similar or even slightly higher in mass than the true ones at all epochs. We can understand at a qualitatively level this systematical behavior, associated also to the flattening of the age gradients discussed in subsection \ref{age-gradients}, by examining the respective radial SFHs (see subsection\ref{GLSFH}). For the intermedium radial regions, which assemble the stellar mass more constantly in time, the recovered MGHs are similar to the true ones, excepting at large LBTs.  

\begin{figure*}
\begin{center}
\includegraphics[width=0.48\linewidth]{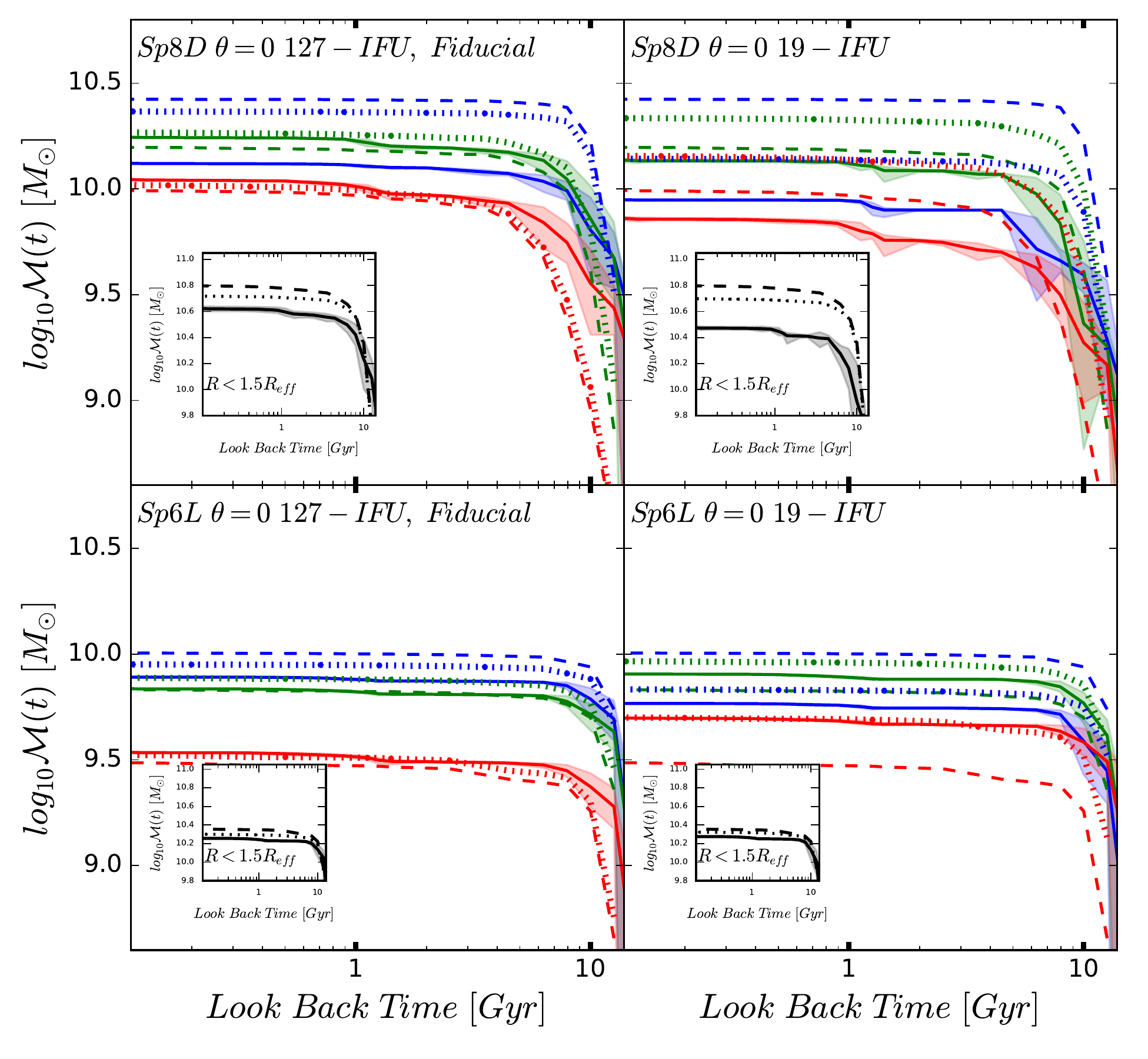}
\includegraphics[width=0.48\linewidth]{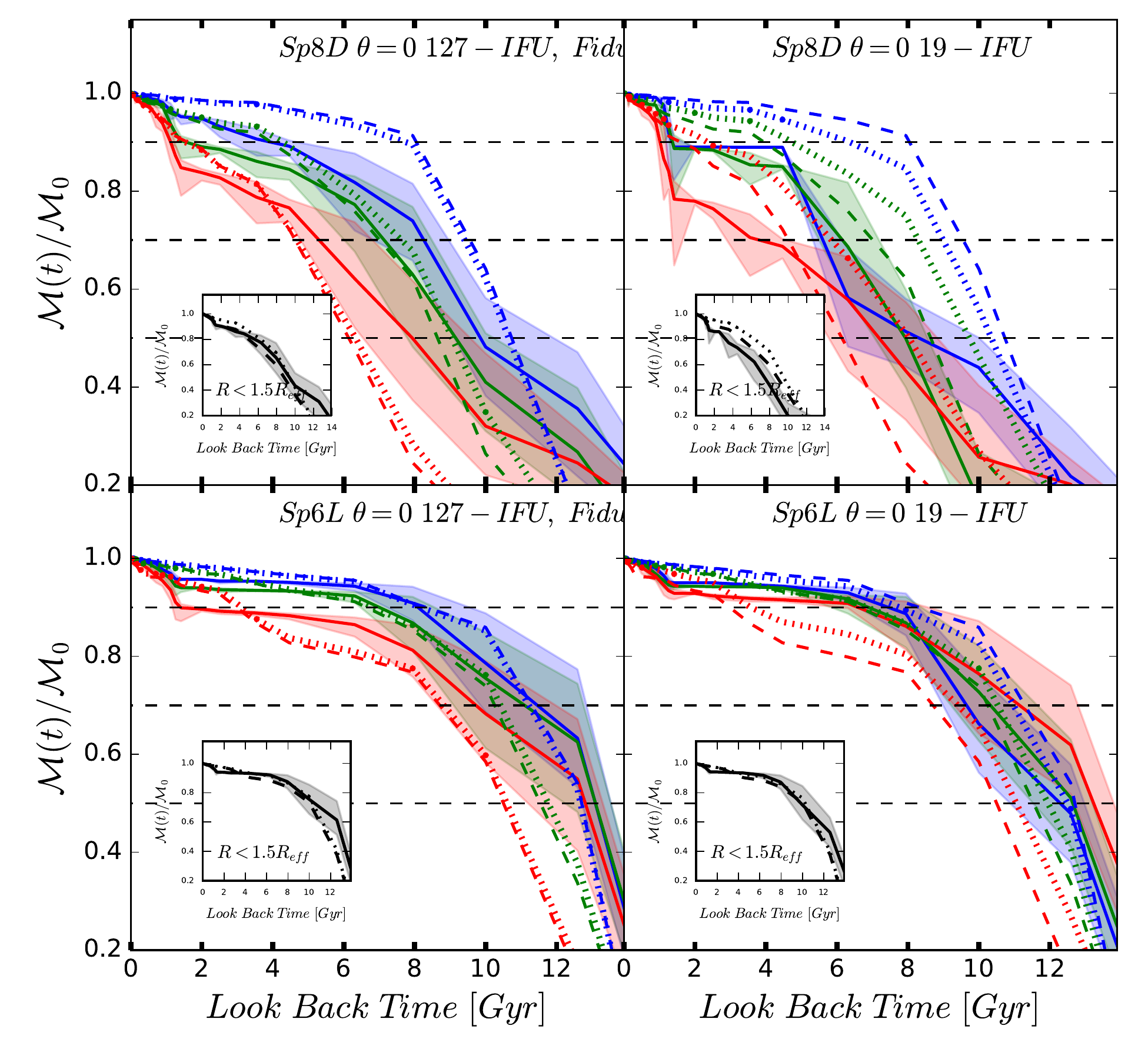}
\end{center}
\caption{As panels b) and d) (face-on observations with 127 and 19 fibers) from Figures \ref{sp8_re_mgh} and \ref{sp6_re_mgh} but showing now the respective radial absolute MGHS (right block of panels) and normalized  MGHs (left block of panels) as obtained from the simulations after imposing the instrumental/observational conditions (dotted lines). For the well resolved/sampled case of 127 fibers, these conditions do not significantly affect the ``observed'' radial MGHs (as well as the global MGHs, shown in the insets), so that the differences between the true (dashed lines) and recovered by {\sc Pipe3D} (solid lines) absolute/normalized MGHs are mainly due to the fossil record method. For the case of 19 fibers, both the poor resolution/sampling and the fossil record method affect the recovered MGHs.  
}
\label{obs-vs-Pipe3D}
\end{figure*}

{\it The inside-out growth mode,} i.e., the later assembly as outer are the regions of the simulated galaxies, is recovered with {\sc Pipe3D} at a qualitative level, at least when using three radial bins. This can be better appreciated in the plots of the normalized MGHs (panel b in the lower block of Figures \ref{sp8_re_mgh} and \ref{sp6_re_mgh}).  
For the early-assembled galaxy Sp6L, the three inferred normalized MGHs (solid color lines) are nearly similar to the true ones (dashed color lines) with differences $\lesssim 10\%$ in the last $\sim 10$ Gyr. For the galaxy with an extended assembly history, Sp8D, these differences are $\lesssim 10\%$ only in the last $\sim 5$ Gyr for the outermost radial bin, and in the last $\sim 7-8$ Gyr for the inner and intermediate radial bin. As seen, the recovered radial normalized MGHs, in general, present a {\it less pronounced inside-out mass growth than the true ones}  both for the Sp8D and Sp6L galaxies.   
At the largest LBTs, where the age resolution of the SSP templates is very poor, the recovered normalized MGHs show apparent faster early growth rates than the true ones, but at smaller LBTs, the normalized MGHs become slower or similar to the true ones. The latter effect is partially by construction given that both the recovered and true normalized MGHs should tend to 1 at $t_{\rm lb}\rightarrow0$.  However, there are clear differences in this trend with radius: the transit from faster to slower growth rates with respect to the true one happens at larger LBTs as inner are the radii, in such a way that the original (inside-out) differences among the three radial normalized MGHs result diminished, more in the case of Sp8D than in the case of Sp6L. As we will see below (Fig. \ref{obs-vs-Pipe3D}) this diminish of the inside-out trend is basically produced by the fossil record method and not by the observational conditions (dithering, seeing, etc.). For a discussion on the reasons of this result, the reader is again referred to subsection \ref{GLSFH}.

\subsubsection{The effects of spatial sampling and resolution} 
\label{resolution-inclination}

In the upper rows of both blocks of panels of Figure \ref{sp8_re_mgh}, we present the radial/global absolute and normalized MGHs of the galaxy Sp8D inferred with {\sc Pipe3D} for different bundle configurations: 547, 127, 61, and 19 fibers, while keeping the same inclination ($\theta=0$), atmospheric PSF seeing, dithering procedure, and noise limit $\sigma_{\rm lim}$ (see Table \ref{tab2}). By decreasing the number of fibers, the bundle size decreases (hence the spatial resolution decreases) and the spatial sampling get poorer. As seen, the absolute and normalized MGHs are better recovered for the 547 fibers configuration. However, even in this case, the recovered radial normalized MGHs show a less pronounced inside-out mass growth than the true one. As the spatial resolution/sampling decreases, the radial MGHs are recovered with less precision, being in general of lower amplitude (lower masses) and with growth rate histories delayed with respect to the true ones. 

The upper rows of both block of panels of Figure \ref{sp6_re_mgh} show the same sequence of bundle configurations as in Figure \ref{sp8_re_mgh} but for the galaxy Sp6L.  As seen, the absolute and normalized radial MGHs are equally well recovered with 547 and 127 fibers. The recovery of the normalized MGHs is reasonable even for 19 fibers, excepting for the innermost radial bin, which is largely smoothed out due to the lack of resolution.  The original inside-out trend tends to be diminished, more as worse is the spatial sampling. 

To explore what does influence more on the recovered radial MGHs, either the spatial resolutions/sampling or the fossil record method, we measure the MGHs directly from the simulated galaxies but after applying the instrumental/observational conditions, taking into account the dithering process. In particular, we explore two IFU bundle configurations corresponding to 127 and 19 fibers. In Figure \ref{obs-vs-Pipe3D}, the dotted lines show the results (upper panels for Sp8D and lower panels for Sp6L); the dashed lines, as in Figures \ref{sp8_re_mgh} and \ref{sp6_re_mgh}, are the true archaeological normalized radial MGHs, while the solid lines correspond to the inferred MGHs with Pipe3D. The instrumental/observational conditions almost do not affect the radial MGHs in the case of 127 fibers and face on for both Sp8D and Sp6L. Therefore, for face-on, well resolved/sampled cases, the differences between the recovered MGHs with {\sc Pipe3D} and the true ones are mainly due to the fossil record method, that is, the spectral inversion method and the SSP age binning, see subsection \ref{GLSFH}. 

For only 19 fibers, the radial MGHs measured from the simulation are clearly distorted by the observational conditions:  the inside-out growth trends are significantly diminished. The poor spatial sampling and dithering contribute to mix the stellar populations of different regions, producing this radial MGHs more similar among them than the true ones. In particular, the innermost radial bin, which is dominated by old stellar populations in both galaxies, is strongly smoothed out and mixed with younger populations from the more external radii. These populations have lower M/L ratios, therefore, the global recovered masses result lower than the true ones. 
In addition, the MGHs calculated with the fossil record method from the observed spectra, suffer further distortions due to the inversion method and age binning; all the radial MGHs are shifted to later epochs and lower masses, specially for Sp8D.    
As discussed in the previous subsection, as poorer is the spatial sampling, the lower is the possibility to have spectra not affected by young populations. The inversion method is very sensitive to features produced by young populations, even if they make intrinsically a small fraction in mass, in such a way that the inferred MGHs result biased to younger populations.

\begin{figure*}
\begin{center}
\includegraphics[width=0.48\linewidth]{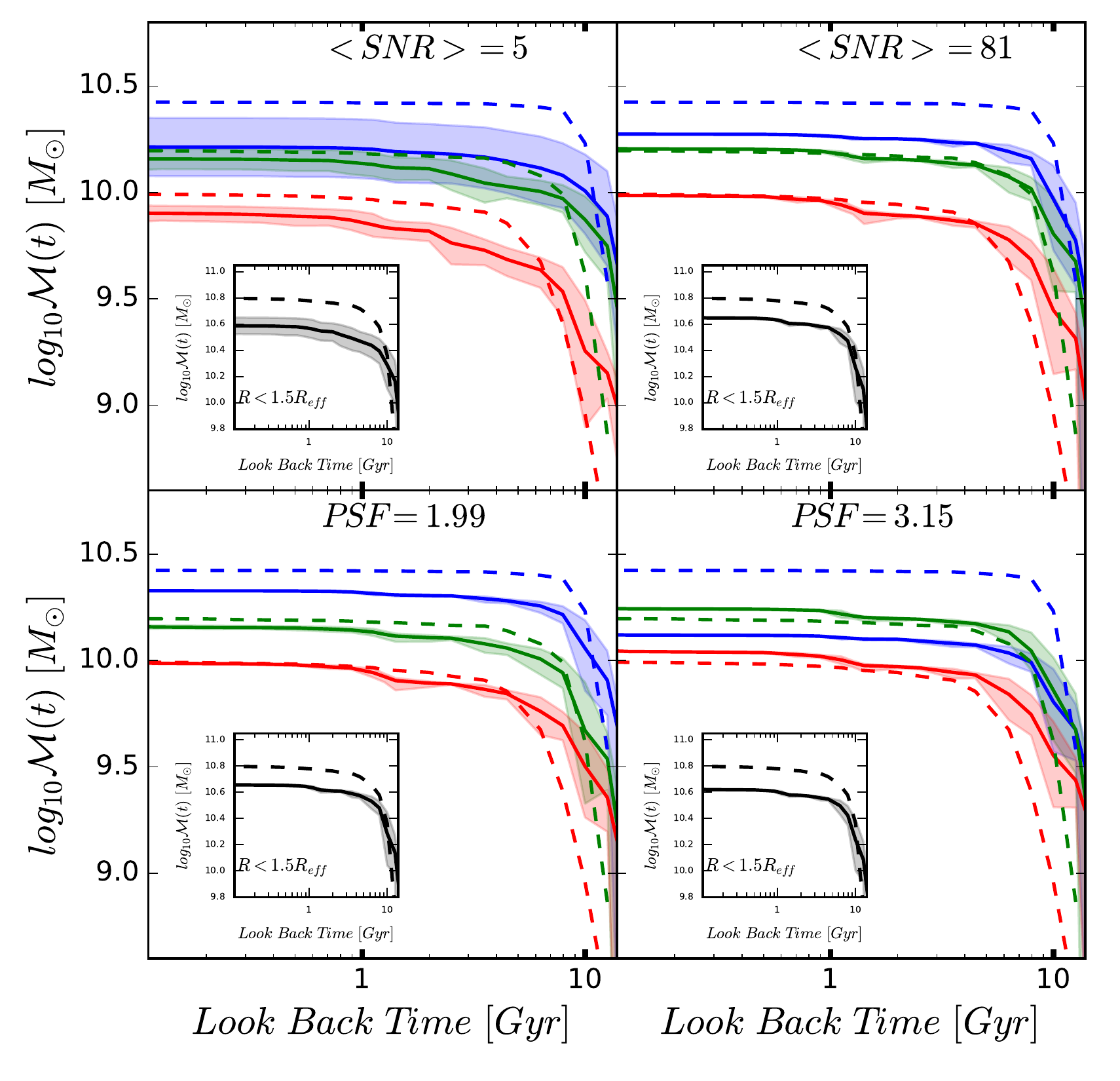}
\includegraphics[width=0.48\linewidth]{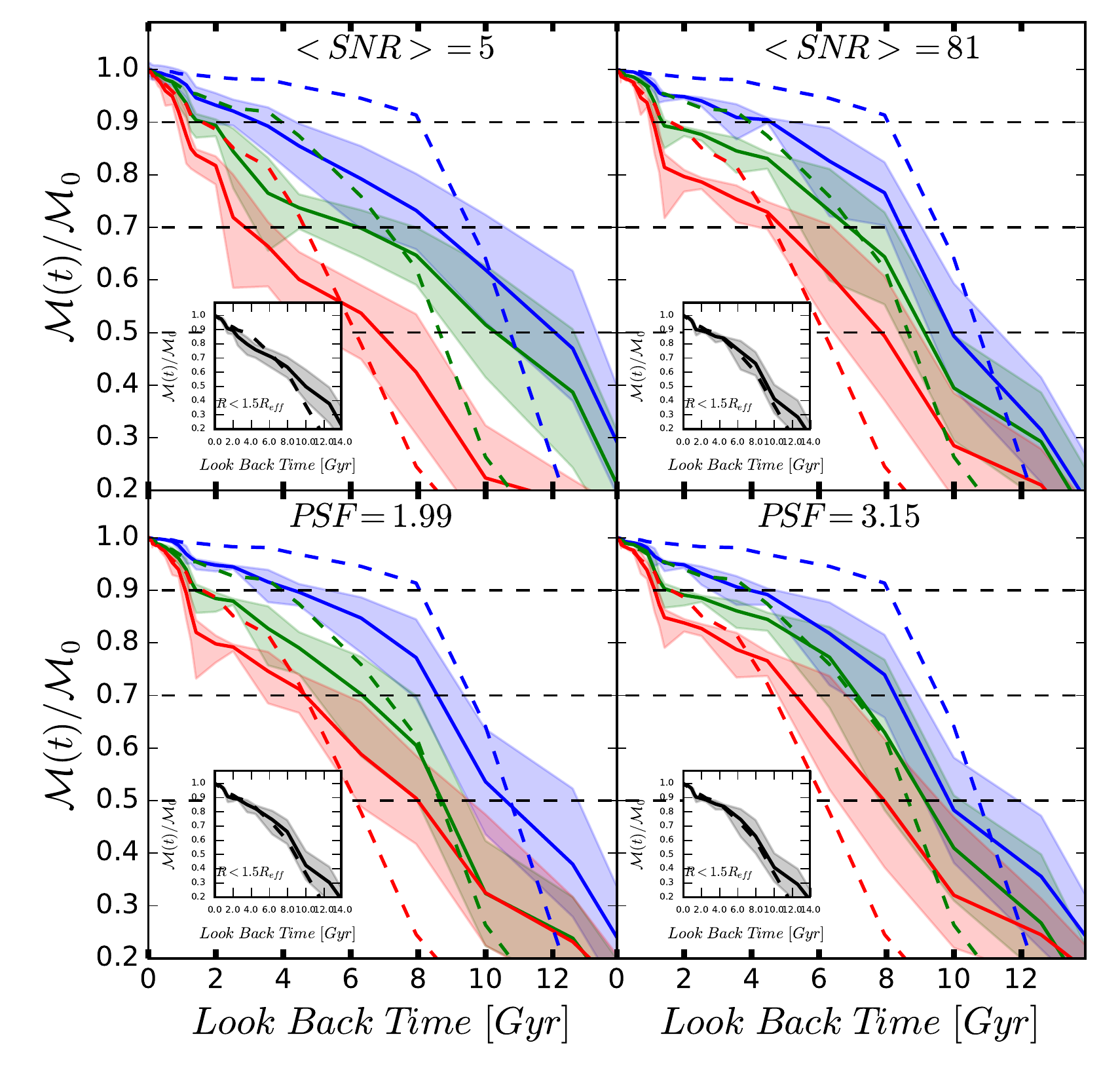}
\end{center}
\caption{Effects of changing the average SNR and PSF dispersion on the recovered absolute MGHs (left block of panels) and normalized MGHs (right block of panels) for the galaxy Sp8D. The fiducial instrumental/observational setting is applied excepting that (i) in the upper panels the average SNR is degraded to a value of 5 or improved to a value of 81 (the fiducial value is 24, see panel b of Fig. \ref{sp8_re_mgh}), and (ii) in the lower panels the reconstructed PSF size is lowered to 3.15 arcsec or improved to 1.99 arcsec (the fiducial value es 2.21, see again panel b of Fig. \ref{sp8_re_mgh}). The line styles and color are as in Fig. \ref{sp8_re_mgh}.}
\label{mgh_sn_psf}
\end{figure*}


\subsubsection{The effects of inclination and extinction}
\label{inclination-MGHs}

We explore now the effects of inclination and extinction on the reconstruction of the global and radial MGHs. In the lower panels of the two blocks of Figures \ref{sp8_re_mgh} and \ref{sp6_re_mgh}, we show these histories as inferred with {\sc Pipe3D} (solid lines) for the simulated galaxies Sp8D and Sp6L, respectively, keeping the fiducial observational setup but varying the inclination ($\theta=45^{\circ}$, $75^{\circ}$ and $90^{\circ}$; the case of $\theta=0^{\circ}$ is shown in panels b). For comparison, we plot with dashed lines the MGHs as measured directly from the simulated galaxies. We have seen already in Figure~\ref{simu_an} (Appendix \ref{projection}) that the measured radial normalized MGHs tend to present smaller differences among them than the true ones when the inclination is high just due to a superposition of the inner (older) and outer (younger) stellar particles along the LOS. As discussed in subsection \ref{resolution-inclination2}, besides this geometrical effect, as more inclined is the galaxy, the extinction increases but the ``observed'' extinction along the LOS results more underestimated, again due to a geometrical effect (see Fig. \ref{extinction}).  Though {\sc Pipe3D} recovers well the dust attenuation along the LOS, this attenuation is highly underestimated by the observation when the galaxies are highly inclined, and more towards the center. The incomplete recovery of the true extinction has two effects on the spectrum associated to the stellar populations: it remains with a lower flux and this happens more efficiently at the shorter wavelengths due to the extinction law (``reddening''; see Eq. \ref{stellar_ext}). Consequently, the inversion method recovers lower stellar masses and older stellar populations than the real ones.

The combination of the effects mentioned above for high inclinations explains the trends seen in Figures \ref{sp8_re_mgh} and \ref{sp6_re_mgh}. For the innermost radial bins, due to the geometrical effect, the old intrinsic populations are contaminated by the younger ones seen along the LOS, but at the same time the accumulation of extinction and its underestimation along the LOS, produces that the reconstructed stellar populations result biased to older ages and lower masses. As the result of these compensatory effects, the innermost normalized MGH tends to resemble the true one when the inclination is very high, both for Sp8D and Sp6L. However, the recovered stellar masses are much lower than the true ones. Since most of the mass in galaxies is in the central regions, then the global masses results also lower than the true ones. For the outermost radial bin, the effect of the incorrect recovery of extinction is less important (Fig. \ref{extinction}) and in combination with the geometrical effect, tend to bias the recovered stellar populations to older ones, resulting in more extended normalized MGHs (shifted to larger LBTs).  

\subsubsection{The effects of the average SNR and the PSF size}

Further, we explore the effects of the average SNR and PSF size on the reconstructed MGHs for the galaxy Sp8D. Since the inside-out mass growth of this galaxy and its diversity of stellar populations are larger than the ones of the Sp6L galaxy, any potential effect of the SNR or PSF size is expected to be more relevant for it. 
The upper panels of Figure~\ref{mgh_sn_psf} show the absolute and normalized MGHs for the fiducial setup but with two different average SNRs, a lower one (=5) and higher one (=81). The fiducial case is for SNR=24 and it is shown in panels b of Figure \ref{sp8_re_mgh}. The SNR is not an input in the synthetic observation but rather an output. To operate over the SNR, we actually vary the $\sigma_{\rm lim}$ noise level (see subsection \ref{mock-obs} and Appendix \ref{noise}). We impose three different values of $\sigma_{\rm lim}$ (see Table \ref{tab2}) to attain average SNRs of 5, 24, and 81. 

The global absolute MGHs (insets in Fig. \ref{mgh_sn_psf}) show slightly more under-estimation in the mass of intermediate and young populations as the average SNR decreases, while the normalized MGHs imply similar growth rates at all epochs independently of the SNR value. For the radial normalized MGHs, the recovered inside-out trend is less pronounced than the true one as the SNR decreases. 
The differences between the recovered radial MGHs with a SNR of 81 and 24 are actually small, while these differences between a SNR of 24 and 5 are more significant; in the latter case, the inside-out mass growth mode is strongly diminished, and the masses of the innermost and outermost radial bins are recovered with significantly lower values. 
By construction, {\sc Pipe3D} sums the spaxels within the segments to achieve a specific SNR. If the average SNR of the galaxy is low, {\sc Pipe3D} generates few segments with larger areas. In contrast, if the SNR is high enough, it produces a large number of segments with very small areas. In the best case, the minimum area per segment should be limited by the area of a single spaxel. The number of segments is the final number of spectra that {\sc Pipe3D} analyses \citep{Sanchez:2016ab}.  Therefore, {\it a low SNR works in the same direction of a poor spatial sampling} (see subsection \ref{resolution-inclination}). For poor SNRs, the number of segments is so small that the differences among the radial MGHs tend to disappear, and the light of some localized young stellar populations appears in these extended segments,  biasing the inversion method to an excess of young populations, which have low M/L ratios. For higher SNR, the number of segments will reach a maximum. Therefore, the MGHs are well resolved radially, and in many of these segments there are not young populations.

The lower panels of Figure~\ref{mgh_sn_psf} are as the upper panels but now we fix the SNR to its fiducial value (24), and vary the atmospheric PSF sizes to values of $2.86$ and $0.2$ arcsec. Hence for the diameter of the core fiber fixed (see subsection \ref{mock-obs}), the reconstructed PSFs are 3.15 and 1.99 arcsec, respectively. Recall that in the fiducial case the atmospheric and recovered PSF sizes are 1.43 and 2.21 arcsec, respectively; this case is shown in panels b of Figure \ref{sp8_re_mgh}. For the values of the PSF size explored here, we see that the recovered global and radial MGHs change only slightly. The PSF effect works in the direction of diminishing more the inside-out growth mode and lowering the recovered mass as larger is the PSF dispersion. This is because the PSF dispersion causes a mix of the light along the radial bins, and affects the effective spatial resolution element of an observation. 

We note that the setup with a reconstructed PSF value of 3.1 arcsec and 127 fibers has a roughly similar spatial resolution as the setup with a PSF value of 2.21 arcsec and 61 fibers.  The average SNR is similar in both cases, but the MGHs recovered in the latter setup are shifted to later epochs with respect to those recovered in the former one,  and with respect to the true MGHs (compare the MGHs in Figs. \ref{sp8_re_mgh} and \ref{mgh_sn_psf}). Note that in the former setup (127 fibers) {\sc Pipe3D} may generate more segments than in the latter one (61 fibers) since it contains more spaxels (on average) with the same SNR. As we discussed above, as larger the number of segments, more stable is the recovery of the MGHs.


\section{Conclusions and Discussion}
\label{conclusions}

Post-processed high-resolution N-body + Hydrodynamics simulations of two Milky Way-sized galaxies \citep{Colin+2016,Avila-Reese+2017} were used to explore the ability of the fossil record method to recover the mass- and ligth-weighted age profiles, as well as the global and radial SFHs and MGHs. To each stellar particle of the simulations at $z=0$, according to its age and metallicity, an SSP spectrum was assigned, and using the spatial cold gas distribution, the respective dust distribution and optical extinction along the LOS were estimated. Different realistic instrumental/observational settings, mainly emulating the MaNGA/SDSS-IV optical observations \citep{Bundy+2015}, were established to generate IFU datacubes. For any selected IFU bundle configuration and inclination, the studied galaxy is accommodated to cover up to 1.5 R$_{\rm eff}$ within the hexagonal bundle aperture ($\sim$70\% of the objects observed by the MaNGA survey). Thus, as small is the bundle aperture (less the number of fibers), the further away is located the galaxy. The generated datacubes were analyzed with the {\sc Pipe3D} fossil record tool \citep{Sanchez:2016aa,Sanchez:2016ab} to reconstruct the stellar population and dust extinction properties of the galaxies. 

For this study, we selected two galaxies: a disk-dominated one, with an extended in time and strong radial inside-out stellar mass growth (Sp8D), and a bulge-dominated (lenticular-like) one, with an early and weak radial inside-out mass growth (Sp6L). The main results and conclusions from our explorations are as follow:


$\bullet$ For our fiducial instrumental/observational setting,\footnote{Our fiducial setting is for a face on orientation covering 1.5 R$_{\rm eff}$ of the galaxy with a SNR=24, and it emulates a MaNGA bundle of 127 fibers. Note that for the MaNGA survey, $\sim$30\%\ of the objects are being observed with 127 fibers.} the global mass- and luminosity weighted ages are very well recovered within the uncertainties of the method, which are $\lesssim 0.1$ dex. The remaining ones are observed with smaller fiber bundles. Regarding the radial age profiles, {\it the fossil record method tends to slightly flatten them}, specially for the Sp8D galaxy. The global masses (within 1.5 R$_{\rm eff}$) are slightly underestimated, by less than 0.15 dex. The recovered relative rates of mass growth of the radial regions (radial normalized MGHs) tend to present smaller differences among them than the true ones, that is, {\it the inside-out trend is diminished }, specially for Sp8D, which presents a true prominent inside-out mode. The inversion method, in general, recovers younger stellar populations for the innermost (early formed) regions in such a way that their masses result lower than the true ones and with later assembly histories. The opposite happens for the outermost, lately formed regions, while the recovered MGHs of the intermediate regions agree better with the true ones. 

$\bullet$ The behaviors described above can be understood by comparing the true and recovered SFHs (Fig. \ref{sfh_sp8}); the SFHs are equivalent to the  mass fraction distributions by age. First,  the shape of the recovered SFHs is significantly smoothed (flattened) at old ages just due to the applied SSP age binning, which is very coarse for ages older than $\sim 8$ Gyr; further, the inversion method tends in general to spread out the radial SFHs. This could be understood as a renormalization of the cosmological look-back times when sampled by the ages of the SSPs. Second, the way the radial SFHs are spreaded out depends on the shape of the true SFH (or mass fraction distribution by age). For the innermost radial SFHs, which are peaked at very old ages (close to the oldest SSP template), the spread out results naturally biased to younger ages, while for the outermost SFHs, which are more extended and with a peak at intermediate ages (for Sp8D), the spread out is biased to slightly older ages. The true (and real galaxies) SFHs are actually more complex than simple smooth peaked functions, therefore, several effects may combine to bias the reconstructed SFHs, and these effects depend on the given instrumental/observational setting. 

$\bullet$ The recovery of the age profiles and the global/radial MGHs {\it worsens as the spatial sampling/resolution is lowered} when the number of fibers is diminished (and hence the distance is increased due to the smaller bundle aperture), specially for the Sp8D galaxy. 
As poorer is the spatial sampling and resolution, the more are mixed radially the stellar populations, producing this slightly flatter age gradients and less different radial normalized MGHs among them. However, the main effects are from the inversion method. As less and more extended are the binned regions, the more probable is that the spectra of these regions will include localized young stellar populations if they exist (the Sp8D galaxy contains more young populations than the Sp6L one). The young populations are bright and, even if their fraction in mass is small, they ``contaminate'' significantly the given spectrum in such a way that the inversion method tends to overestimate the fraction of these populations, with the consequent recovery of younger ages and lower stellar masses than the true ones. This effect affects more the radial bin that contains intrinsically larger fractions of old populations, that is, the innermost one, lowering significantly the recovered mass and delaying its growth history. Since more mass is contained in the innermost regions of the galaxies, this effect dominates in the recovered {\it global} mass and MGH. The outermost regions, have intrinsically larger fractions of younger populations, so that they suffer less the mentioned effect or even, depending on the true mass distribution by age, the bias of recovering older stellar populations (see the previous item) may dominate. In summary, we advice the reader against over-interpreting the age gradients in the stellar populations of poorly sampled IFS data. This is the case of the smaller MaNGA bundles (that comprises 19 and 37 fibers, used in $\sim$35\%\ of the survey), or when the full optical extension of the galaxy up to 2.5 R$_{eff}$ is sampled by even larger bundle.

$\bullet$ The effects of inclination/extinction are to flatten the age profiles, to diminish the radial differences among the normalized MGHs, and to lower the recovered mass from the innermost regions. However, these effects are significant only for high inclinations ($\ga 70^\circ$). The more affected regions by the inclination are the central ones just due to the accumulation of dust attenuation  along the LOS  and its underestimation.  The latter effect is geometrical rather than a result of the fossil record method. Indeed, {\sc Pipe3D} reconstructs well the ``observed'' extinctions. However, these are already strongly underestimated with respect to the true ones, in particular for high inclinations and towards the central regions. 
This introduces a bias in the observed spectra, and in particular in the properties derived by any fossil record method,  implying the recovery of lower stellar masses and older stellar populations than the true ones. There is also a geometrical effect of superposition of the (younger) stellar populations from the outer regions along the LOS that works in the opposite direction regarding the fraction of stellar populations (it overestimates young populations towards the center). For the outermost regions of the highly inclined galaxies, the dust attenuation effect is less severe and the measure along the LOS results contaminated by inner stars, which tend to be older. Therefore, the recovered outer radial MGHs look slightly older than the respective ones for the face-on case. In summary, we discourage the study of spatial resolved properties of the stellar populations of highly-inclined galaxies or, at least, to be cautious on the interpretation of the results.

$\bullet$ Low average SNRs (e.g., $\sim 5$) produce a flattening in the recovered age profiles, less differences in the radial MGHs among them, and younger stellar populations than the true ones. The latter applies specially in the innermost (older) regions, with the consequent under-estimation of the stellar masses. The effects of a low SNR are similar to the ones of poor spatial sampling because with a low SNR the sizes of the spatially binned segments result large and its number low. The effects of a large PSF size work in the same direction as poor spatial resolution. 
 
The exploration presented in this paper allows us to evaluate the level of reliability of many inferences from the IFU observations --in particular the MaNGA ones-- regarding the spatially-resolved stellar population analysis through fossil record tools as {\sc Pipe3D}. The most systematical trend found from our analysis is that the differences with radius in the recovered ages, SFHs, and MGHs become less pronounced than they actually are. The main bias is in recovering (i) younger stellar populations in the innermost regions, which have intrinsically old populations, and (ii) slightly older populations in the outermost regions, which can have a mix of old, intermediate, and young populations. Since young stellar populations have lower M/L ratios, the recovered stellar masses in the inner regions (hence the whole galaxy because they dominate in the total mass) result smaller than the true ones. These effects are more significant for a galaxy that intrinsically has a steeper age gradient and strong inside-out mass growth, with a large diversity of stellar populations (as our disk-dominated galaxy Sp8D) than for a galaxy formed more coherently radially and at earlier epochs (as our bulge-dominated galaxy Sp6L). The former is representative of late-type galaxies, while the latter is representative of earlier-type galaxies.

The mentioned above deviations become stronger as lower is the spatial sampling/resolution (less is the number of fibers), more inclined is the observed galaxy, lower is the SNR, and worse is the atmospheric seeing. 
{ To explore whether some of these systematical deviations affect, in a statistical way, real observations, we calculated the MW and LW age radial profiles for Milky Way-sized spiral and lenticular galaxies using the daraproducts publically distributed as part of the {\sc Pipe3D} Value Added Catalog \citep{sanchez17b}\footnote{\url{https://www.sdss.org/dr14/manga/manga-data/manga-pipe3d-value-added-catalog/}} for the MaNGA survey galaxies, included in the SDSS-DR14 \citep{abol18}. The average radial profiles for the low-inclined galaxies observed with high and low spatial sampling are presented in Appendix \ref{observa}. We also present therein the average age profiles of highly-inclined spiral and lenticular galaxies. The results confirm the trends with spatial sampling/resolution and inclination found here. 
}

From the analysis of at least two very different galaxies, we can conclude that based on MaNGA-SDSS/IV observations with high spatial sampling (bundles of 61 or more fibers) and a SNR around 25, we can recover with {\sc Pipe3D} reasonably well the stellar masses and main radial trends in ages, extinction, SFHs, and MGHs of normal galaxies observed at inclinations less than $70^\circ$.  On the other hand, the recovered SFHs and MGHs are reasonably well resolved in time up to ages $\sim 5$ Gyr. For ages larger than $\sim 8$ Gyr, the temporal resolution becomes very poor and the SFHs are strongly smoothed out. 

We would like to stress-out that the previous conclusions were derived for a particular fossil record method ({\sc Pipe3D}) simulating a particular instrumental setup in the optical spectral range (MaNGA; 3600-10300 \AA). However, some of the results may be valid for other methods and setups. For example, the bias towards a sharper MGH and a slightly larger fraction of young stellar populations discussed here have been reported also in recent studies with other methods and setups. \citet{lopfer16} analyzed CALIFA datacubes (covering the wavelength range between 3745-7000\AA) using {\sc Starlight}. They have found that including photometric information from the far-ultraviolet in the fitting process decreases the contribution of young stellar populations. More recently, Bitsakis et al. (submitted), have shown that the SFHs derived using multi-band photometric data, including both far-ultraviolet and infrared information, are smoother (with a lower contribution of very old stellar populations) than those derived by {\sc Pipe3D} using only optical spectroscopic data. Therefore, the wavelength range seems to contribute somehow to the observed biases, irrespective of the adopted fitting code. In future studies we will try to compare the results using photometric information at larger wavelength regimes and implement other inversion codes in the analysis to explore these results.

Finally, it is important to note that along this study we have not explored the recovery of many other relevant properties of the galaxies based on our mock observations. Among them we can list (i) the stellar kinematics properties, including the recovery of the rotational curve, velocity dispersion distribution and asymmetries in the line-of-sight velocity distribution; (ii) the emission line properties from the ionized gas, including the line intensities, line ratios, dust estimation, and the kinematics properties; (iii) other relevant properties of the stellar population, like the stellar metallicity. We exclude them from the current study for clarity, but we will address them in future studies using the same mock observations of the simulated galaxies presented here.

We make publicly available the code used to generate the mock IFU datacubes\footnote{\url{https://github.com/hjibarram/mock_ifu}} analyzed in this paper and its docker container as a software tool.\footnote{\url{https://hub.docker.com/r/hjibarram/ifu_mocks/}} Also, we plan to provide most of the mock IFU datacubes used in this paper,\footnote{Contact the authors} corresponding to our two post-processed simulated galaxies under different instrumental/observational settings. These datacubes can be analyzed with a fossil record tool different to {\sc Pipe3D}. Both the code and the datacubes are delivered freely with the requirement of properly acknowledge to the current study, citing this article as reference.

\section*{Acknowledgments}

HIM received financial support from the Postdoctoral fellowship ``Luc Binette'' and then a Postdoctoral fellowship provided by DGAPA-UNAM. AGS acknowledges support from a DGAPA-UNAM Postdoctoral fellowship.
The authors acknowledge CONACyT grant (Ciencia B\'asica) 285080 and PAPIIT-DGAPA-IA104118 (UNAM) project for partial funding.
SFS thanks funding from the PAPIIT-DGAPA-IA101217 (UNAM) project.

This project makes use of the MaNGA-Pipe3D dataproducts. We thank the IA-UNAM MaNGA team for creating this catalogue, and the ConaCyt-180125 project for supporting them.

Funding for the Sloan Digital Sky Survey IV has been provided by the Alfred P. Sloan Foundation, the U.S. Department of Energy Office of Science, and the Participating Institutions. SDSS-IV acknowledges
support and resources from the Center for High-Performance Computing at
the University of Utah. The SDSS web site is www.sdss.org.

SDSS-IV is managed by the Astrophysical Research Consortium for the 
Participating Institutions of the SDSS Collaboration including the 
Brazilian Participation Group, the Carnegie Institution for Science, 
Carnegie Mellon University, the Chilean Participation Group, the French Participation Group, Harvard-Smithsonian Center for Astrophysics, 
Instituto de Astrof\'isica de Canarias, The Johns Hopkins University, 
Kavli Institute for the Physics and Mathematics of the Universe (IPMU) / 
University of Tokyo, the Korean Participation Group, Lawrence Berkeley National Laboratory, 
Leibniz Institut f\"ur Astrophysik Potsdam (AIP),  
Max-Planck-Institut f\"ur Astronomie (MPIA Heidelberg), 
Max-Planck-Institut f\"ur Astrophysik (MPA Garching), 
Max-Planck-Institut f\"ur Extraterrestrische Physik (MPE), 
National Astronomical Observatories of China, New Mexico State University, 
New York University, University of Notre Dame, 
Observat\'ario Nacional / MCTI, The Ohio State University, 
Pennsylvania State University, Shanghai Astronomical Observatory, 
United Kingdom Participation Group,
Universidad Nacional Aut\'onoma de M\'exico, University of Arizona, 
University of Colorado Boulder, University of Oxford, University of Portsmouth, 
University of Utah, University of Virginia, University of Washington, University of Wisconsin, 
Vanderbilt University, and Yale University.

\bibliography{paper,references,CALIFAI}{}

\begin{thebibliography}{}
\makeatletter
\relax
\def\mn@urlcharsother{\let\do\@makeother \do\$\do\&\do\#\do\^\do\_\do\%\do\~}
\def\mn@doi{\begingroup\mn@urlcharsother \@ifnextchar [ {\mn@doi@}
  {\mn@doi@[]}}
\def\mn@doi@[#1]#2{\def\@tempa{#1}\ifx\@tempa\@empty \href
  {http://dx.doi.org/#2} {doi:#2}\else \href {http://dx.doi.org/#2} {#1}\fi
  \endgroup}
\def\mn@eprint#1#2{\mn@eprint@#1:#2::\@nil}
\def\mn@eprint@arXiv#1{\href {http://arxiv.org/abs/#1} {{\tt arXiv:#1}}}
\def\mn@eprint@dblp#1{\href {http://dblp.uni-trier.de/rec/bibtex/#1.xml}
  {dblp:#1}}
\def\mn@eprint@#1:#2:#3:#4\@nil{\def\@tempa {#1}\def\@tempb {#2}\def\@tempc
  {#3}\ifx \@tempc \@empty \let \@tempc \@tempb \let \@tempb \@tempa \fi \ifx
  \@tempb \@empty \def\@tempb {arXiv}\fi \@ifundefined
  {mn@eprint@\@tempb}{\@tempb:\@tempc}{\expandafter \expandafter \csname
  mn@eprint@\@tempb\endcsname \expandafter{\@tempc}}}

\bibitem[\protect\citeauthoryear{{Abolfathi} et~al.,}{{Abolfathi}
  et~al.}{2018}]{abol18}
{Abolfathi} B.,  et~al., 2018, \mn@doi [\apjs] {10.3847/1538-4365/aa9e8a},
  \href {http://adsabs.harvard.edu/abs/2018ApJS..235...42A} {235, 42}

\bibitem[\protect\citeauthoryear{{Albareti} et~al.,}{{Albareti}
  et~al.}{2017}]{SDSS:2016aa}
{Albareti} F.~D.,  et~al., 2017, \mn@doi [\apjs] {10.3847/1538-4365/aa8992},
  \href {http://adsabs.harvard.edu/abs/2017ApJS..233...25A} {233, 25}

\bibitem[\protect\citeauthoryear{{Allen} et~al.,}{{Allen}
  et~al.}{2015}]{Allen:2015aa}
{Allen} J.~T.,  et~al., 2015, \mn@doi [\mnras] {10.1093/mnras/stu2057}, \href
  {http://adsabs.harvard.edu/abs/2015MNRAS.446.1567A} {446, 1567}

\bibitem[\protect\citeauthoryear{{Aquino-Ort{\'{\i}}z}
  et~al.,}{{Aquino-Ort{\'{\i}}z} et~al.}{2018}]{Aquino-Ortiz+2018}
{Aquino-Ort{\'{\i}}z} E.,  et~al., 2018, \mn@doi [\mnras]
  {10.1093/mnras/sty1522}, \href
  {http://adsabs.harvard.edu/abs/2018MNRAS.479.2133A} {479, 2133}

\bibitem[\protect\citeauthoryear{{Avila-Reese}, {Col{\'{\i}}n},
  {Gonz{\'a}lez-Samaniego}, {Valenzuela}, {Firmani}, {Vel{\'a}zquez}  \&
  {Ceverino}}{{Avila-Reese} et~al.}{2011}]{Avila-Reese+2011}
{Avila-Reese} V.,  {Col{\'{\i}}n} P.,  {Gonz{\'a}lez-Samaniego} A.,
  {Valenzuela} O.,  {Firmani} C.,  {Vel{\'a}zquez} H.,   {Ceverino} D.,  2011,
  \mn@doi [\apj] {10.1088/0004-637X/736/2/134}, \href
  {http://adsabs.harvard.edu/abs/2011ApJ...736..134A} {736, 134}

\bibitem[\protect\citeauthoryear{{Avila-Reese}, {Gonz{\'a}lez-Samaniego},
  {Col{\'{\i}}n}, {Ibarra-Medel}  \& {Rodr{\'{\i}}guez-Puebla}}{{Avila-Reese}
  et~al.}{2018}]{Avila-Reese+2017}
{Avila-Reese} V.,  {Gonz{\'a}lez-Samaniego} A.,  {Col{\'{\i}}n} P.,
  {Ibarra-Medel} H.,   {Rodr{\'{\i}}guez-Puebla} A.,  2018, \mn@doi [\apj]
  {10.3847/1538-4357/aaab69}, \href
  {http://adsabs.harvard.edu/abs/2018ApJ...854..152A} {854, 152}

\bibitem[\protect\citeauthoryear{{Bacon} et~al.,}{{Bacon}
  et~al.}{2001}]{Bacon:2001aa}
{Bacon} R.,  et~al., 2001, \mn@doi [\mnras] {10.1046/j.1365-8711.2001.04612.x},
  \href {http://adsabs.harvard.edu/abs/2001MNRAS.326...23B} {326, 23}

\bibitem[\protect\citeauthoryear{{Battisti}, {Calzetti}  \& {Chary}}{{Battisti}
  et~al.}{2017}]{batt17}
{Battisti} A.~J.,  {Calzetti} D.,   {Chary} R.-R.,  2017, \mn@doi [\apj]
  {10.3847/1538-4357/aa9a43}, \href
  {http://adsabs.harvard.edu/abs/2017ApJ...851...90B} {851, 90}

\bibitem[\protect\citeauthoryear{{Bell} \& {de Jong}}{{Bell} \& {de
  Jong}}{2001}]{bell01}
{Bell} E.~F.,  {de Jong} R.~S.,  2001, \mn@doi [\apj] {10.1086/319728}, \href
  {http://adsabs.harvard.edu/abs/2001ApJ...550..212B} {550, 212}

\bibitem[\protect\citeauthoryear{{Bershady}, {Verheijen}, {Swaters},
  {Andersen}, {Westfall}  \& {Martinsson}}{{Bershady}
  et~al.}{2010}]{Bershady+2010}
{Bershady} M.~A.,  {Verheijen} M.~A.~W.,  {Swaters} R.~A.,  {Andersen} D.~R.,
  {Westfall} K.~B.,   {Martinsson} T.,  2010, \mn@doi [\apj]
  {10.1088/0004-637X/716/1/198}, \href
  {http://adsabs.harvard.edu/abs/2010ApJ...716..198B} {716, 198}

\bibitem[\protect\citeauthoryear{{Bruzual} \& {Charlot}}{{Bruzual} \&
  {Charlot}}{2003}]{Bruzual:2003aa}
{Bruzual} G.,  {Charlot} S.,  2003, \mn@doi [\mnras]
  {10.1046/j.1365-8711.2003.06897.x}, \href
  {http://adsabs.harvard.edu/abs/2003MNRAS.344.1000B} {344, 1000}

\bibitem[\protect\citeauthoryear{{Bundy} et~al.,}{{Bundy}
  et~al.}{2015}]{Bundy+2015}
{Bundy} K.,  et~al., 2015, \mn@doi [\apj] {10.1088/0004-637X/798/1/7}, \href
  {http://adsabs.harvard.edu/abs/2015ApJ...798....7B} {798, 7}

\bibitem[\protect\citeauthoryear{{Buzzoni}}{{Buzzoni}}{1989}]{Buzzoni:1989aa}
{Buzzoni} A.,  1989, \mn@doi [\apjs] {10.1086/191399}, \href
  {http://adsabs.harvard.edu/abs/1989ApJS...71..817B} {71, 817}

\bibitem[\protect\citeauthoryear{{Cano-D{\'{\i}}az} et~al.,}{{Cano-D{\'{\i}}az}
  et~al.}{2016}]{Cano-Diaz+2016}
{Cano-D{\'{\i}}az} M.,  et~al., 2016, \mn@doi [\apjl]
  {10.3847/2041-8205/821/2/L26}, \href
  {http://adsabs.harvard.edu/abs/2016ApJ...821L..26C} {821, L26}

\bibitem[\protect\citeauthoryear{{Cappellari}}{{Cappellari}}{2017}]{cappe17}
{Cappellari} M.,  2017, \mn@doi [\mnras] {10.1093/mnras/stw3020}, \href
  {http://adsabs.harvard.edu/abs/2017MNRAS.466..798C} {466, 798}

\bibitem[\protect\citeauthoryear{{Cappellari} et~al.,}{{Cappellari}
  et~al.}{2011}]{Cappellari+2011}
{Cappellari} M.,  et~al., 2011, \mn@doi [\mnras]
  {10.1111/j.1365-2966.2010.18174.x}, \href
  {http://adsabs.harvard.edu/abs/2011MNRAS.413..813C} {413, 813}

\bibitem[\protect\citeauthoryear{{Cappellari} et~al.,}{{Cappellari}
  et~al.}{2012}]{Cappellari+2012}
{Cappellari} M.,  et~al., 2012, \mn@doi [\nat] {10.1038/nature10972}, \href
  {http://adsabs.harvard.edu/abs/2012Natur.484..485C} {484, 485}

\bibitem[\protect\citeauthoryear{{Cardelli}, {Clayton}  \& {Mathis}}{{Cardelli}
  et~al.}{1989}]{Cardelli:1989aa}
{Cardelli} J.~A.,  {Clayton} G.~C.,   {Mathis} J.~S.,  1989, \mn@doi [\apj]
  {10.1086/167900}, \href {http://adsabs.harvard.edu/abs/1989ApJ...345..245C}
  {345, 245}

\bibitem[\protect\citeauthoryear{{Catal{\'a}n-Torrecilla}
  et~al.,}{{Catal{\'a}n-Torrecilla} et~al.}{2015}]{catalan15}
{Catal{\'a}n-Torrecilla} C.,  et~al., 2015, \mn@doi [\aap]
  {10.1051/0004-6361/201526023}, \href
  {http://adsabs.harvard.edu/abs/2015A%26A...584A..87C} {584, A87}

\bibitem[\protect\citeauthoryear{{Cid Fernandes}, {Mateus}, {Sodr{\'e}},
  {Stasi{\'n}ska}  \& {Gomes}}{{Cid Fernandes}
  et~al.}{2005}]{Cid-Fernandes:2005aa}
{Cid Fernandes} R.,  {Mateus} A.,  {Sodr{\'e}} L.,  {Stasi{\'n}ska} G.,
  {Gomes} J.~M.,  2005, \mn@doi [\mnras] {10.1111/j.1365-2966.2005.08752.x},
  \href {http://adsabs.harvard.edu/abs/2005MNRAS.358..363C} {358, 363}

\bibitem[\protect\citeauthoryear{{Cid Fernandes} et~al.,}{{Cid Fernandes}
  et~al.}{2013}]{Cid-Fernandes:2013aa}
{Cid Fernandes} R.,  et~al., 2013, \mn@doi [\aap]
  {10.1051/0004-6361/201220616}, \href
  {http://adsabs.harvard.edu/abs/2013A%26A...557A..86C} {557, A86}

\bibitem[\protect\citeauthoryear{{Col{\'{\i}}n}, {Avila-Reese},
  {Roca-F{\`a}brega}  \& {Valenzuela}}{{Col{\'{\i}}n}
  et~al.}{2016}]{Colin+2016}
{Col{\'{\i}}n} P.,  {Avila-Reese} V.,  {Roca-F{\`a}brega} S.,   {Valenzuela}
  O.,  2016, \mn@doi [\apj] {10.3847/0004-637X/829/2/98}, \href
  {http://adsabs.harvard.edu/abs/2016ApJ...829...98C} {829, 98}

\bibitem[\protect\citeauthoryear{{Croom} et~al.,}{{Croom}
  et~al.}{2012}]{Croom:2012aa}
{Croom} S.~M.,  et~al., 2012, \mn@doi [\mnras]
  {10.1111/j.1365-2966.2011.20365.x}, \href
  {http://adsabs.harvard.edu/abs/2012MNRAS.421..872C} {421, 872}

\bibitem[\protect\citeauthoryear{{Ellison}, {S{\'a}nchez}, {Ibarra-Medel},
  {Antonio}, {Mendel}  \& {Barrera-Ballesteros}}{{Ellison}
  et~al.}{2018}]{Ellison:2018aa}
{Ellison} S.~L.,  {S{\'a}nchez} S.~F.,  {Ibarra-Medel} H.,  {Antonio} B.,
  {Mendel} J.~T.,   {Barrera-Ballesteros} J.,  2018, \mn@doi [\mnras]
  {10.1093/mnras/stx2882}, \href
  {http://adsabs.harvard.edu/abs/2018MNRAS.474.2039E} {474, 2039}

\bibitem[\protect\citeauthoryear{{Falc{\'o}n-Barroso},
  {S{\'a}nchez-Bl{\'a}zquez}, {Vazdekis}, {Ricciardelli}, {Cardiel}, {Cenarro},
  {Gorgas}  \& {Peletier}}{{Falc{\'o}n-Barroso}
  et~al.}{2011}]{Falcon-Barroso:2011aa}
{Falc{\'o}n-Barroso} J.,  {S{\'a}nchez-Bl{\'a}zquez} P.,  {Vazdekis} A.,
  {Ricciardelli} E.,  {Cardiel} N.,  {Cenarro} A.~J.,  {Gorgas} J.,
  {Peletier} R.~F.,  2011, \mn@doi [\aap] {10.1051/0004-6361/201116842}, \href
  {http://adsabs.harvard.edu/abs/2011A%26A...532A..95F} {532, A95}

\bibitem[\protect\citeauthoryear{{Gallazzi}, {Charlot}, {Brinchmann}, {White}
  \& {Tremonti}}{{Gallazzi} et~al.}{2005}]{Gallazzi:2005aa}
{Gallazzi} A.,  {Charlot} S.,  {Brinchmann} J.,  {White} S.~D.~M.,   {Tremonti}
  C.~A.,  2005, \mn@doi [\mnras] {10.1111/j.1365-2966.2005.09321.x}, \href
  {http://adsabs.harvard.edu/abs/2005MNRAS.362...41G} {362, 41}

\bibitem[\protect\citeauthoryear{{Garc{\'{\i}}a-Benito}
  et~al.,}{{Garc{\'{\i}}a-Benito} et~al.}{2017}]{Garcia-Benito:2017aa}
{Garc{\'{\i}}a-Benito} R.,  et~al., 2017, \mn@doi [\aap]
  {10.1051/0004-6361/201731357}, \href
  {http://adsabs.harvard.edu/abs/2017A%26A...608A..27G} {608, A27}

\bibitem[\protect\citeauthoryear{{Goddard} et~al.,}{{Goddard}
  et~al.}{2017a}]{Goddard+2017}
{Goddard} D.,  et~al., 2017a, \mn@doi [\mnras] {10.1093/mnras/stw3371}, \href
  {http://adsabs.harvard.edu/abs/2017MNRAS.466.4731G} {466, 4731}

\bibitem[\protect\citeauthoryear{{Goddard} et~al.,}{{Goddard}
  et~al.}{2017b}]{Goddard:2017aa}
{Goddard} D.,  et~al., 2017b, \mn@doi [\mnras] {10.1093/mnras/stw3371}, \href
  {http://adsabs.harvard.edu/abs/2017MNRAS.466.4731G} {466, 4731}

\bibitem[\protect\citeauthoryear{{Gonz{\'a}lez Delgado} et~al.,}{{Gonz{\'a}lez
  Delgado} et~al.}{2014a}]{GonzalezDelgado+2014}
{Gonz{\'a}lez Delgado} R.~M.,  et~al., 2014a, \mn@doi [\aap]
  {10.1051/0004-6361/201322011}, \href
  {http://adsabs.harvard.edu/abs/2014A%26A...562A..47G} {562, A47}

\bibitem[\protect\citeauthoryear{{Gonz{\'a}lez Delgado} et~al.,}{{Gonz{\'a}lez
  Delgado} et~al.}{2014b}]{Gonzalez-Delgado:2014ab}
{Gonz{\'a}lez Delgado} R.~M.,  et~al., 2014b, \mn@doi [\apjl]
  {10.1088/2041-8205/791/1/L16}, \href
  {http://adsabs.harvard.edu/abs/2014ApJ...791L..16G} {791, L16}

\bibitem[\protect\citeauthoryear{{Gonz{\'a}lez Delgado} et~al.,}{{Gonz{\'a}lez
  Delgado} et~al.}{2015}]{Gonzalez-Delgado:2015aa}
{Gonz{\'a}lez Delgado} R.~M.,  et~al., 2015, \mn@doi [\aap]
  {10.1051/0004-6361/201525938}, \href
  {http://adsabs.harvard.edu/abs/2015A%26A...581A.103G} {581, A103}

\bibitem[\protect\citeauthoryear{{Gonz{\'a}lez Delgado} et~al.,}{{Gonz{\'a}lez
  Delgado} et~al.}{2016}]{Gonzalez-Delgado:2016aa}
{Gonz{\'a}lez Delgado} R.~M.,  et~al., 2016, \mn@doi [\aap]
  {10.1051/0004-6361/201628174}, \href
  {http://adsabs.harvard.edu/abs/2016A%26A...590A..44G} {590, A44}

\bibitem[\protect\citeauthoryear{{Gonz{\'a}lez Delgado} et~al.,}{{Gonz{\'a}lez
  Delgado} et~al.}{2017}]{Gonzalez-Delgado:2017aa}
{Gonz{\'a}lez Delgado} R.~M.,  et~al., 2017, \mn@doi [\aap]
  {10.1051/0004-6361/201730883}, \href
  {http://adsabs.harvard.edu/abs/2017A%26A...607A.128G} {607, A128}

\bibitem[\protect\citeauthoryear{{Guidi} et~al.,}{{Guidi}
  et~al.}{2018}]{guidi18}
{Guidi} G.,  et~al., 2018, \mn@doi [\mnras] {10.1093/mnras/sty1480}, \href
  {http://adsabs.harvard.edu/abs/2018MNRAS.479..917G} {479, 917}

\bibitem[\protect\citeauthoryear{{Gunn} et~al.,}{{Gunn}
  et~al.}{2006}]{Gunn:2006aa}
{Gunn} J.~E.,  et~al., 2006, \mn@doi [\aj] {10.1086/500975}, \href
  {http://adsabs.harvard.edu/abs/2006AJ....131.2332G} {131, 2332}

\bibitem[\protect\citeauthoryear{{Haardt} \& {Madau}}{{Haardt} \&
  {Madau}}{1996}]{HM96}
{Haardt} F.,  {Madau} P.,  1996, \mn@doi [\apj] {10.1086/177035}, \href
  {http://adsabs.harvard.edu/abs/1996ApJ...461...20H} {461, 20}

\bibitem[\protect\citeauthoryear{{Hsieh} et~al.,}{{Hsieh}
  et~al.}{2017}]{Hsieh+2017}
{Hsieh} B.~C.,  et~al., 2017, \mn@doi [\apjl] {10.3847/2041-8213/aa9d80}, \href
  {http://adsabs.harvard.edu/abs/2017ApJ...851L..24H} {851, L24}

\bibitem[\protect\citeauthoryear{{Ibarra-Medel} et~al.,}{{Ibarra-Medel}
  et~al.}{2016}]{Ibarra-Medel+2016}
{Ibarra-Medel} H.~J.,  et~al., 2016, \mn@doi [\mnras] {10.1093/mnras/stw2126},
  \href {http://adsabs.harvard.edu/abs/2016MNRAS.463.2799I} {463, 2799}

\bibitem[\protect\citeauthoryear{{Jonsson}}{{Jonsson}}{2006}]{Jonsson:2006aa}
{Jonsson} P.,  2006, \mn@doi [\mnras] {10.1111/j.1365-2966.2006.10884.x}, \href
  {http://adsabs.harvard.edu/abs/2006MNRAS.372....2J} {372, 2}

\bibitem[\protect\citeauthoryear{{Kauffmann} et~al.,}{{Kauffmann}
  et~al.}{2003a}]{Kauffmann:2003aa}
{Kauffmann} G.,  et~al., 2003a, \mn@doi [\mnras]
  {10.1046/j.1365-8711.2003.06291.x}, \href
  {http://adsabs.harvard.edu/abs/2003MNRAS.341...33K} {341, 33}

\bibitem[\protect\citeauthoryear{{Kauffmann} et~al.,}{{Kauffmann}
  et~al.}{2003b}]{Kauffmann:2003ab}
{Kauffmann} G.,  et~al., 2003b, \mn@doi [\mnras]
  {10.1046/j.1365-8711.2003.06292.x}, \href
  {http://adsabs.harvard.edu/abs/2003MNRAS.341...54K} {341, 54}

\bibitem[\protect\citeauthoryear{{Kravtsov}}{{Kravtsov}}{2003}]{Kravtsov:2003aa}
{Kravtsov} A.~V.,  2003, \mn@doi [\apjl] {10.1086/376674}, \href
  {http://adsabs.harvard.edu/abs/2003ApJ...590L...1K} {590, L1}

\bibitem[\protect\citeauthoryear{{Kravtsov}, {Klypin}  \&
  {Khokhlov}}{{Kravtsov} et~al.}{1997}]{Kravtsov:1997aa}
{Kravtsov} A.~V.,  {Klypin} A.~A.,   {Khokhlov} A.~M.,  1997, \mn@doi [\apjs]
  {10.1086/313015}, \href {http://adsabs.harvard.edu/abs/1997ApJS..111...73K}
  {111, 73}

\bibitem[\protect\citeauthoryear{{Kravtsov}, {Nagai}  \&
  {Vikhlinin}}{{Kravtsov} et~al.}{2005}]{Kravtsov+2005}
{Kravtsov} A.~V.,  {Nagai} D.,   {Vikhlinin} A.~A.,  2005, \mn@doi [\apj]
  {10.1086/429796}, \href {http://adsabs.harvard.edu/abs/2005ApJ...625..588K}
  {625, 588}

\bibitem[\protect\citeauthoryear{{Law} et~al.,}{{Law}
  et~al.}{2015}]{Law:2015aa}
{Law} D.~R.,  et~al., 2015, \mn@doi [\aj] {10.1088/0004-6256/150/1/19}, \href
  {http://adsabs.harvard.edu/abs/2015AJ....150...19L} {150, 19}

\bibitem[\protect\citeauthoryear{{Law} et~al.,}{{Law}
  et~al.}{2016}]{Law:2016aa}
{Law} D.~R.,  et~al., 2016, \mn@doi [\aj] {10.3847/0004-6256/152/4/83}, \href
  {http://adsabs.harvard.edu/abs/2016AJ....152...83L} {152, 83}

\bibitem[\protect\citeauthoryear{{Leitner}}{{Leitner}}{2012}]{Leitner2012}
{Leitner} S.~N.,  2012, \mn@doi [\apj] {10.1088/0004-637X/745/2/149}, \href
  {http://adsabs.harvard.edu/abs/2012ApJ...745..149L} {745, 149}

\bibitem[\protect\citeauthoryear{{Leitner} \& {Kravtsov}}{{Leitner} \&
  {Kravtsov}}{2011}]{Leitner+2011}
{Leitner} S.~N.,  {Kravtsov} A.~V.,  2011, \mn@doi [\apj]
  {10.1088/0004-637X/734/1/48}, \href
  {http://adsabs.harvard.edu/abs/2011ApJ...734...48L} {734, 48}

\bibitem[\protect\citeauthoryear{{L{\'o}pez Fern{\'a}ndez} et~al.,}{{L{\'o}pez
  Fern{\'a}ndez} et~al.}{2016}]{lopfer16}
{L{\'o}pez Fern{\'a}ndez} R.,  et~al., 2016, \mn@doi [\mnras]
  {10.1093/mnras/stw260}, \href
  {http://adsabs.harvard.edu/abs/2016MNRAS.458..184L} {458, 184}

\bibitem[\protect\citeauthoryear{{L{\'o}pez Fern{\'a}ndez} et~al.,}{{L{\'o}pez
  Fern{\'a}ndez} et~al.}{2018}]{lopfer18}
{L{\'o}pez Fern{\'a}ndez} R.,  et~al., 2018, \mn@doi [\aap]
  {10.1051/0004-6361/201732358}, \href
  {http://adsabs.harvard.edu/abs/2018A%26A...615A..27L} {615, A27}

\bibitem[\protect\citeauthoryear{{Martins}, {Gonz{\'a}lez Delgado},
  {Leitherer}, {Cervi{\~n}o}  \& {Hauschildt}}{{Martins}
  et~al.}{2005}]{Martins:2005aa}
{Martins} L.~P.,  {Gonz{\'a}lez Delgado} R.~M.,  {Leitherer} C.,  {Cervi{\~n}o}
  M.,   {Hauschildt} P.,  2005, \mn@doi [\mnras]
  {10.1111/j.1365-2966.2005.08703.x}, \href
  {http://adsabs.harvard.edu/abs/2005MNRAS.358...49M} {358, 49}

\bibitem[\protect\citeauthoryear{{Miller} \& {Scalo}}{{Miller} \&
  {Scalo}}{1979}]{MS79}
{Miller} G.~E.,  {Scalo} J.~M.,  1979, \mn@doi [\apjs] {10.1086/190629}, \href
  {http://adsabs.harvard.edu/abs/1979ApJS...41..513M} {41, 513}

\bibitem[\protect\citeauthoryear{{Pan} et~al.,}{{Pan} et~al.}{2018}]{Pan+2018}
{Pan} H.-A.,  et~al., 2018, \mn@doi [\apj] {10.3847/1538-4357/aaa9bc}, \href
  {http://adsabs.harvard.edu/abs/2018ApJ...854..159P} {854, 159}

\bibitem[\protect\citeauthoryear{{Peng}, {Ho}, {Impey}  \& {Rix}}{{Peng}
  et~al.}{2002}]{Peng:2002aa}
{Peng} C.~Y.,  {Ho} L.~C.,  {Impey} C.~D.,   {Rix} H.-W.,  2002, \mn@doi [\aj]
  {10.1086/340952}, \href {http://adsabs.harvard.edu/abs/2002AJ....124..266P}
  {124, 266}

\bibitem[\protect\citeauthoryear{{P{\'e}rez} et~al.,}{{P{\'e}rez}
  et~al.}{2013}]{Perez+2013}
{P{\'e}rez} E.,  et~al., 2013, \mn@doi [\apjl] {10.1088/2041-8205/764/1/L1},
  \href {http://adsabs.harvard.edu/abs/2013ApJ...764L...1P} {764, L1}

\bibitem[\protect\citeauthoryear{{R{\'e}my-Ruyer} et~al.,}{{R{\'e}my-Ruyer}
  et~al.}{2014}]{Remy-Ruyer:2014aa}
{R{\'e}my-Ruyer} A.,  et~al., 2014, \mn@doi [\aap]
  {10.1051/0004-6361/201322803}, \href
  {http://adsabs.harvard.edu/abs/2014A%26A...563A..31R} {563, A31}

\bibitem[\protect\citeauthoryear{{Roca-F{\`a}brega}, {Valenzuela},
  {Col{\'{\i}}n}, {Figueras}, {Krongold}, {Vel{\'a}zquez}, {Avila-Reese}  \&
  {Ibarra-Medel}}{{Roca-F{\`a}brega} et~al.}{2016}]{Roca-Fabrega:2016aa}
{Roca-F{\`a}brega} S.,  {Valenzuela} O.,  {Col{\'{\i}}n} P.,  {Figueras} F.,
  {Krongold} Y.,  {Vel{\'a}zquez} H.,  {Avila-Reese} V.,   {Ibarra-Medel} H.,
  2016, \mn@doi [\apj] {10.3847/0004-637X/824/2/94}, \href
  {http://adsabs.harvard.edu/abs/2016ApJ...824...94R} {824, 94}

\bibitem[\protect\citeauthoryear{{Rowlands} et~al.,}{{Rowlands}
  et~al.}{2018}]{Rowlands+2018}
{Rowlands} K.,  et~al., 2018, \mn@doi [\mnras] {10.1093/mnras/sty1916}, \href
  {http://adsabs.harvard.edu/abs/2018MNRAS.tmp.1823R} {}

\bibitem[\protect\citeauthoryear{{S{\'a}nchez}}{{S{\'a}nchez}}{2015}]{Sanchez+215IAU}
{S{\'a}nchez} S.~F.,  2015, in {Ziegler} B.~L.,  {Combes} F.,  {Dannerbauer}
  H.,   {Verdugo} M.,  eds,  IAU Symposium Vol. 309, Galaxies in 3D across the
  Universe. pp 85--92 (\mn@eprint {arXiv} {1410.0295}),
  \mn@doi{10.1017/S1743921314009375}

\bibitem[\protect\citeauthoryear{{S{\'a}nchez-Bl{\'a}zquez}
  et~al.,}{{S{\'a}nchez-Bl{\'a}zquez} et~al.}{2006}]{Sanchez-Blazquez:2006aa}
{S{\'a}nchez-Bl{\'a}zquez} P.,  et~al., 2006, \mn@doi [\mnras]
  {10.1111/j.1365-2966.2006.10699.x}, \href
  {http://adsabs.harvard.edu/abs/2006MNRAS.371..703S} {371, 703}

\bibitem[\protect\citeauthoryear{{S{\'a}nchez} et~al.,}{{S{\'a}nchez}
  et~al.}{2012}]{Sanchez:2012aa}
{S{\'a}nchez} S.~F.,  et~al., 2012, \mn@doi [\aap]
  {10.1051/0004-6361/201117353}, \href
  {http://adsabs.harvard.edu/abs/2012A%26A...538A...8S} {538, A8}

\bibitem[\protect\citeauthoryear{{S{\'a}nchez} et~al.,}{{S{\'a}nchez}
  et~al.}{2014}]{Sanchez:2014aa}
{S{\'a}nchez} S.~F.,  et~al., 2014, \mn@doi [\aap]
  {10.1051/0004-6361/201322343}, \href
  {http://adsabs.harvard.edu/abs/2014A%26A...563A..49S} {563, A49}

\bibitem[\protect\citeauthoryear{{S{\'a}nchez} et~al.,}{{S{\'a}nchez}
  et~al.}{2016a}]{Sanchez:2016ab}
{S{\'a}nchez} S.~F.,  et~al., 2016a, \rmxaa, \href
  {http://adsabs.harvard.edu/abs/2016RMxAA..52...21S} {52, 21}

\bibitem[\protect\citeauthoryear{{S{\'a}nchez} et~al.,}{{S{\'a}nchez}
  et~al.}{2016b}]{Sanchez:2016aa}
{S{\'a}nchez} S.~F.,  et~al., 2016b, \rmxaa, \href
  {http://adsabs.harvard.edu/abs/2016RMxAA..52..171S} {52, 171}

\bibitem[\protect\citeauthoryear{{S{\'a}nchez} et~al.,}{{S{\'a}nchez}
  et~al.}{2018a}]{sanchez18a}
{S{\'a}nchez} S.~F.,  et~al., 2018a, \rmxaa, \href
  {http://adsabs.harvard.edu/abs/2018RMxAA..54..217S} {54, 217}

\bibitem[\protect\citeauthoryear{{S{\'a}nchez} et~al.,}{{S{\'a}nchez}
  et~al.}{2018b}]{sanchez17b}
{S{\'a}nchez} S.~F.,  et~al., 2018b, \rmxaa, \href
  {http://adsabs.harvard.edu/abs/2018RMxAA..54..217S} {54, 217}

\bibitem[\protect\citeauthoryear{{S{\'a}nchez} et~al.,}{{S{\'a}nchez}
  et~al.}{2019}]{sanchez18b}
{S{\'a}nchez} S.~F.,  et~al., 2019, \mn@doi [\mnras] {10.1093/mnras/sty2730},
  \href {http://adsabs.harvard.edu/abs/2019MNRAS.482.1557S} {482, 1557}

\bibitem[\protect\citeauthoryear{{Stoughton} et~al.,}{{Stoughton}
  et~al.}{2002}]{Stoughton:2002ab}
{Stoughton} C.,  et~al., 2002, \mn@doi [\aj] {10.1086/324741}, \href
  {http://adsabs.harvard.edu/abs/2002AJ....123..485S} {123, 485}

\bibitem[\protect\citeauthoryear{{Szomoru}, {Franx}, {van Dokkum}, {Trenti},
  {Illingworth}, {Labb{\'e}}  \& {Oesch}}{{Szomoru}
  et~al.}{2013}]{Szomoru:2013aa}
{Szomoru} D.,  {Franx} M.,  {van Dokkum} P.~G.,  {Trenti} M.,  {Illingworth}
  G.~D.,  {Labb{\'e}} I.,   {Oesch} P.,  2013, \mn@doi [\apj]
  {10.1088/0004-637X/763/2/73}, \href
  {http://adsabs.harvard.edu/abs/2013ApJ...763...73S} {763, 73}

\bibitem[\protect\citeauthoryear{{Tinsley}}{{Tinsley}}{1980}]{Tinsley:1980aa}
{Tinsley} B.~M.,  1980, \fcp, \href
  {http://adsabs.harvard.edu/abs/1980FCPh....5..287T} {5, 287}

\bibitem[\protect\citeauthoryear{{Tissera}, {Machado}, {Sanchez-Blazquez},
  {Pedrosa}, {S{\'a}nchez}, {Snaith}  \& {Vilchez}}{{Tissera}
  et~al.}{2016}]{Tissera+2016}
{Tissera} P.~B.,  {Machado} R.~E.~G.,  {Sanchez-Blazquez} P.,  {Pedrosa} S.~E.,
   {S{\'a}nchez} S.~F.,  {Snaith} O.,   {Vilchez} J.,  2016, \mn@doi [\aap]
  {10.1051/0004-6361/201628188}, \href
  {http://adsabs.harvard.edu/abs/2016A%26A...592A..93T} {592, A93}

\bibitem[\protect\citeauthoryear{{Tojeiro}, {Heavens}, {Jimenez}  \&
  {Panter}}{{Tojeiro} et~al.}{2007}]{Tojeiro+2007}
{Tojeiro} R.,  {Heavens} A.~F.,  {Jimenez} R.,   {Panter} B.,  2007, \mn@doi
  [\mnras] {10.1111/j.1365-2966.2007.12323.x}, \href
  {http://adsabs.harvard.edu/abs/2007MNRAS.381.1252T} {381, 1252}

\bibitem[\protect\citeauthoryear{{Tuffs}, {Popescu}, {V{\"o}lk}, {Kylafis}  \&
  {Dopita}}{{Tuffs} et~al.}{2004}]{tuff04}
{Tuffs} R.~J.,  {Popescu} C.~C.,  {V{\"o}lk} H.~J.,  {Kylafis} N.~D.,
  {Dopita} M.~A.,  2004, \mn@doi [\aap] {10.1051/0004-6361:20035689}, \href
  {http://adsabs.harvard.edu/abs/2004A%26A...419..821T} {419, 821}

\bibitem[\protect\citeauthoryear{{Vazdekis}, {S{\'a}nchez-Bl{\'a}zquez},
  {Falc{\'o}n-Barroso}, {Cenarro}, {Beasley}, {Cardiel}, {Gorgas}  \&
  {Peletier}}{{Vazdekis} et~al.}{2010}]{Vazdekis:2010aa}
{Vazdekis} A.,  {S{\'a}nchez-Bl{\'a}zquez} P.,  {Falc{\'o}n-Barroso} J.,
  {Cenarro} A.~J.,  {Beasley} M.~A.,  {Cardiel} N.,  {Gorgas} J.,   {Peletier}
  R.~F.,  2010, \mn@doi [\mnras] {10.1111/j.1365-2966.2010.16407.x}, \href
  {http://adsabs.harvard.edu/abs/2010MNRAS.404.1639V} {404, 1639}

\bibitem[\protect\citeauthoryear{{Walcher}, {Groves}, {Budav{\'a}ri}  \&
  {Dale}}{{Walcher} et~al.}{2011}]{Walcher:2011aa}
{Walcher} J.,  {Groves} B.,  {Budav{\'a}ri} T.,   {Dale} D.,  2011, \mn@doi
  [\apss] {10.1007/s10509-010-0458-z}, \href
  {http://adsabs.harvard.edu/abs/2011Ap%26SS.331....1W} {331, 1}

\bibitem[\protect\citeauthoryear{{Wild}, {Charlot}, {Brinchmann}, {Heckman},
  {Vince}, {Pacifici}  \& {Chevallard}}{{Wild} et~al.}{2011}]{wild11}
{Wild} V.,  {Charlot} S.,  {Brinchmann} J.,  {Heckman} T.,  {Vince} O.,
  {Pacifici} C.,   {Chevallard} J.,  2011, \mn@doi [\mnras]
  {10.1111/j.1365-2966.2011.19367.x}, \href
  {http://adsabs.harvard.edu/abs/2011MNRAS.417.1760W} {417, 1760}

\bibitem[\protect\citeauthoryear{{Yip}, {Szalay}, {Wyse}, {Dobos},
  {Budav{\'a}ri}  \& {Csabai}}{{Yip} et~al.}{2010}]{yip10}
{Yip} C.-W.,  {Szalay} A.~S.,  {Wyse} R.~F.~G.,  {Dobos} L.,  {Budav{\'a}ri}
  T.,   {Csabai} I.,  2010, \mn@doi [\apj] {10.1088/0004-637X/709/2/780}, \href
  {http://adsabs.harvard.edu/abs/2010ApJ...709..780Y} {709, 780}

\bibitem[\protect\citeauthoryear{{Zibetti} et~al.,}{{Zibetti}
  et~al.}{2017}]{Zibetti:2017aa}
{Zibetti} S.,  et~al., 2017, \mn@doi [\mnras] {10.1093/mnras/stx251}, \href
  {http://adsabs.harvard.edu/abs/2017MNRAS.468.1902Z} {468, 1902}

\makeatother
\end{thebibliography}
\bibliographystyle{mnras}
\clearpage
\appendix

\section{Post-processing and construction of the synthetic spectra from the simulations}
\label{post-processing}

\begin{figure*}
\begin{center}
\includegraphics[width=0.28\linewidth]{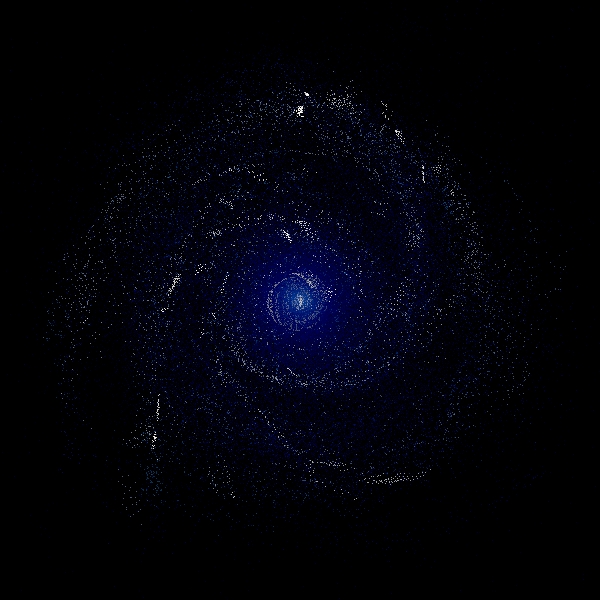}
\includegraphics[width=0.28\linewidth]{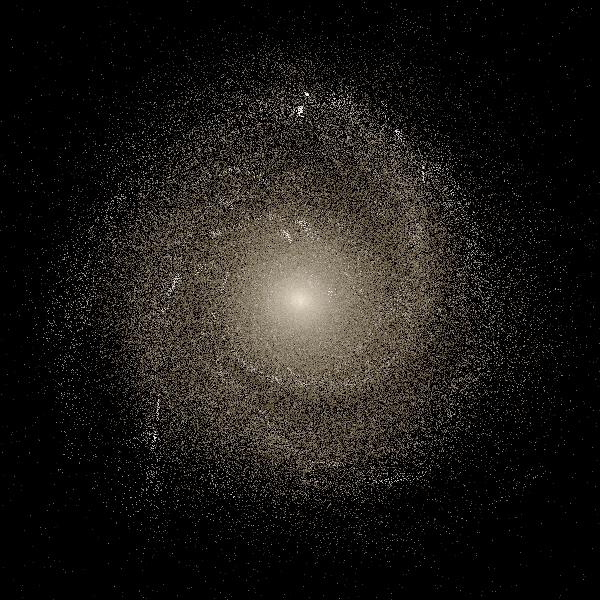}
\includegraphics[width=0.28\linewidth]{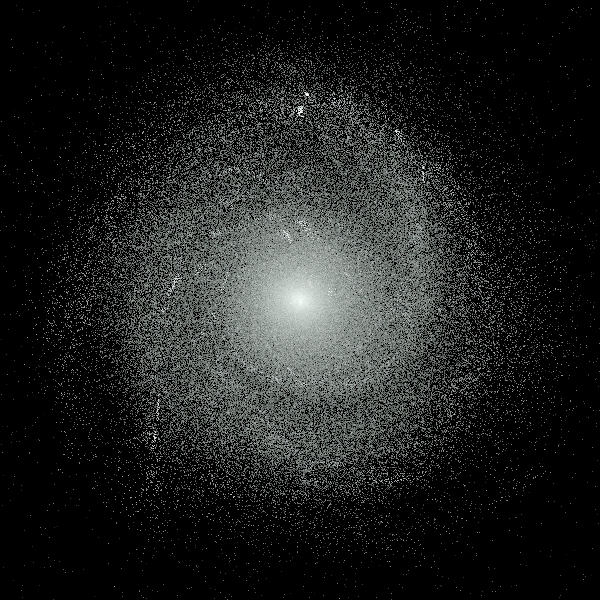}
\includegraphics[width=0.28\linewidth]{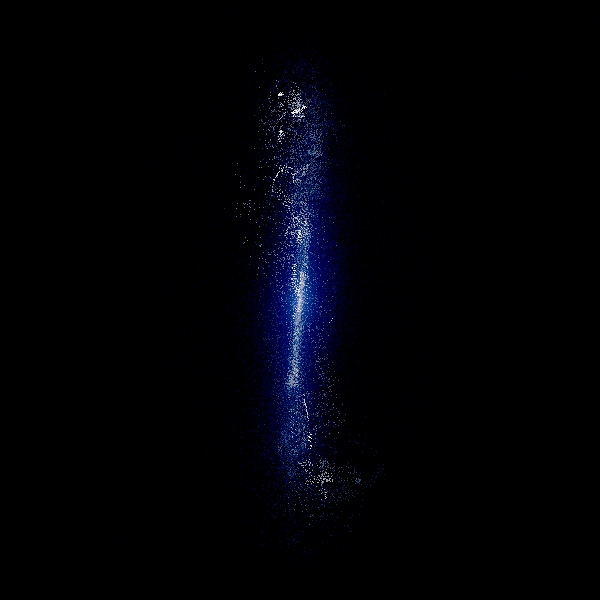}
\includegraphics[width=0.28\linewidth]{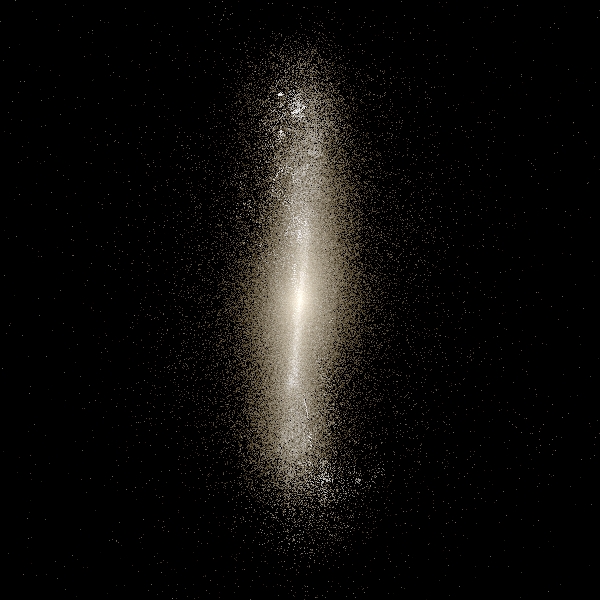}
\includegraphics[width=0.28\linewidth]{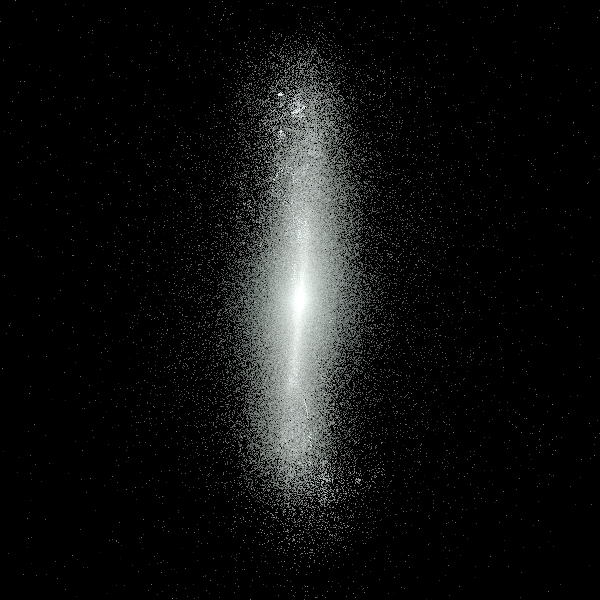}
\end{center}
\caption{Synthetic stellar flux emission of the simulated galaxy Sp8D in several bands. Upper panels are the face-on view and lower panels are the edge-on view. For these reconstructions, we do not take into account the dust extinction nor the effects of a PSF dispersion. {\it Left panels:} Flux emission in GALEX $FUV$, GALEX $NUV$, and Gunn $u$ bands with the RGB channels in this order. Note that the spiral arms and thin disk are well traced by these three UV bands, and that the $u$ band (blue color) dominates among them.   {\it Central panels:} Flux emission in the optical Gunn $i,r,g$ bands with the RGB channels in this order.  The spiral arms are yet seen in these bands; in these bands it can be seen the thick disk. {\it Right panels:} Flux emission in the infrared $K, H, J$ bands with the RGB channels in this order. The dominant white color is because the three bands (RGB channels) contribute similarly, and they trace all the stellar disk. In these bands the oldest stars can be seen, including the most distant ones from the plane. }
\label{photo_sp8}
\end{figure*}
\begin{figure*}
\begin{center}
\includegraphics[width=0.28\linewidth]{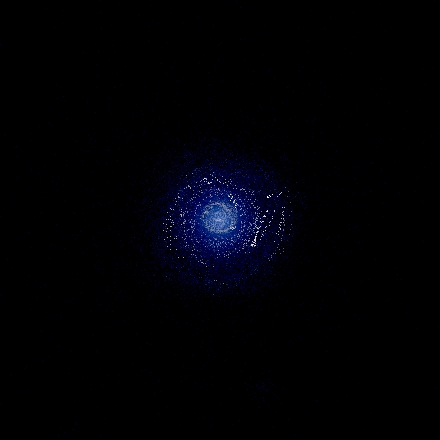}
\includegraphics[width=0.28\linewidth]{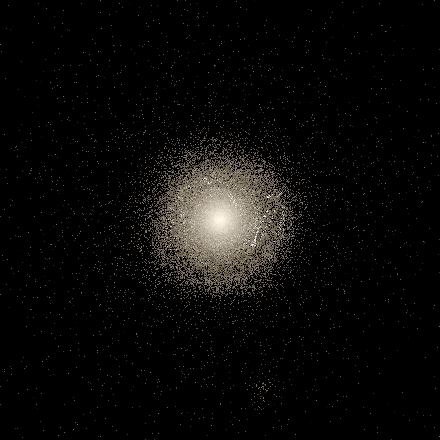}
\includegraphics[width=0.28\linewidth]{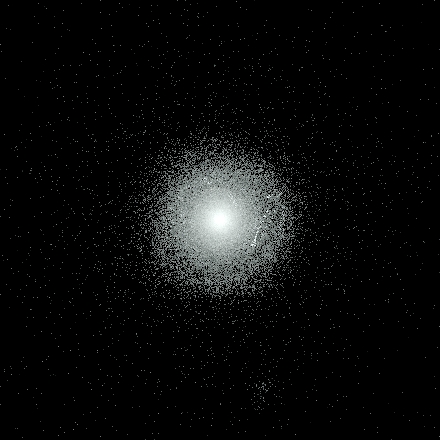}
\includegraphics[width=0.28\linewidth]{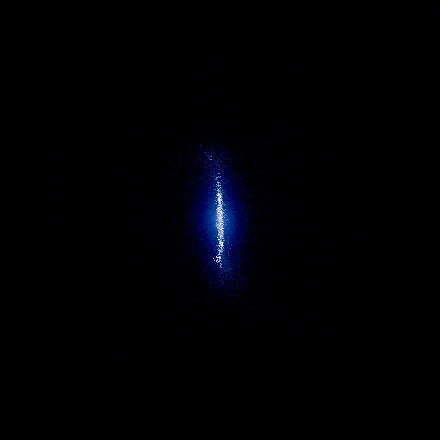}
\includegraphics[width=0.28\linewidth]{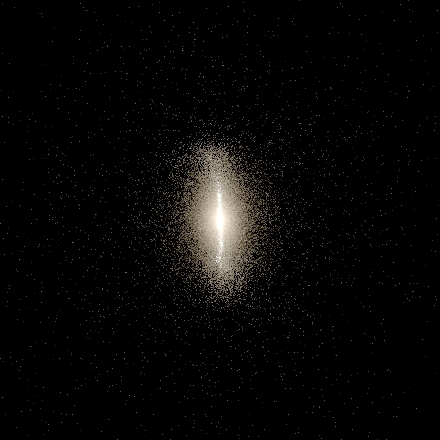}
\includegraphics[width=0.28\linewidth]{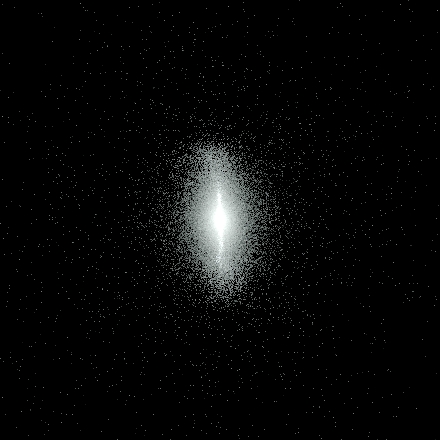}
\end{center}
\caption{As Figure \ref{photo_sp8} but for the simulated galaxy Sp6L. This galaxy is located at the same distance as the galaxy Sp8D. }
\label{photo_sp6}
\end{figure*}

In this Appendix we describe in detail the post-processing and construction of the synthetic optical stellar spectra (\S\S \ref{synthetic-spectrum}) and gas emission spectra (\S\S \ref{gas-spectrum}) from the simulations towards the observing line of sight (LOS), taking into account dust extinction (\S\S \ref{extinction}). We require a set of properties from each stellar particle and gas cell from the last snapshot ($z=0$) of the simulations. For the stellar particles, we need the position vectors $\overrightarrow{\textbf{r}}_s$ in kpc, the velocity vectors $\overrightarrow{\textbf{v}}_s$ in km/s, the actual mass $m_s$ of the particle in $M_{\odot}$, the age $t_s$ of the particle in Gyr, and the metallicity $Z_{s}$ in solar units. For the gas cells, we require the positions vectors $\overrightarrow{\textbf{r}}_g$ in kpc, the velocity vectors $\overrightarrow{\textbf{v}}_g$ in km/s, the actual mass $m_g$ of the gas cells in $M_{\odot}$, the number density $nh_g$ of the cells in cm$^{-3}$, the temperature $T_g$ in K, the volume $V_g$ of the cells in cm$^3$, and the metallicities $Z_{g}$ in solar units. With these quantities, we derive the total radius of the gas cell as $R_g=(3V_g/4\pi)^{1/3}$.  

The positions and velocities of both the stellar particles and the gas cells are referenced at the point of view of an imaginary observer located at $\overrightarrow{\textbf{r}_{obs}}$.  
To define the coordinate reference system, first we estimate the center of mass of the simulated galaxy as $X_c,Y_c,Z_c$. Then, we redefine the coordinate system as: $\overrightarrow{\textbf{r}_0}=\overrightarrow{\textbf{r}_b}-\overrightarrow{\textbf{r}_c}$, where $\overrightarrow{\textbf{r}_c}=(X_c,Y_c,Z_c)$, $\overrightarrow{\textbf{r}_b}=(X_b,Y_b,Z_b)$, $\overrightarrow{\textbf{r}}=(X,Y,Z)$,  and $X_b,Y_b,Z_b$ are the original coordinates of the stellar particles or gas cells within the box of the simulation. Therefore,  $\overrightarrow{\textbf{r}_0}$ are the coordinates referenced to the center of mass of the simulated galaxy for both stellar particles and gas cells. Then, we define the coordinate system referenced at the point of view of the imaginary observer located at $\overrightarrow{\textbf{r}_{obs}}$ (referenced to the simulated galaxy center of mass) as: $\overrightarrow{\textbf{r}}=\textbf{R}_1\cdot\textbf{R}_2\cdot\overrightarrow{\textbf{r}_{0}}-(0,0,|\overrightarrow{\textbf{r}_{obs}}|),$ were $\textbf{R}_1$ and $\textbf{R}_2$ are rotational matrix defined as: $$\textbf{R}_1=\begin{pmatrix} 1 & 0 & 0 \\  0 & \cos(A1) & -\sin(A1) \\ 0 & \sin(A1) & \cos(A1) \end{pmatrix},$$ and $$\textbf{R}_2=\begin{pmatrix} \cos(A2) & 0 & -\sin(A2) \\  0 & 1 & 0 \\ \sin(A2) & 0 & \cos(A2) \end{pmatrix}.$$ The values of $A1$ and $A2$ are the Euler angles to transform the coordinates $\overrightarrow{\textbf{r}_0}$ to the observer point of view. We define those angles as $A1=\atantwo(Y_{obs},Z_{obs})$ and $A2=\arcsin(X_{obs}/|\overrightarrow{\textbf{r}_{obs}}|)$. Also, we redefine the velocity vectors  $\overrightarrow{\textbf{v}_b}$ for all particles and gas cells in the observer  point of view with $\overrightarrow{\textbf{v}}=\textbf{R}_1\cdot\textbf{R}_2\cdot\overrightarrow{\textbf{v}_{b}},$ assuming that the observer has a zero velocity. In this case, we use the original velocity vector from the  simulation in order to take in to account the peculiar velocity of the hole simulated galaxy. 
Finally, the projected velocity for each stellar particle or gas cell towards the LOS is $v_{los,s,g}=\overrightarrow{\textbf{v}}_{s,g}\cdot\overrightarrow{\textbf{r}}_{s,g}/|\overrightarrow{\textbf{r}}_{s,g}|$.

For taking into account the observer position and inclination, we define the three characteristic vectors of the simulated galaxy as $\textbf{e}_1,\textbf{e}_2,\textbf{e}_3$, were $\textbf{e}_1$ is a unitary vector along the disk spin vector and therefore, orthogonal to the disk plane of the galaxy. $\textbf{e}_2$ and $\textbf{e}_3$ are orthogonal unitary vectors within the disk plane. Hence, for simplicity, we put the imaginary observer in the direction of $\textbf{e}_1$ at a given distance $r_{obs}$. To change the observer perspective as a function of the galaxy inclination we use the next expression: $$\overrightarrow{\textbf{r}_{obs}}=\left(\textbf{e}_1,\textbf{e}_2,\textbf{e}_3 \right)\cdot \textbf{R}_3(\theta)\cdot \textbf{e}_x \times r_{obs},$$  were $\textbf{e}_x$ is a unitary vector along the $x$ axis of a new coordinate system which basis vectors are $\textbf{e}_1,\textbf{e}_2,\textbf{e}_3$ (on the coordinates of the box simulation) and $\textbf{R}_3(\theta)$ is the rotation matrix along the $z$ axis (in the direction of $\textbf{e}_3$ and within the galaxy disk plane). Therefore, $\theta$ is the inclination of the galaxy disk plane with respect of the observer, where $\theta=0$ represents a face on projection and $\theta=90$ represents an edge on projection.

\subsection{Optical synthetic spectrum}
\label{synthetic-spectrum}

With the quantities and coordinates of the observer reference system mentioned above, we construct a synthetic observed spectrum within a given solid angle $\Delta\Omega$ by considering each stellar particle as a single stellar population (SSP) of mass $m_{s}$, age $t_s$, and metalicity $Z_s$. Therefore, we can assign an spectrum of a SSP from a stellar library. In this work we use a combination of the synthetic stellar spectra from the GRANADA library \citep{Martins:2005aa} and the MILES project \citep{Sanchez-Blazquez:2006aa, Vazdekis:2010aa, Falcon-Barroso:2011aa}, which is also used in the  {\sc Pipe3D} code. 

We are interested in the optical-infrared spectral region (3,800 to 10,000 \AA.) Since the $t_s$ and $Z_s$ values of the stellar particles are not discrete, we group the values of $t_s$ and $Z_s$ in bins centered in ages and metallicities ($t_b$, $Z_b$) of the stellar libraries. Hence, we assign to all stellar particles that lie within a given mass, age and metallicity bin an SSP spectrum that has the corresponding ($t_b$, $Z_b$) values from the library\footnote{We do not interpolate values from the stellar libraries because this operation should be performed in all the spectral range and for each stellar particle. This increases largely the computational time making too expensive the generation of many exploratory mock observations.}. Then, we define a pure synthetic stellar spectrum within a solid angle $\Delta\Omega$ (in the LOS direction) as the total contribution of all associated SSP spectra $L_{SSP}(\lambda,Z_{SSP},t_{SSP})$ within $\Delta\Omega$: 
  \begin{multline}
F_{\alpha,\delta,s}(\lambda)=\sum_{i=0}^{ns} L_{SSP,i}(\lambda(1+z_0+v_{los,s,i}/c),Z_{SSP,i},t_{SSP,i}) \\ \times\frac{m_{s,i}}{ML_{SSP,i}} \times\frac{L_{\odot}}{4\pi |\overrightarrow{\textbf{r}}_{s,i}|^2},\label{stellar_only}
\end{multline} 
for all $ns$ stellar particles that fall within $\sqrt{(\alpha_{i,s}-\alpha)^2+(\delta_{i,s}-\delta)^2}<\Delta\Omega$, where $\alpha_{i,s}$ and $\delta_{i,s}$ are the angular projected position of the stellar particle $i$ in the observer point of view. $\alpha$ and $\delta$  are the projected angular positions of the LOS, and $ML_{SSP,i}$ is the mass-to-light ratio of the associated SSP spectra for the $i-th$ stellar particle.
The final  synthetic spectrum along the LOS, $F_{\alpha,\delta,s}(\lambda)$, takes into accounts the effects of  galaxy kinematics and cosmological expansion by shifting each SSP spectrum $L_{SSP,i}$ by $z_0+v_{los,i}/c$. Figures~\ref{photo_sp8} and \ref{photo_sp6} show RGB images of the reconstructed photometric ultraviolet, optic and infrared ``observations'' of the Sp8D and Sp6L runs from the synthetic stellar flux only (Eq.~\ref{stellar_only}), without any other consideration.

\subsection{Extinction}
\label{extinction}

The dust extinction affects the spectrophotometric observations of galaxies depending on the inclination at which they are seen. The simulations do not include dust production, so we introduce it in an approximate way using the gas properties given by the simulation. Recall that our goal here is not to make predictions for comparing with real galaxies, but just to emulate spectro-photometric galaxy properties and explore how well they are recovered by our mock IFS and photometric observations. Therefore, we use a simple model based on a given gas-to-dust scaling law instead of modeling the dust extinction by the implementation of a radiative transfer code as SUNRISE \citep[][]{Jonsson:2006aa}. 

To model the extinction we use the dust-to-gas mass ratio defined as: 
\begin{equation}\frac{D}{G}=\frac{4\pi a^3\rho_D}{3 m_H}\frac{n_D}{n_H},\label{A1}
\end{equation} 
where $a$ is the dust grain size, $\rho_D$ is the dust grain density, $m_H$ is the hydrogen atom mass, and $n_D$ and $n_H$ are the dust and hydrogen number densities. We can calculate the extinction at a given wavelength as $$A_{\lambda}=\frac{1}{0.4\log(10)}\int_0^R\sigma_{\lambda}n_D(r)dr,$$ 
where $\sigma_{\lambda}$ is the effective cross-section of the dust grain at a wavelength $\lambda$. If one considers only a very small section along the LOS of the extinction contribution, then $n_D(r)$ can be assumed as a constant value, and the above equation can be simplified: $$A_{\lambda}\simeq\frac{\sigma_{\lambda}n_D\Delta R}{0.4\log(10)},$$ 
where $\Delta R$ is the longitude of the section of the LOS, where we can consider $n_D(r)$ constant. On the other hand, we can use the scaling relation between the total extinction $A_{\lambda}$ and the column density of Hydrogen gas along the LOS: 
\begin{equation}
A_{\lambda}C_1=N_H\simeq n_H\Delta R,\label{A2}
\end{equation} where $C_1$ is a constant. Therefore, we can obtain the next relation: 
\begin{equation}
\frac{n_D}{n_H}=\frac{0.4\log(10)}{\sigma_{\lambda}C_1}.\label{A3}
\end{equation} 
Using Equations \ref{A1}, \ref{A2} and \ref{A3}, and defining the cross-section as $\sigma_{\lambda}=2\pi a_{\lambda}^2$, we obtain the next expression: $$A_{\lambda}=\frac{3m_{H}n_{H}\Delta R}{2a_{\lambda}\rho_D 0.4 \log(10)}\frac{D}{G}.$$ Since, at first approximation, the dust grain size must be of the order of the wavelength to scatter the light and therefore, produce extinction at that wavelength, we can use the relation $a_{\lambda}=\frac{\lambda}{2\pi}$. Thus, the final formula that relates the extinction at $\lambda$ with the gas-to-dust ratio is as follows: 
\begin{equation} 
A_{\lambda}=\frac{3m_H\pi n_H \Delta R}{0.4 \log(10)\lambda \rho_{D}}\times (G/D)^{-1}.
\end{equation}
Using this Equation, we calculate the extinction in a given cell for the effective wavelength, $\lambda_V$, of the Johnson $V$ photometric band, $A_{Vcell}$. Here, we assume $\Delta R=R_g$. For the $G/D$ ratio, which is metallicity dependent, we use the scaling relations given by \cite{Remy-Ruyer:2014aa}.

The total extinction in the spectrum of a single stellar particle is the total contribution of the extinction of all individual gas cells along the LOS within a solid angle $\Delta\Omega$. Thus, we redefine Eq. (\ref{stellar_only}) to take into account extinction as: 
\begin{multline}
F_{\alpha,\delta,s}(\lambda)=\sum_{i=0}^{ns} L_{SSP,i}(\lambda(1+z_0+v_{los,s,i}/c),Z_{SSP,i},t_{SSP,i}) \\  \times10^{-0.4A_{Vcell,s}(|\overrightarrow{\textbf{r}}_{s,i}|,\Delta\Omega)A_{\lambda}/A_V} \times\frac{m_{s,i}}{ML_{SSP,i}} \times\frac{L_{\odot}}{4\pi |\overrightarrow{\textbf{r}}_{s,i}|^2},\label{stellar_ext}
\end{multline} 
where $A_{Vcell,s}(|\overrightarrow{\textbf{r}}_{s,i}|,\Delta\Omega)=\sum_{j=0}^{ng}A_{V,g,j}$ for all $ng$ gas cells that have a $|\overrightarrow{\textbf{r}}_{g,j}| \leq |\overrightarrow{\textbf{r}}_{s,i}|$ and have a location $\sqrt{(\alpha_{j,g}-\alpha)^2+(\delta_{j,g}-\delta)^2}<\Delta\Omega$. As in the case of stellar particles, $\alpha_{j,g}$ and $\delta_{j,g}$ are the angular projected position of the gas cell $j$ in the observer point of view. We define the function $A_{\lambda}/A_V$ as the \cite{Cardelli:1989aa} extinction law with a extinction factor $R_v=3.1$. Again, different galaxies or regions within a galaxy may present different extinction curves and/or different extinction factors. However, it is beyond the goal of this study to explore these possibilities.

\subsection{Photoionized gas spectrum}
\label{gas-spectrum}

To complete the mock observations, we model the gas emission spectrum. The model that we use is not totally realistic; however, in this study we use the gas emission spectrum only to add it to the total mock spectrum. We do not perform any further analysis of the emission lines. For a future study, we will implement a more realistic modelling of the emission line spectrum.
We model the gas emission spectrum $L_{gas}(\lambda)$ per unit of mass using the CLOUDY code. For this objective, we generate a grid of 524 gas emission spectra that cover five black-body temperatures in the temperature range $3\times10^7 \leq T/K \leq 5\times10^3$, three number densities in the range $1\leq nh/cm^{-3} \leq 2.5\times10^{-4}$, five gas metallicities in the range $1.75\leq Z\leq0.001$, and seven photo-ionization parameters in the range $10^{53}\leq Q(H)/s^{-1}\leq10^{47}$. Then, we integrate all the associated SSP spectra, $L_{SSP,i}$, for all stellar particles within each gas cell. With this integrated stellar spectra we calculate the number of ionizing photons than can contain each gas cell, defining it as $Q(H)_g=\int_{\nu_0}^{\infty}\sum_i^gL_{SSP,i}d\nu/h\nu$. We associate a gas emission spectrum for each gas cell when its values $Z_g$, $nh_g$, $T_g$ and $Q(H)_g$ lie within the $Z$, $nh$, $T$ and $Q(H)$ grid of the CLOUDY gas emission models. Therefore, we model the total gas emission spectrum towards the LOS within a solid angle $\Delta\Omega$ as \begin{multline}
F_{\alpha,\delta,g}(\lambda)=\sum_{i=0}^{ng} L_{gas,i}(\lambda(1+z_0+v_{los,g,i}/c))\times m_{g,i} \\ \times10^{-0.4A_{Vcell,g}(|\overrightarrow{\textbf{r}}_{g,i}|,\Delta\Omega)A_{\lambda}/A_V} \times\frac{L_{\odot}}{4\pi |\overrightarrow{\textbf{r}}_{g,i}|^2}.\label{gas_ext}
\end{multline} 
With this oversimplified implementation, we can obtain an emission gas spectrum related to the properties of the gas and stellar particles of the simulation. This model is not entirely realistic but for the purpouse of generating mock observations is enough.


\section{Modeling the Noise of the detector}
\label{noise}

\begin{figure}
\begin{center}
\includegraphics[width=0.93\linewidth]{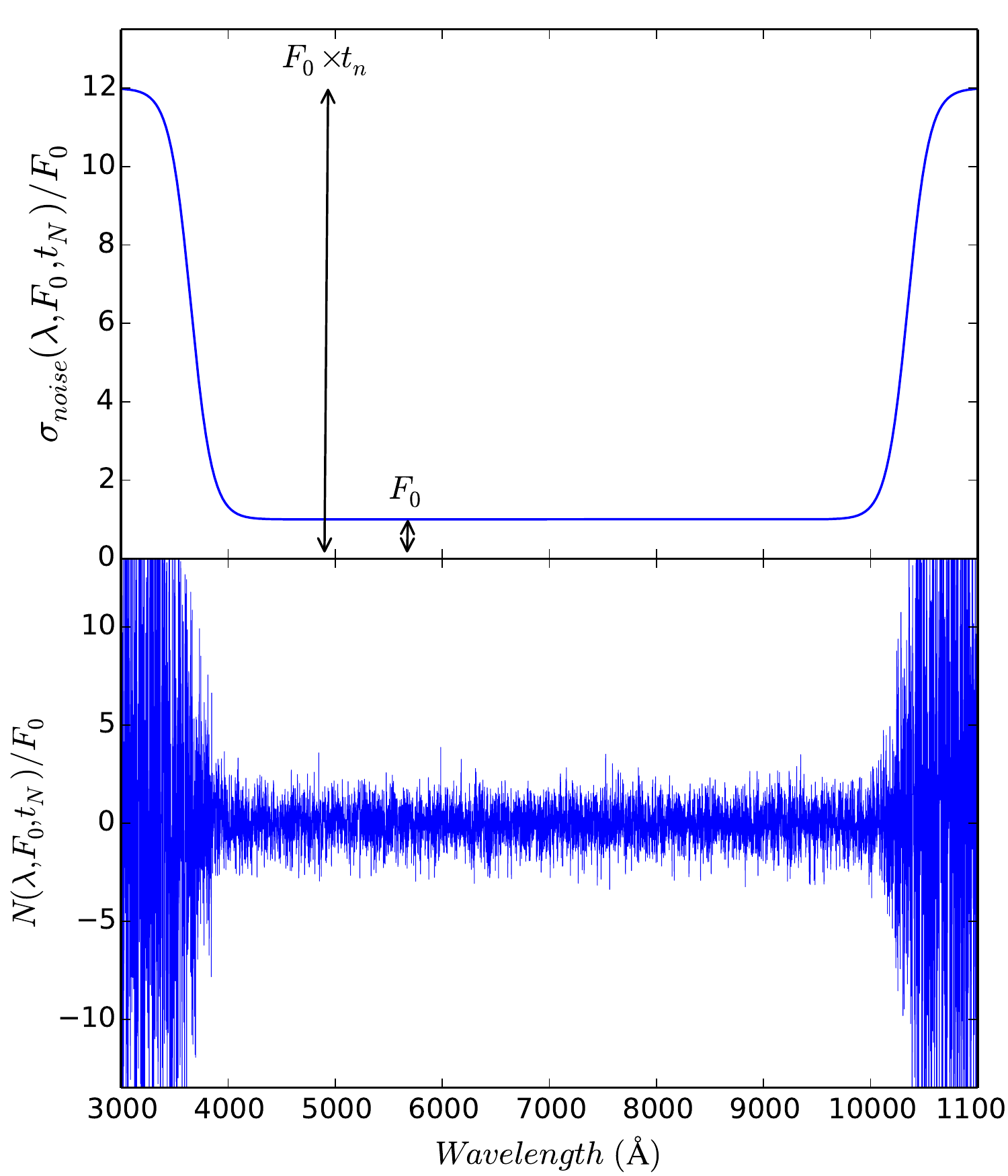}
\end{center}
\caption{Example of the noise spectrum that models the sensibility of an imaginary detector. {\it Upper panel:} The noise scatter as a function of the wavelength. {\it Lower panel:} The final noise spectrum.}
\label{NOISE}
\end{figure}

To obtain an accurate mock observation, we intoduce in Equation \ref{full-spectrum} a random noise spectrum function of an imaginary detector, $N_{\alpha_{obs},\delta_{obs}}(\lambda,F_0;t_N)$. The random noise is described as a zero mean normal distribution with a dispersion $\sigma_{noise}$ equal to $F_0\times t_N/f(\lambda;t_N).$ The function $f(\lambda;t_N)$ is the inverse of a normalized response function that is defined as $$f(\lambda;t_N)=1+(t_N-1)/[S(\lambda,k_1,\lambda_1)\times S(\lambda,k_2,\lambda_2)],$$ where $S(\lambda,k_i,\lambda_i)$ is a Sigmoid function with a  steepness value of $k_i$ and a midpoint value of $\lambda_i$. We select the values of $\lambda_1=3,900$ \AA, $k_1=100$ \AA, $\lambda_2=10,100$ \AA, and $k_2=100$ \AA\ to mimic as best as possible the detector noise response function of a MaNGA IFS observation at the Apache Point Observatory (APO). This function is similar to the surface brightness sensitivity presented by \citet{Law:2016aa}. Therefore, at wavelengths larger than $\lambda_2$ or lower than $\lambda_1$, the noise scatter, $\sigma_{noise}$, increases exponentially until a plateau of $F_0\times t_N$. Between $\lambda_1$ and $\lambda_2$ the noise scatter is almost constant with a value of $F_0$, we refer hereafter to this value as $\sigma_{\rm lim}$. We select a constant value of $t_N=15$, that gives $3$ magnitude difference between $\sigma_{\rm lim}$ and the plateau value. This noise simulates a detector that is efficient in the optical wavelengths and is not so efficient in the ultraviolet or infrared part of the spectrum.



\section{Definitions of mass growth histories and the results from the simulations}
\label{simulations}

\begin{figure}
\begin{center}
\includegraphics[width=1.0\linewidth]{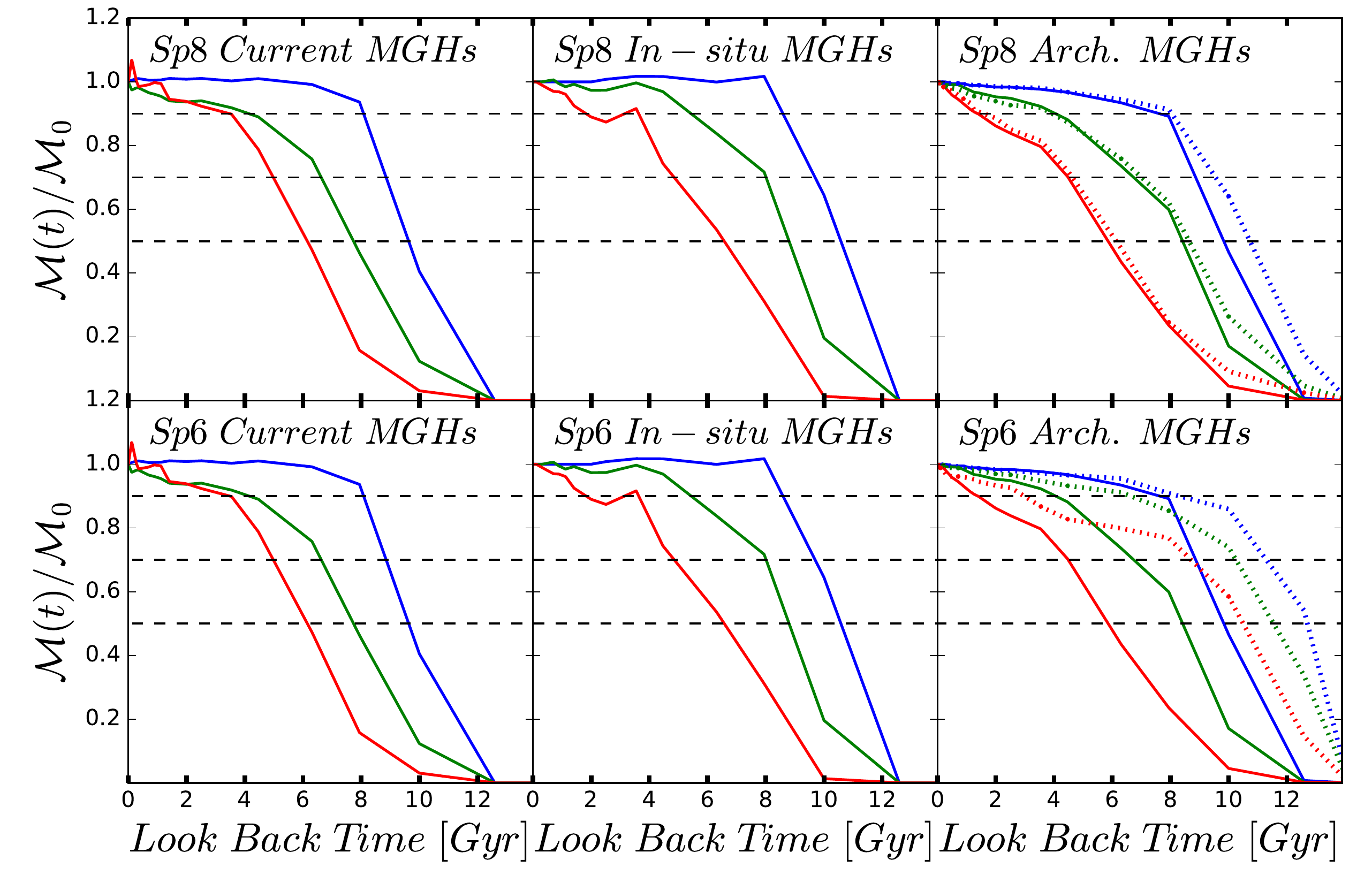}
\end{center}
\caption{Mean mass growth histories in three radial bins measured directly from the Sp8D (upper panels) and Sp6L (lower panels) simulations and normalized to their corresponding $z=0$ masses. From left to to right, the panels show the current, in-situ, and archaeological MGHs. Blue, green, and red lines are for the inner ($R/R_{\rm eff}<0.5$), intermediate ($0.5<R/R_{\rm eff}<1$), and external ($1<R/R_{\rm eff}<1.5$) regions, respectively.  The solid lines represent the MGHs measured within $\pm2$ kpc from the disk plane. For the archaeological MGHs, the dotted lines represent the MGHs taking into account all the stellar particles along the LOS. The LBT grid is the same of the stellar library templates used in Section \ref{reconstruction} for the fossil record method. } 
\label{simu_mghs}
\end{figure}

In \citet{Avila-Reese+2017} the authors obtained from their simulations the stellar mass growth rates (MGHs) in different ways. We summarize below how are defined these MGHs, and present the corresponding true determinations for the simulated Sp8D and Sp6L galaxies. These MGHs are compared with those inferred from the respective mock IFU data cubes in subsection \ref{radialMGHs}.  

The {\it current} MGH measures the total accumulated stellar mass at each snapshot (redshift) and radial bin of the simulation. This MGH takes into account the stellar mass between two snapshots that is added by in-situ star formation ($M_{\star,in-situ}$), acquired from outside ($M_{\star,in}$), lost (by radial mass transport, $M_{\star,out}$), and the mass lost by stellar winds ($M_{\star,ml}$), that is, $\Delta M_{\star}=M_{\star,in-situ}+M_{\star,in}-M_{\star,out}-M_{\star,ml}$ \citep[Eq. 1 in ][]{Avila-Reese+2017}. The {\it in-situ} MGH takes into account only the stellar mass formed in-situ and the mass lost by stellar winds, without considering the amount of stellar mass acquired from outside or lost previously, i.e., it does not take into account the terms $M_{\star,in}$ and $M_{\star,out}$. Finally, the {\it archaeological} MGH is calculated by using the age distribution of the stellar particles at $z=0$ (within a given radial bin). With this distribution, the fractions of mass contained in particles of different ages are used to calculate the cumulative mass as a function of age (or LBT), taking into account the stellar mass returned to the insterstellar medium \citep[for the stellar mass-loss function, see][]{Leitner+2011}.  The archaeological MGHs are conceptually similar to the MGHs obtained by the fossil record methods \citep{Perez+2013,Ibarra-Medel+2016}.

In Figure~\ref{simu_mghs} we plot the current, in-situ, and archaeological {\it normalized} MGHs for the Sp8D and Sp6L galaxies just to get a sense of the differences among these different definitions of the radial MGHs \citep[for more details, see][]{Avila-Reese+2017}. The MGHs are normalized to their respective masses attained at $z=0$.  Therefore, rather than the absolute MGHs what we compare among them are the {\it relative mass growth rates in the different radial bins}. In Figure~\ref{simu_mghs} we use the the same LBT grid as in the mock observations presented in Section \ref{reconstruction}, conditioned by the available stellar library SSP templates (see subsection \ref{datacube}).  
We take into account all the $z=0$ stellar particles within each one of the three radial bins defined in Section \ref{simulation-results} and within a height of $\pm2$ kpc from the disk plane. The Sp8D galaxy has an extended in time global MGH with a pronounced inside-out radial trend, while Sp6L presents an earlier global MGH than Sp8D and a less pronounced inside-out radial mode.

Comparing the in-situ and the current radial MGHs, we find little differences among them, both for the Sp8D and Sp6L galaxies. This shows that the stellar mass at each radial bin grows in average by in-situ star formation, without significant contribution of mergers and/or large-scale radial mass transport \citep[][]{Avila-Reese+2017}. In more detail, for the Sp8D run, the current inner normalized MGHs ($<1 R_{\rm eff}$) are slightly shifted to lower fractions at a given time than the corresponding 
in-situ MGHs. This is because, besides the growth by in-situ SF, the (inner) stellar mass grows by some accretion events (minor mergers).  For both galaxies, the in-situ MGHs show an evident inside-out growth mode at all epochs, with an abrupt slowdown of their star formation (quenching) at their inner regions.

The archaeological MGHs (within $\pm2$ kpc of the disk plane) show a similar trend to the current and in-situ MGHs. We calculate also the archaeological MGHs taking into account all the stellar particles along the LOS (dotted lines). For Sp8D, the inner MGH calculated along the LOS is $\sim 1$ Gyr earlier than the disk MGH. This is likely due to the contribution of old stars from the bulge component. In the case of run Sp6, since it has a prominent bulge, the differences between the disk and LOS MGHs are more significant (only at early epochs). Since real galaxies are observed along the LOS, here we will use the LOS archaeological MGHs.

\subsection{Projection effects} 
\label{projection}

\begin{figure}
\begin{center}
\includegraphics[width=\linewidth]{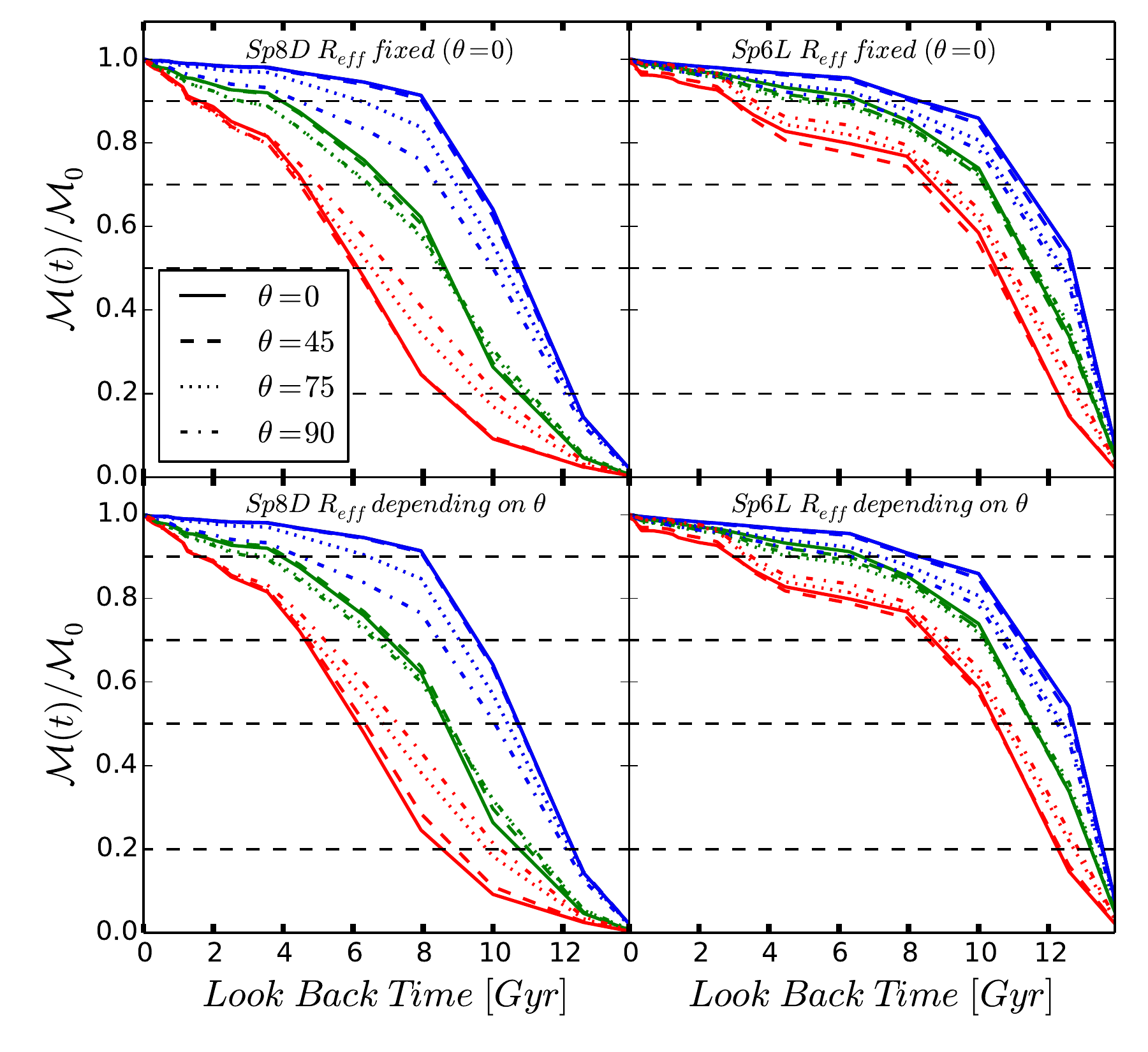}
\end{center}
\caption{Projection effects on the archaeological normalized radial MGHs. {\it Upper panels:} The MGHs of Sp8D and Sp6L for three radial bins (indicated in the inset) and at four different inclinations: $\theta=0^{\circ}, 45^{\circ}, 75^{\circ}$ and $90^{\circ}$ (solid, dashed, dotted, and dot-dashed lines, respectively). The $R_{\rm eff}$ used for the radial binning in all the cases is the one determined for the face-on position. At very high inclinations, the radial MGHs become less different among them than the true ones, i.e., the inside-out trend is diminished.
{\it Lower panels:} The same as in the upper panels but now using for the radial binning the $R_{\rm eff}$ corresponding to each inclination. The results are very similar to those shown in the upper panels.}
\label{simu_an}
\end{figure}

We explore the effects of projection on the measurement of the archaeological normalized radial MGHs. The MGHs are calculated using the stellar particles along the LOS of the galaxy projected in four directions:  $\theta=0^{\circ}$ (face-on), $\theta=45^{\circ}$, $\theta=75^{\circ}$ and $\theta=90^{\circ}$ (edge-on). First, we measure the MGHs in fixed radial bins (defined in terms of the $R_{\rm eff}$ calculated for the face-on case, as above) for all the inclinations.  As seen in the upper panels of Figure~\ref{simu_an}, due to the projection, the inner regions appear to have a later mass growth while the outer parts appear to have an earlier growth. This is because external (young) and internal (old) stellar particles contaminate the inner and outer MGHs along the LOS, respectively. 
Therefore, the projection effects tend to dimish the differences among the radial MGHs. However, the effect is important only for inclinations larger than $\sim 75^{\circ}$. For run Sp6L, the projection effects are even less important since this is a galaxy rounder than Sp8D. 

Next, we measure the radial MGHs defined in terms of the photometric $R_{\rm eff}$ calculated {\it for each} inclination. These radii are reported in Table \ref{tab2}.  As more inclined the galaxy, the $R_{\rm eff}$ fluctuate for the Sp6L and is smaller for the Sp8D. Therefore, the sizes of the radial bins now change with inclination. The lower panels of Figure~\ref{simu_an} show the corresponding radial MGHs. Comparing with the upper panels, we see that the differences are actually very small. Therefore, the radial binning for all the inclinations can be applied in terms of a same nominal radius, for example, the face-on radius $R_{\rm eff}$. 

\begin{figure*}
\begin{center}
\includegraphics[width=\linewidth]{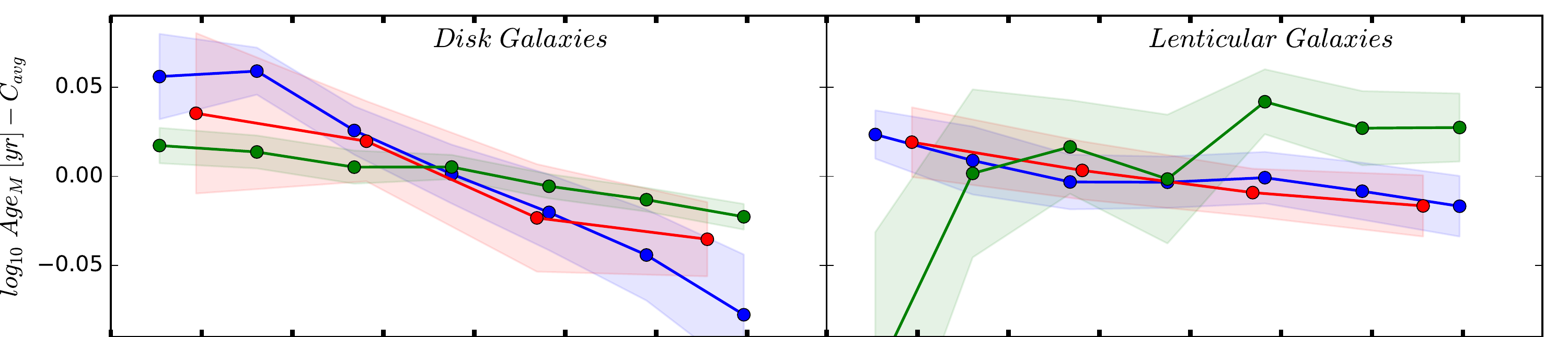}
\includegraphics[width=\linewidth]{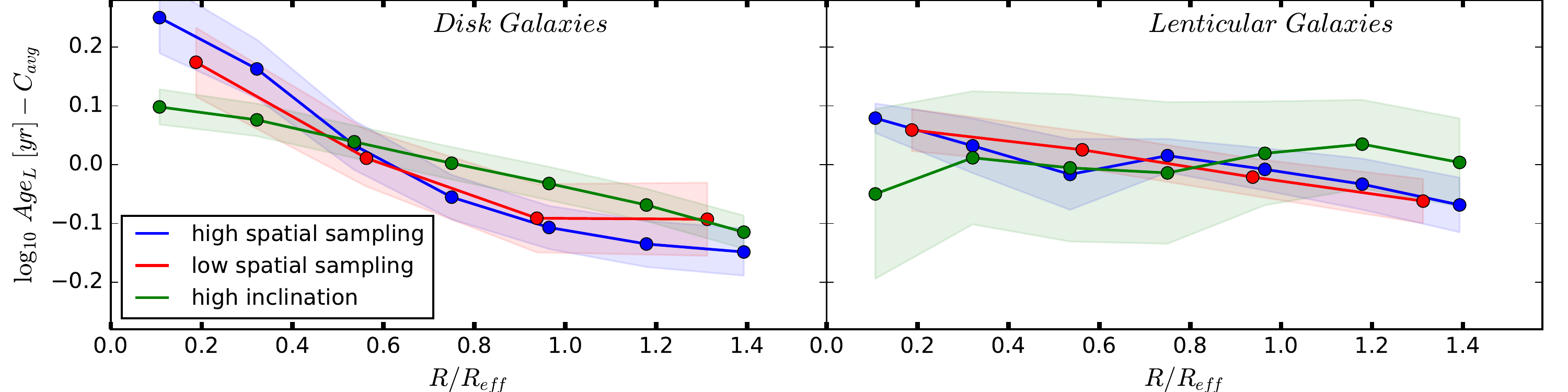}
\end{center}
\caption{{\it Upper panels:} Average and standard deviation of the mass-weighted age normalized profiles of MaNGA Milky Way-sized spiral (left) and lenticular (right) galaxies. Blue, red and green lines correspond to the subsamples  A, B, and C, respectively (see text). The shaded areas represent the standard deviations.   
{\it Lower panels:} As the upper panels but for the luminosity-weighted age profiles.}
\label{obs_M}
\end{figure*}


{
\section{Sampling and inclination effects on real observed galaxies}
\label{observa}

To explore the effects of spatial resolution/sampling and inclination on the archaeological inferences from real observed galaxies, we use here the inferred mass- and luminosity-weighted age radial profiles of Milky Way-sized galaxies taken from the publicly available MaNGA survey corresponding to the SDSS-DR14 \citep{abol18}, distributed as part of the {\sc Pipe3D} Value Added Catalog \citep{sanchez17b}\footnote{\url{https://www.sdss.org/dr14/manga/manga-data/manga-pipe3d-value-added-catalog/}}. From the original sample of 2,812 galaxies, we select those that have similar morphologies and stellar masses as the simulated Sp8D and Sp6L galaxies, respectively.  We select Sb-Sc (S0-S0a ) galaxies as spiral (lenticular) galaxies within a stellar mass range of $10<\log_{10}\mathcal{M}_{\star}<11$. For each group, we define three subsamples: A) galaxies observed with high spatial sampling (IFU bundles of 127 and 91 fibers) and low inclinations ($0^{\circ}<\theta<30^{\circ}$);  B)  galaxies observed with low spatial sampling  (IFU bundles of 37 and 19 fibers) and low inclinations;  and C) galaxies with high inclinations ($70^{\circ}<\theta<90^{\circ}$) and observed with high spatial sampling.  Each of these subsamples contain 27 (17), 11 (31) and 94 (8) disk (lenticular) galaxies. In Figure \ref{obs_M} we show the average radial profiles of the LW and MW ages and their standard deviations for each subsample. To make clearer the comparison the profiles were normalized to their respective global ages calculated within 1.5 $R_{eff}$ ($C_{avg}$). 

We observe that the age gradients are on average flatter for the low spatial-sampling observations than for the high spatial-sampling ones (blue vs. red lines), and for the highly-inclined galaxies as compared to the low-inclined ones (blue vs. green lines). This result is in qualitative agreement with the results obtained along this article, for the case of disk galaxies.
On the other hand, for our sub-sample of lenticular galaxies, its difficult to see differences on average values since these galaxies are dominated by old stellar populations at any galactocentric distance. Therefore, they already present nearly flat age gradients, as we have shown along this article. We observe that the gradients of high-inclined galaxies present a large scatter and even an apparent positive gradient on average. This behavior can be explained as a consequence of the projections effects as higher is the inclination, what again, is in agreement with the results based on the exploration of our simulations.}

\end{document}